\documentclass[journal=jacsat,layout=traditional]{achemso}

\usepackage[version=3]{mhchem} % Formula subscripts using \ce{}
\usepackage{caption}
\usepackage{subcaption}
\usepackage{array}
\newcolumntype{M}[1]{>{\centering\arraybackslash}m{#1}}
\usepackage{tabularx}
\usepackage[table]{xcolor}
\usepackage{multirow}
\usepackage{textcomp}
\usepackage[normalem]{ulem}
\usepackage{hyperref}
\hypersetup{
    colorlinks=true,
    linkcolor=blue,
    citecolor=blue}
\setkeys{acs}{doi = true}

\usepackage{tikz}
\usepackage{pgfplots}
\usepackage{pgfplotstable}
\usetikzlibrary{shapes.geometric, arrows}
\usetikzlibrary{patterns}
\usetikzlibrary{positioning}
\usetikzlibrary{matrix,calc,fit}

\pgfplotsset{compat=1.7}
\tikzstyle{startstop} = [rectangle, rounded corners, line width = 0.6pt, minimum width=3cm, minimum height=1cm,text centered, draw=black, fill=red!30]
\tikzstyle{long_startstop} = [rectangle, rounded corners, line width = 0.6pt, minimum width=4cm, minimum height=1cm, text width=4cm, text centered, draw=black, fill=red!30]
\tikzstyle{io} = [trapezium, trapezium left angle=70, trapezium right angle=110, line width = 0.6pt, minimum width=3cm, minimum height=1cm, text centered, draw=black, fill=blue!30]
\tikzstyle{process} = [rectangle, line width = 0.6pt, minimum width=3cm,  text width=3cm, minimum height=1cm, text centered, draw=black, fill=orange!30]
\tikzstyle{short_process} = [rectangle, line width = 0.6pt, minimum width=2.5cm,  text width=2.5cm, minimum height=1cm, text centered, draw=black, fill=orange!30]
\tikzstyle{long_process} = [rectangle, minimum width=4cm, line width = 0.6pt, minimum height=1cm, text centered, text width=4cm, draw=black, fill=orange!30]
\tikzstyle{very_long_process} = [rectangle, minimum width=5cm, line width = 0.6pt, minimum height=1cm, text centered, text width=5cm, draw=black, fill=orange!30]
\tikzstyle{decision} = [diamond, line width = 0.6pt, minimum width=3cm, minimum height=1cm, text centered, aspect=3, draw=black, fill=green!30]
\tikzstyle{arrow} = [thick,->,>=stealth]

\mciteErrorOnUnknownfalse

\newcommand{\rev}[1]{{\color{black} #1}}

\author{Shivang Agarwal}
\author{Clarice D. Aiello}
\affiliation[ucla-ece]
{Department of Electrical and Computer Engineering, University of California, Los Angeles}
\author{Daniel R. Kattnig}
\affiliation[uoe]
{Department of Physics and Living Systems Institute, University of Exeter}
\author{Amartya S. Banerjee}
\email{asbanerjee@ucla.edu}
\affiliation[ucla-mse]
{Department of Materials Science and Engineering, University of California, Los Angeles}

\SectionNumbersOn

\title[Posner]{The Dynamical Ensemble of the Posner Molecule is not Symmetric}

\keywords{Posner Molecule, Symmetry, Biological Qubit, Density Functional Theory}

\begin{document}

%\begin{tocentry}

% Some journals require a graphical entry for the Table of Contents.
% This should be laid out ``print ready'' so that the sizing of the

% Inside the \texttt{tocentry} environment, the font used is Helvetica
% 8\,pt, as required by \emph{Journal of the American Chemical

% The surrounding frame is 9\,cm by 3.5\,cm, which is the maximum
% permitted for  \emph{Journal of the American Chemical Society}
% graphical table of content entries. The box will not resize if the
% content is too big: instead it will overflow the edge of the box.

% This box and the associated title will always be printed on a
% separate page at the end of the document.

%\includegraphics[height=4.45cm,width=\textwidth]{TOC.pdf}

%\end{tocentry}

\begin{abstract}
The {Posner molecule}, \ce{Ca9(PO4)6}, has long been recognized to have biochemical relevance in various physiological processes. It has found recent attention for its possible role as a biological quantum information processor, whereby the molecule purportedly maintains long-lived nuclear spin coherences among its \ce{^{31}P} nuclei (presumed to be symmetrically arranged), allowing it to function as a room temperature qubit. The structure of the molecule has been of much dispute in the literature, although the \ce{S6} point group symmetry has often been assumed and exploited in calculations. Using a variety of simulation techniques (including \textit{ab initio} molecular dynamics and structural relaxation), rigorous data analysis tools and by exploring thousands of individual configurations, we establish that the molecule predominantly assumes low symmetry structures (\ce{C_s} and \ce{C_i}) at room temperature, as opposed to the higher symmetry configurations explored previously. Our findings have important implications on the viability of this molecule as a qubit.
\end{abstract}

%\section{Introduction}
%\attn{As per JPCL submission guidelines: preferred maximum length for each Letter is 2500 words or the equivalent (8 --10 double-spaced typewritten pages of text, 3–4 figures, and 1–2 schemes/illustrations). Are we satisfying these requirements closely enough ? The abstract should be 100-150 words, and I think we are slightly over at 158 words. Finally, the guidelines ask to avoid the use of section headers. This has to be fixed. See \href{https://publish.acs.org/publish/author_guidelines?coden=jpclcd}{here}.}. 

The calcium phosphate trimer, \ce{Ca9(PO4)6}, is of special biological interest. First discovered in the bone mineral hydroxyapatite in 1975 by Betts and Posner \cite{posner1975synthetic}, and henceforth coined as the \textit{Posner molecule} (PM), it is thought to form the structural building block of amorphous calcium phosphate \cite{treboux2000symmetry}. Its presence in simulated body fluids was confirmed by Onuma and Ito \cite{onuma1998cluster}, and its aggregation has been hypothesized to underpin bone growth \cite{yin2003biological, du2013structure, dey2010g, wang2012posner}. More recently, it has been proposed that the \ce{^{31}P} nuclear spins within PMs can maintain long-lived entanglement, and that this could play an important role in nervous excitation via synaptic \ce{Ca^{2+}} ion release \cite{fisher2015quantum, weingarten2016new, swift2018posner}. These and other studies \cite{player2018posner} have subsequently explored PMs as potential ``neural qubits'', drawing upon the fact that nuclear spin coherence times associated with these systems have been found to be exceptionally large per theoretical estimates. Such studies have suggested or assumed that the prototypical structure for the PM is one with an \ce{S_6} molecular point group symmetry, at least on the average \cite{swift2018posner}. Furthermore, in the presence of a well-defined rotation axis of the cluster (such as the three-fold \ce{C3} rotational symmetry of the supposed \ce{S6} symmetric cluster), the binding and unbinding of PMs could arguably act as a ``pseudospin" entangler of the nuclear spin states of multiple PMs, which is a necessary precondition for the ``quantum brain" concept, as suggested in Ref.\ \citenum{swift2018posner}.

In the context of the aforementioned mechanism, molecular point group symmetries for PMs are important, because they dictate the form of the spin-spin coupling network. The number of independent components in this network is directly related to the point group symmetry of the structure \cite{buckingham1970theory}. Certain molecular symmetries can render the six \ce{^{31}P} nuclei magnetically equivalent (\textit{e.g.},~\ce{S_6}), resulting in a small number of unique scalar ($J$) couplings (\textit{e.g.}, three unique coupling constants for \ce{S6}) \cite{player2018posner}. Understandably, other molecular symmetries could treat groups of \ce{^{31}P} nuclei as distinct, thus, resulting in a larger number of pertinent scalar couplings \cite{perras2013symmetry}. As the ability of the system to sustain long-lived spin coherences is starkly related to the asymmetry in the coupling network \cite{annabestani2018dipolar, lidar2012review, feng2014long, vinogradov2007long, stevanato2015long}, the spin physics of PMs is inherently linked to the molecule's point group symmetries. Further, the presence of an inversion center in the PM, as found for \ce{S6}, would render the intra-cluster dipolar coupling block-diagonal in a basis of well-defined parity under exchange of two \ce{^{31}P} nuclei related by inversion. \rev{As a consequence, the inclusion of a singlet-polarized diphosphate molecule in such a cluster could generate long-lived spin population differences, spared from fast spin relaxation by the dipolar coupling within the PM.}

The structure of the isolated PM is unknown. However, a series of studies \cite{treboux2000symmetry, treboux2000existence, swift2018posner, player2018posner} suggest a high degree of symmetry, e.g.~\ce{S6} and beyond, which was later exploited \cite{swift2018posner} in deriving the molecule's hypothetical entanglement-driven interaction mechanisms. \cite{lin2018structures,player2018posner}. That the symmetry of the cluster might in fact be lower, has, on the other hand, been pointed out as early as 2003 in the work of Yin et al.\ \cite{yin2003biological}, and then later in Refs.~\citenum{lin2018structures} and \citenum{ player2018posner}. \rev{Based on studies of the PM so far, two possibilities ought to be addressed: a) could the molecule exist as a stable entity, i.e.\ corresponding to a minimum on the Potential Energy Surface (PES), of high symmetry that is dominantly populated at physiological temperatures, as motivated by Refs.\ \citenum{treboux2000existence,treboux2000symmetry}, or b) could thermal fluctuations average the molecule's geometry to an effective structure of high symmetry, i.e.\ of an overall \ce{S6} symmetry, despite the ensemble comprising predominantly low symmetry states, as advocated in \citenum{swift2018posner} ? Ref.\ \citenum{swift2018posner} reports the energy difference between \ce{S6} and \ce{C1} to be $0.06$ eV = $1.53$ meV/atom = $2.33\;k_{B}T$, while Ref.\ \citenum{treboux2000existence} suggests an alternate value of $0.13$ eV = $3.33$ meV/atom = $5.06\;k_{B}T$ between \ce{S6} and \ce{C2}, suggesting that these structures are thermally accessible, thereby highlighting the importance of probing hypothesis b) by studying the dynamical ensemble properties.} Indeed, Swift et al.\cite{swift2018posner}, while utilizing the \ce{S6}-symmetry in deriving the specifics of the quantum brain hypothesis, acknowledge the existence of multiple more stable structures of lower symmetries. %\rev{\sout{ and the energy differences between the higher and lower symmetry structures reported by them are greater than what would be provided by thermal fluctuations at room temperature.}} 
Moreover, it was understood that they considered non-equilibrated structures, assumed to exist in the average, for their calculations \cite{Swift2020pvtcomms}.  Naturally, considering a more symmetric yet less stable molecular structure has important implications on its spin properties, and prior works \cite{treboux2000symmetry, treboux2000existence, kanzaki2001calcium} could have suffered from the use of poor basis sets available at the time and the use of molecular force fields rather than \textit{ab initio} methods. Molecular dynamics-based studies \cite{mancardi2017detection} using improved force-fields \cite{ainsworth2012polarizable, demichelis2018simulation} suggest the less symmetric point group \ce{C3} for the molecule as well. Thus, given the conflicting state of the literature on the one the hand, and the importance of the existence of highly symmetric Posner clusters in support of recent quantum biological hypotheses on the other hand, we were prompted to conclusively re-examine the structure of the PM.

%\section{Methodology and Results}

%\subsection{Outline}
Since the PM is a calcium phosphate trimer, we performed initial simulations on the monomer \ce{Ca3(PO4)2} and dimer \ce{Ca6(PO4)4} configurations to validate our simulation setup. As detailed in the Supporting Information (SI), the symmetry of the optimized monomer structure agreed with previous studies \cite{treboux2000existence, kanzaki2001calcium}. For the dimer, the work of Kanzaki et al.\ \cite{kanzaki2001calcium} was followed closely and many of their structures were replicated. These results gave us confidence to pursue the structural symmetry of the PM, which is discussed below.

As the PM is hypothesized to exist in a variety of molecular symmetries \cite{swift2018posner, kanzaki2001calcium, treboux2000existence}, we set out to identify symmetric minima on the PES of the molecule. To cover an appreciable portion of the PES of the PM, we used various techniques to set up the initial atomic positions, and different analysis techniques to extract the relevant data, as detailed below.
%\rev{\sout{Firstly, as suggested in previous studies \cite{kanzaki2001calcium}, all atoms constituting the PM were arranged on a cube of appropriate dimensions. The nine \ce{Ca} atoms were placed in a body-centred cubic arrangement and the six \ce{PO4} groups in a face-centred cubic arrangement. The size of the cube was chosen such that the length of the diagonal was close to $9$\AA, which is the approximate diameter of the PM~ \cite{player2018posner,swift2018posner,roohani2021build}. Countless configurations of the cluster can be realized by rotating each \ce{PO4} group around its center. In order to extensively sample the configuration space, the phosphate groups were rotated in steps of $30$\textdegree\ in 3 dimensions to create over $2,800$ structures. Additionally, the coordinates of the calcium atoms and the \ce{PO_4} groups were scaled with reference to the central \ce{Ca} atom to account for the possibility of some previously built structures being over-strained. This resulted in over $10,000$ viable structures. A schematic representation of the method can be seen in Fig.~\ref{fig:rotation_scaling_of_PM}. In all cases, once the molecular structure was optimized using \textit{ab initio} structural relaxation, the resultant configuration had a very low symmetry, \textit{i.e.}, either \ce{C_s}, \ce{C_i}, or no symmetry (\ce{C_1}). Fig.~\ref{fig:30k_bar} shows the distribution of the observed symmetries for the relaxed structures obtained by the aforementioned strategy. It is evident that the PES of the PM is dominated by low symmetry structures.}}

\begin{figure}
    \centering
    \begin{subfigure}[b]{0.48\textwidth}
    \centering
    \includegraphics[width=0.9\textwidth]{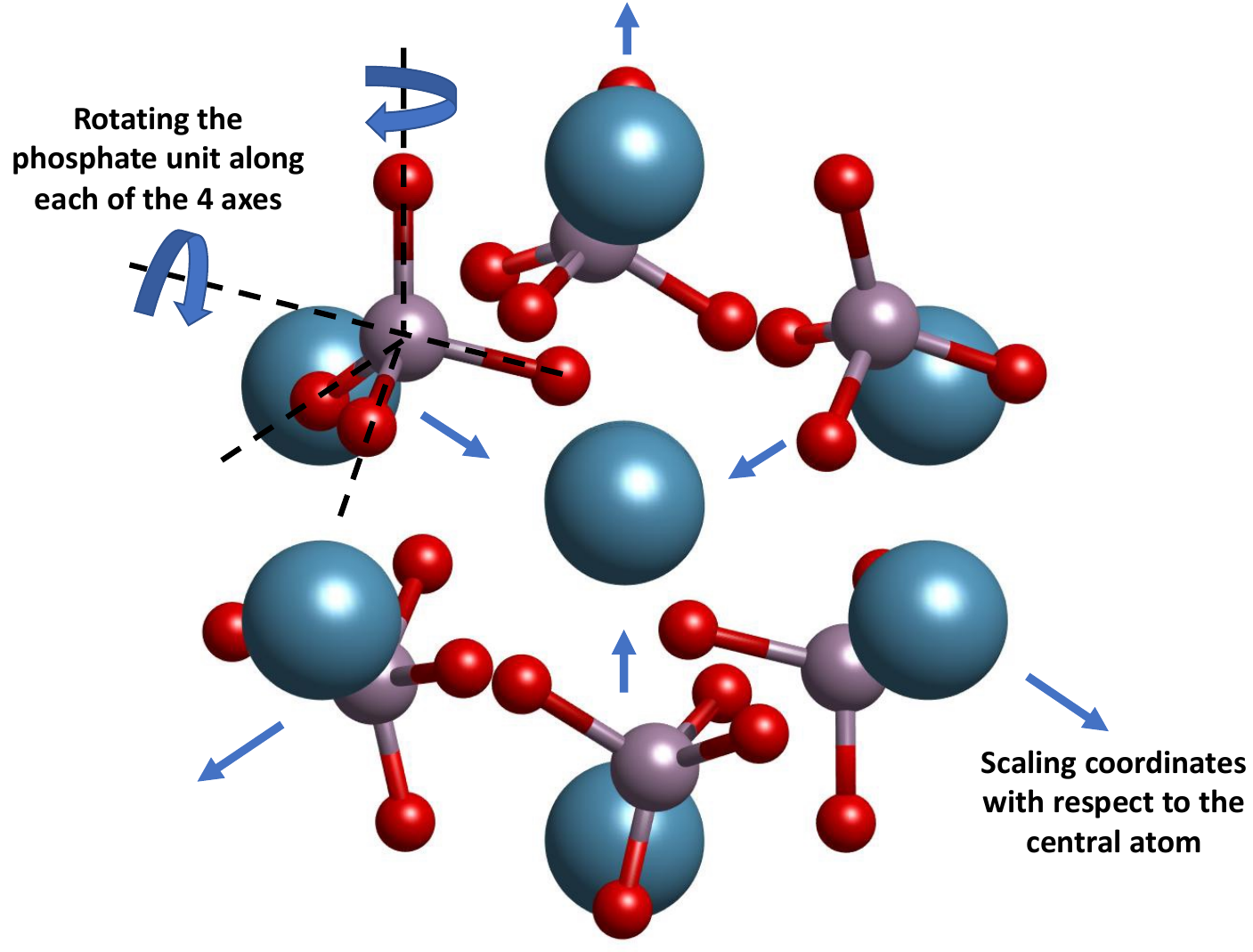}
    \vspace{0.8cm}
    \caption{}
    \label{fig:rotation_scaling_of_PM}
    \end{subfigure}
    \begin{subfigure}[b]{0.48\textwidth}
    \centering
    \begin{tikzpicture}
    \begin{axis}[ybar,
    symbolic x coords={\ce{C_1}, \ce{C_s}, \ce{C_i}, \ce{C_2}, \ce{C_{2h}}, \ce{D_{2h}}},
    xtick=data,
    xlabel = {Resultant symmetry of the relaxed structure},
    ylabel = {Percentage occurrence},
    xlabel style = {font=\footnotesize},
    ylabel style = {font=\footnotesize},
    yticklabel style = {font=\footnotesize},
    xticklabel style = {font=\footnotesize},
    width = 7cm,
    height = 6.5cm,
    ymin = 0,
    bar width = 0.7cm,
    point meta = explicit symbolic,
    visualization depends on=y\as\DataY,
    nodes near coords = {\pgfmathprintnumber[fixed]{\DataY} \pgfplotspointmeta},
    nodes near coords align={vertical},
    every node near coord/.append style={font=\scriptsize}]
        \addplot[blue!20!black,fill=blue!80!white] coordinates {(\ce{C_1},16.5623) (\ce{C_s},79.2984) (\ce{C_i},3.0717) (\ce{C_2},0.0643) (\ce{C_{2h}},0.2398) (\ce{D_{2h}},0.7635)};
    \end{axis}
    \end{tikzpicture}
    \caption{}
    \label{fig:30k_bar}
    \end{subfigure}
    \caption{(a) The scheme used for creating over $10,000$ structures by rotating the phosphate units and scaling all the coordinates with respect to central atom \rev{The blue, purple, and red spheres represent \ce{Ca}, \ce{P}, and \ce{O} atoms respectively.} (b) Percentage occurrence of each point group symmetry after \textit{ab initio} structural relaxation of over $10,000$ initial structures (DFT with B3LYP hybrid functional and 6-311G(d,p) basis set).}
\end{figure}

\rev{Firstly, a wide variety of possible structures were created by modifying some of the techniques used in previous studies \cite{kanzaki2001calcium}. This resulted in over $10,000$ viable structures. A schematic representation of the method can be seen in Fig.~\ref{fig:rotation_scaling_of_PM} and further details are provided in the SI. In all cases, once the molecular structure was optimized using \textit{ab initio} structural relaxation, the resultant configuration had low symmetry, \textit{i.e.}, either \ce{C_s}, \ce{C_i}, or no symmetry (\ce{C_1}). Fig.~\ref{fig:30k_bar} shows the distribution of the observed symmetries of the relaxed structures obtained by the described strategy. It is evident that the PES of the PM is dominated by low symmetry structures.}

\rev{Secondly, the atoms were arranged in relatively high symmetry configurations ``by hand'' without giving any consideration to the existence and initial stability of the structure, or to the forces on the individual atoms. The rationale was that since none of the previous structures resulted in one of the high-symmetry structures reported in earlier studies, the molecule might instead transition into one of these high-symmetry structures if the starting configuration was constrained in symmetry. Structures with symmetries such as \ce{S6}, \ce{T_h}, \ce{C_{3v}}, and \ce{D_{3d}} were constructed. The molecular structures were perturbed slightly from their original sysmetiries and optimized. However, for all the four above-mentioned symmetries, when subjected to structural relaxation, the molecule failed to retain or increase the point-group symmetry and, instead, tumbled down to a low-symmetry structure -- \ce{C_i} or \ce{C_s} -- as was also the case in our previous approach. Structural relaxation with solvent effects included via a Polarizable Continuum Model (PCM) \cite{truong1995new,barone1998quantum} did not result in high-symmetry structures either. Finally, to verify that our results were not an artifact of the particular basis set, exchange correlation functional or simulation software in use, we repeated our calculations using a semi-local exchange correlation functional and a plane-wave DFT code, and obtained very similar results.}

\rev{Next, noting that the \ce{S6} symmetry has been so widely discussed and accepted, an \ce{S6} symmetric structure was built using an alternative technique. The phosphate units were considered as rigid tetrahedrals. The force field developed by Demichelis et al.\cite{demichelis2018simulation} was used to model the atomic interactions. Imposing the \ce{S6} symmetry enables the parameterization of the structure in terms of $10$ parameters. Following a symmetry constrained global minimization of the system's energy based on the above parameters, a unique structure was obtained, details of which can be found in the SI. Using DFT with the B3LYP hybrid functional and a 6-311G(d,p) basis, it was identified as a transition state structure, and subjected to further geometrical relaxation. The optimized structure exhibited a \ce{C_i} symmetry -- starkly lower in symmetry than the initial \ce{S6} symmetry. When performed under constrained symmetry, the same \textit{ab initio} geometrical relaxation calculation failed to reach self-consistency. Thus, the optimization of symmetric structures modeled on existing force fields also failed to produce stable structures with low symmetries.}

\begin{figure}[htb]
    \centering
    \begin{tikzpicture}[image/.style = {text width=0.22\textwidth, inner sep=0pt, outer sep=0pt}, node distance = 1mm and 1mm] 
\node [image] (frame1)
    {\includegraphics[width=\linewidth]{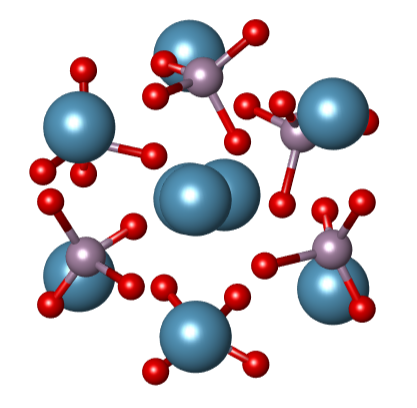}};
\node [draw,above=of frame1] {\textbf{A}};
\node [image,right=of frame1] (frame2) 
    {\includegraphics[width=\linewidth]{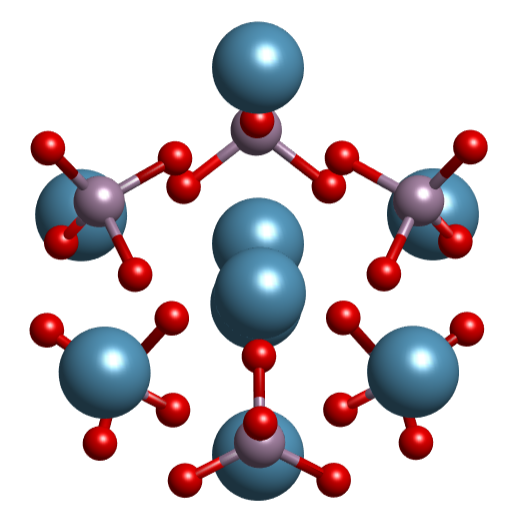}};
\node [draw,above=of frame2] {\textbf{B}};
\node[image,right=of frame2] (frame3)
    {\includegraphics[width=\linewidth]{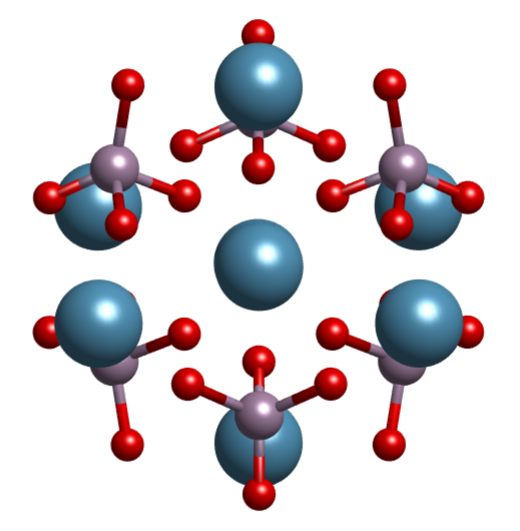}};
\node [draw,above=of frame3] {\textbf{C}};
\node[image,right=of frame3] (frame4)
    {\includegraphics[width=\linewidth]{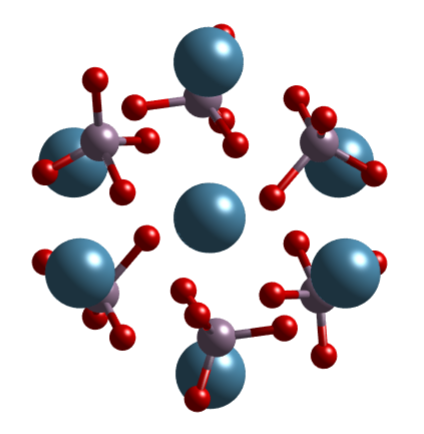}};
\node [draw,above=of frame4] {\textbf{D}};
\node [image, below=of frame1] (frame5)
    {\includegraphics[width=\linewidth]{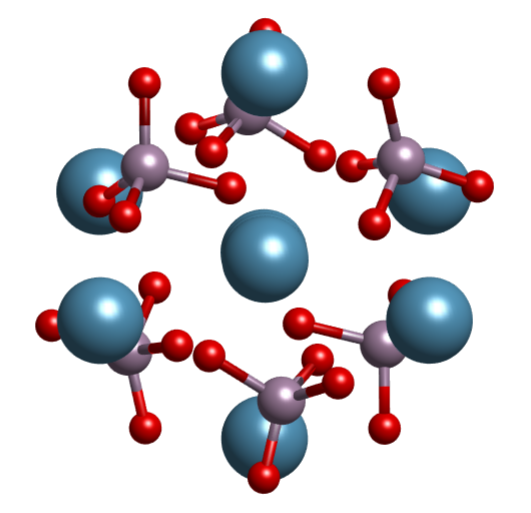}};
\node [draw,below=of frame5] {\textbf{E}};
\node [image,right=of frame5] (frame6) 
    {\includegraphics[width=\linewidth]{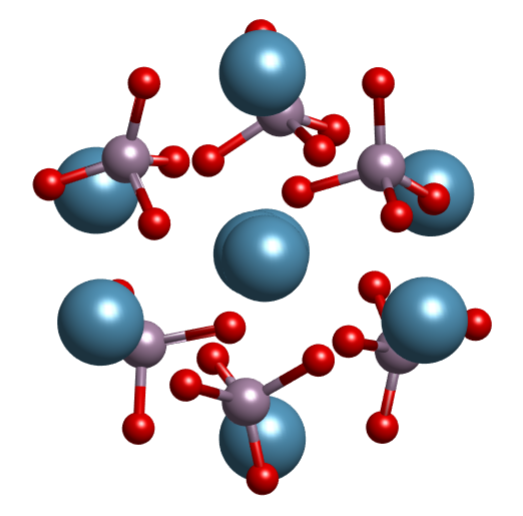}};
\node [draw,below=of frame6] {\textbf{F}};
\node[image,right=of frame6] (frame7)
    {\includegraphics[width=\linewidth]{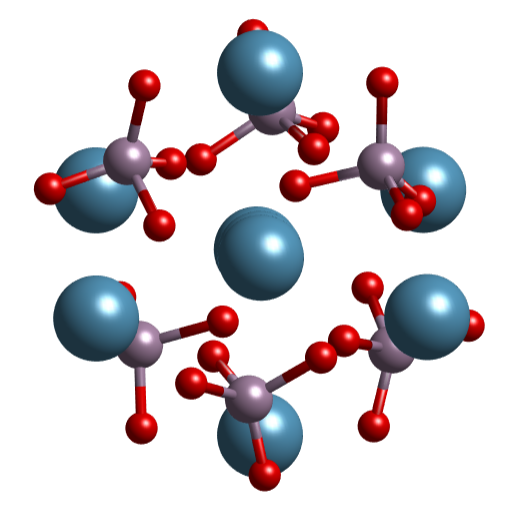}};
\node [draw,below=of frame7] {\textbf{G}};
\node[image,right=of frame7] (frame8)
    {\includegraphics[width=\linewidth]{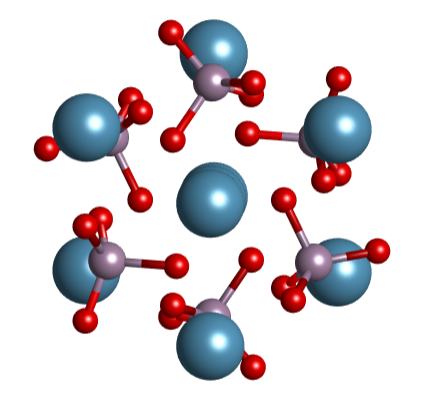}};
\node [draw,below=of frame8] {\textbf{H}};
\end{tikzpicture}
    \caption{The eight different structures that were used as starting geometries for dynamical simulations of the PM. The labels (\textbf{A}--\textbf{H}) used in the manuscript are stated.}
    \label{fig:all_structs}
\end{figure}

As our exhaustive search using structural relaxation failed to identify symmetric species, we then studied the dynamical properties of eight semi-stable Posner structures by \textit{ab initio} molecular dynamics (AIMD). Starting structures were obtained from the supposedly minimum energy structures in Ref.~\citenum{treboux2000existence}, resymmetrized and subjected to \textit{ab initio} structural optimisation (using DFT with the B3LYP hybrid functional and the BP86/Def2TZVPP/W06 basis set). At the end of the relaxation procedure, they had low forces on the atoms (of the order of $10^{-4}\;\text{eV}/$\AA). However, these structures did not correspond to energy minima, but were transition states of higher order instead. Distorting these structures along the normal modes associated with imaginary frequencies followed by further optimization of these structures resulted in the molecule tumbling to lower symmetries, as did the formation of these structures without the symmetry constraints. We considered the 8 unique molecular structures derived through the above procedure. These, along with the corresponding molecular point group symmetries, as obtained by the Visual Molecular Dynamics (VMD) software \cite{HUMP96} are displayed in Fig.~\ref{fig:all_structs}. \rev{We also performed vibrational spectrum calculations on these transition state structures, and compared them with an existing spectrum \cite{swift2018posner}, details of which can be found in the SI.} The structures were then studied by AIMD using the same basis set and DFT functional as above, at two different temperatures, $298K$ (room temperature) and $315K$ (at the higher end of human metabolic temperatures), to see if they maintain their high symmetry or give rise to high-symmetry species in the time average. The molecules were allowed to evolve for a total time of about $24$ ps, in which the first $1.2$ ps -- or the first $5$\% of the total time -- were considered to be the equilibration phase of the molecule, and were not considered for subsequent analyses. \rev{Further details related to this particular choice of total simulation time is available in the SI.}

At the end of each of the eight dynamical runs, and at both the above-mentioned temperatures, the PM exhibited either a \ce{C_s} or a \ce{C_i} point group symmetry. Thus, it was understood that, even if the molecule is forced to exist in a relatively higher symmetry state, as imposed by our method, it will naturally tumble down to one of the lower symmetry states within picoseconds. This leads us to argue that any stable configuration of the PM is likely to exhibit lower symmetries at room temperature.

\begin{center}
\footnotesize
\begin{tabular}{|M{1.9cm}|M{2.9cm}|M{2.5cm}|M{2.8cm}|M{3.9cm}|}
  \hline
  \textbf{Transition structure index} & \textbf{Symmetry prior to the creation of transition structre} & \textbf{Transition structure point group symmetry} & \textbf{Formation energy (in \textit{eV}) of the transition structure} & \textbf{Resultant symmetries over dynamical runs} \\
 \hline
 A & \ce{C_s} & \ce{C_s} & -271.660 & \ce{C_1},\ce{C_2},\ce{C_{s}},\ce{D_{2h}} \\ 
 B & \ce{C_{3v}} & \ce{C_{3v}} & -269.555 & \ce{C_1},\ce{C_2},\ce{C_s},\ce{C_{2h}},\ce{C_{3v}},\ce{D_{2h}} \\ 
 C & \ce{D_{3d}} & \ce{D_{3d}} & -264.997 & \ce{C_1},\ce{C_s},\ce{C_i},\ce{T},\ce{C_{2h}},\ce{D_{2h}},\ce{D_{3d}} \\ 
 D & \ce{D_{3d}} & \ce{C_{2h}} & -269.551 & \ce{C_1},\ce{C_s},\ce{C_i},\ce{T},\ce{C_{2v}},\ce{C_{2h}},\ce{D_{2h}} \\ 
 E & \ce{S6} & \ce{C_i} & -271.552 & \ce{C_1},\ce{C_s},\ce{C_i},\ce{C_2},\ce{T},\ce{C_{2h}},\ce{D_{2h}} \\ 
 F & \ce{S6} & \ce{C_i} & -271.540 & \ce{C_1},\ce{C_s},\ce{C_i},\ce{T},\ce{C_{2h}},\ce{D_{2h}} \\ 
 G & \ce{S6} & \ce{D_{2h}} & -271.239 & \ce{C_1},\ce{C_s},\ce{C_i},\ce{T},\ce{C_{2h}},\ce{D_{2h}},\ce{O_h} \\ 
 H & \ce{T_h} & \ce{D_{2h}} & -269.531 & \ce{C_1},\ce{C_s},\ce{C_i},\ce{C_2},\ce{C_{2v}},\ce{C_{2h}},\ce{D_{2h}} \\ 
 \hline
\end{tabular}
\label{tab:all_structs}
\captionof{table}{The point group symmetries for each of the structures in Fig.~\ref{fig:all_structs}, their formation energies, as well as the point group symmetries displayed by each configuration during a dynamical simulation/evolution over $22.8$ {ps}. In comparison, the formation energy for a monomer calcium phosphate was calculated to be $-84.244$ \textit{eV.} The listed formation energies of all trimer configurations are lower than three times this value.}
\end{center}

\begin{figure}[ht]
    \centering
    \begin{subfigure}[h!]{0.48\textwidth}
    \centering
    \includegraphics[width=0.9\textwidth]{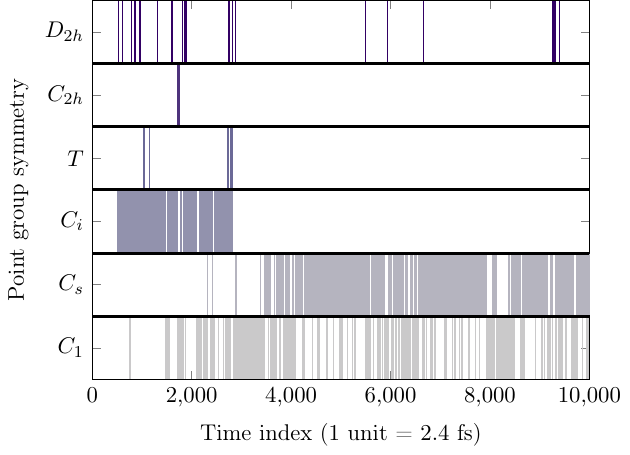}
    \caption{ }
    \label{fig:ts_C}
    \end{subfigure}
    \hfill
    \begin{subfigure}[h!]{0.48\textwidth}
    \centering
    \includegraphics[width=0.9\textwidth]{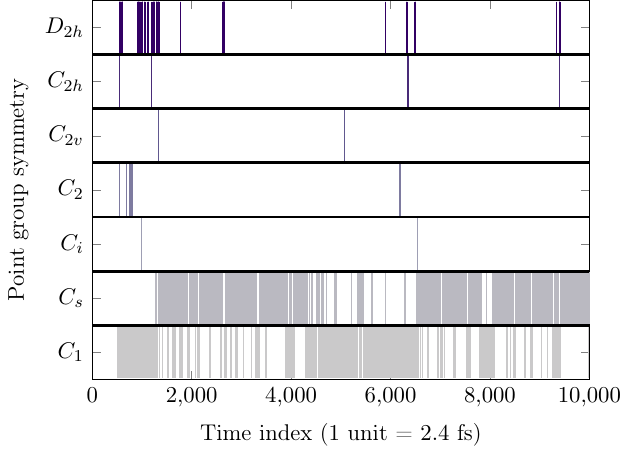}
    \caption{ }
    \label{fig:ts_H}
    \end{subfigure}
    \\
    \begin{subfigure}[b]{0.48\textwidth}
    \centering
    \begin{tikzpicture}
    \begin{axis}[ybar,
    symbolic x coords={\ce{C_{1}}, \ce{C_{s}}, \ce{C_{i}}, \ce{C_{2}}, \ce{T}, \ce{C_{2h}}, \ce{D_{2h}}},
    xtick=data,
    xlabel = {},
    ylabel = {Percentage occurrence},
    xlabel style = {font=\tiny},
    ylabel style = {font=\scriptsize},
    yticklabel style = {font=\tiny},
    xticklabel style = {font=\tiny},
    ymin = 0,
    bar width = 0.3cm,
    width = 7.5cm,
    legend style={font=\tiny}]
        \addplot[red!20!black,fill=red!80!white] coordinates { (\ce{C_{1}},19.78) (\ce{C_{s}},74.88) (\ce{C_{i}},0) (\ce{C_{2}},0.06) (\ce{T},0) (\ce{C_{2h}},0) (\ce{D_{2h}},0.28)};
        \addplot[blue!20!black,fill=blue!80!white] coordinates { (\ce{C_{1}},30.41) (\ce{C_{s}},45.16) (\ce{C_{i}},17.45)  (\ce{C_{2}},0) (\ce{T},0.31)  (\ce{C_{2h}},0.33) (\ce{D_{2h}},1.34)};
        \legend{$T=315\;K$, $T=298\;K$}
    \end{axis}
    \end{tikzpicture}
    \caption{ }
    \label{fig:e_bar_C}
    \end{subfigure}
    \hfill
    \begin{subfigure}[b]{0.48\textwidth}
    \centering
    \begin{tikzpicture}
    \begin{axis}[ybar,
    symbolic x coords={\ce{C_{1}}, \ce{C_{s}}, \ce{C_{i}}, \ce{C_{2}}, \ce{C_{2v}}, \ce{C_{2h}}, \ce{D_{2h}}},
    xtick=data,
    xlabel = {},
    ylabel = {Percentage occurrence},
    xlabel style = {font=\tiny},
    ylabel style = {font=\scriptsize},
    yticklabel style = {font=\tiny},
    xticklabel style = {font=\tiny},
    ymin = 0,
    bar width = 0.3cm,
    width = 7.5cm,
    legend style={font=\tiny}]
        \addplot[red!20!black,fill=red!80!white] coordinates { (\ce{C_{1}},21.21) (\ce{C_{s}},73.04) (\ce{C_{i}},0) (\ce{C_{2}},0.03) (\ce{C_{2v}},0) 
        (\ce{C_{2h}},0) (\ce{D_{2h}},0.72)};
        \addplot[blue!20!black,fill=blue!80!white] coordinates { (\ce{C_{1}},36.6)  (\ce{C_{s}},55.42)  (\ce{C_{i}},0.03)  (\ce{C_{2}},0.51)  (\ce{C_{2v}},0.03)  (\ce{C_{2h}},0.11)  (\ce{D_{2h}},2.27)};
        \legend{$T=315\;K$, $T=298\;K$}
    \end{axis}
    \end{tikzpicture}
    \caption{ }
    \label{fig:e_bar_H}
    \end{subfigure}
        \caption{Time persistence of symmetries and the associated frequency of occurrence of each symmetry for two different starting configurations over a dynamical run. In the above figure, (a) and (c) represent the data for configuration \textbf{C}, whereas (b) and (d) represent data for configuration \textbf{H}. Note that (a) and (b) represent the data at $T=298K$, and thus some additional symmetries which were observed at $T=315K$ are not present in these plots.}
    \label{fig:bar_and_time_for_sym}
\end{figure}

Time averages and temporal variations in the molecule's symmetry were then studied as it appeared plausible that the molecule might exhibit a symmetric structure in the temporal average. Specifically, we studied the point group symmetries and energies of the $\mathcal{N}=9,500$ structures
generated during each AIMD run. This allowed us to infer time persistence of the molecule's symmetry, if present. We observed that the molecule does indeed exhibit a variety of symmetries within the time frame considered, as visualized in Fig.~\ref{fig:ts_C} and Fig.~\ref{fig:ts_H}. However, the higher symmetries were observed only fleetingly, i.e.\ on time scales of the order of $100$ fs --- too short to be significant, both as an independent species, or to markedly determine the average structure. This supports our claim that the PM prefers to exist in low molecular symmetries. Moreover, as exemplified by Fig.~\ref{fig:bar_and_time_for_sym}, we can say that the behavior described above of low symmetry configurations throughout the dynamic evolution is consistent among all the eight unique structures in Fig.~\ref{fig:all_structs}. Further details can be found in the SI.

\begin{figure}
  \centering
  \begin{tabular}{c}
%   \scalebox{0.65}{\input{all_md}}
  \includegraphics[width=0.505\textwidth,height=7.45cm]{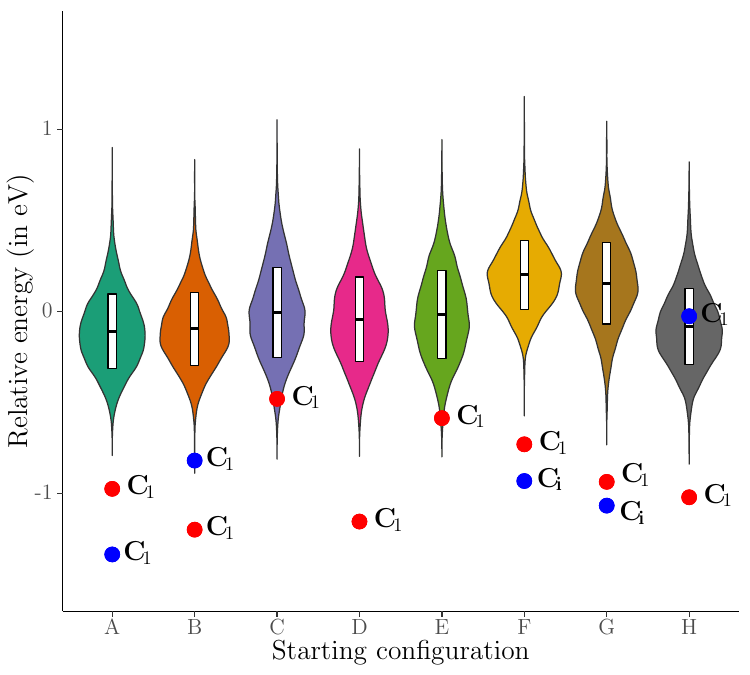}
  \\[-7.60cm]
%   \scalebox{0.65}{\input{seg_md}}
  \includegraphics[width=0.505\textwidth,height=7.45cm]{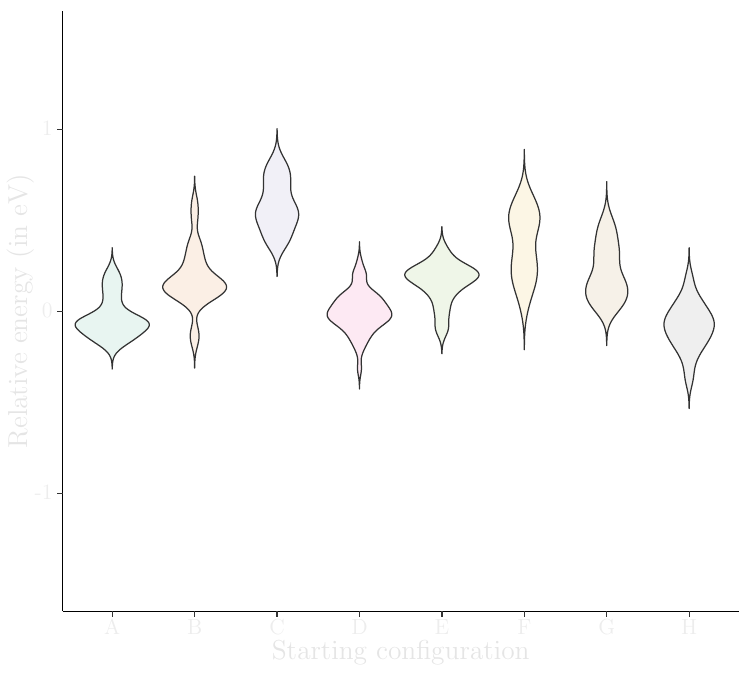}
  \end{tabular}
  \captionsetup{font={small}}
  \caption{The energy spread of the dynamical runs for each of the starting configurations listed in Fig.~\ref{fig:all_structs} represented by kernel density estimations in the form of vertical violin plots. The blue dots represent the energy of the time-average structure over the entire dynamical run, and the red dot represents the energy of the time-average structure of that subset of the high symmetry phase that is present for the longest continuous duration. The point-group symmetry of the averages is indicated next to the symbols. The semi-transparent overlays on each of the violin plots represent the energy distribution for the entire high symmetry phase. The most stable high symmetry phase for each of the cases were  \textbf{\ce{D_{2h}}}, \textbf{\ce{D_{2h}}}, \textbf{\ce{C_{2h}}}, \textbf{\ce{D_{2h}}}, \textbf{\ce{C_{2h}}}, \textbf{\ce{D_{2h}}}, \textbf{\ce{D_{2h}}}, and \textbf{\ce{D_{2h}}}, respectively. Dots missing from the graph were found at energies higher than the scale of the figure. The plot also shows the mean and standard deviation of the energy spread within the representations of probability density functions in the form of white boxes. The energies reported are within the numerical accuracy of the method and basis set used. \cite{weigend2005balanced,weigend2006accurate}.}
  \label{fig:violin}
\end{figure}

Even if the PM exhibits higher symmetries only fleetingly, it may \textit{appear} as a more symmetric structure in the time average. To test this possibility, time-averaged structures for \rev{entire} dynamical runs were created and studied. To this end, the translational and rotational motion of the molecule was eliminated via a rigid-body realignment procedure, and the aligned $\mathcal{N}$ structures averaged. A single point calculation was realized for these time-averaged structures and their energies were compared. \rev{In addition to averaging all the $\mathcal{N}$ structures, an average structure was also obtained for subsets of potentially higher symmetry structures in the dynamic runs, which we define as the most symmetric molecular point group observed in a single dynamical run -- e.g., \ce{D_{2h}} for the cases of Fig.~\ref{fig:ts_C} and Fig.~\ref{fig:ts_H}. Specifically, the subset of structures chosen for averaging from the entire range of these high-symmetry phases of structures were the longest consecutively occuring group of structures. For instance, in Fig.~\ref{fig:ts_H}, this would correspond to the group of \ce{D_{2h}} structures located just after the $1000^{\textrm{th}}$ time index. Note that we refrained from obtaining an average structure for all of the high-symmetry structures because different high-symmetry structures might have emerged thoughout the AIMD runs.}. Fig.~4 provides a representation of of the energy distributions over the AIMD runs together with a comparison of single-point energies and point group symmetries of the time-averaged structures. It can be seen that the energy spread between all averaged structures is relatively small -- about $1.36$ eV on average. \rev{Whenever the $\mathcal{N}$ intermediate structures were averaged, regardless of the initial configuration of the molecule, the point group symmetry of the averaged structure was either \ce{C_1}, \ce{C_i}, or \ce{C_s}, which, for the purposes of this study, have been considered as ``low symmetries''}. Their energies are shown as blue dots.  This further reinforces the hypothesis that the PM prefers to exist in lower point group symmetries at any time as well as on average, at biologically relevant temperatures. \rev{Additionally, when looking at the energies of only the temporary high-symmetry phases of the PM which, in Fig.~\ref{fig:violin}, have been represented by the semi-transparent violin plots overlayed on top of the opaque plots which represent the energies of the entire dynamical run, one can see that the energy of the structures with a higher symmetry is higher than the average energy %\sout{those with a lower symmetry}, 
which agrees with our claim.} However, seemingly in contrast to the above observation, we observe that the time-average structure of the higher symmetry phase can have a lower energy than the time-average structure of the entire dynamical run. This, however, can be explained by the fact that in every studied case, the time-averaging of even a high-symmetry phase yielded a structure with low (\ce{C_i} or \ce{C_s}) symmetry. Thus, the lower energy individual structures always possess low symmetries and the data do not invalidate the claim that lower symmetries are preferred. We also mention in passing that we have carried out our point group symmetry analysis using more than one software package (VMD and WebMO \cite{schmidt2020webmo}) and obtained similar results. Further details can be found in the SI.

Based on the above analyses, we identify the most stable structure of the PM to be the overall time-average structure of the configuration \textbf{A}, which can be seen as a blue dot in Fig.~\ref{fig:violin}. Notably, the point group symmetry for this structure is \ce{C1}, \textit{i.e.},~no symmetry. At the same time, we reiterate that the PMs possibly exists in an ensemble of structures, and that identifying a singular structure as the dominant PM structure is incorrect. More information about configuration \textbf{A} and its atomic coordinates can be found in the SI.

\begin{figure}
    \centering
    \begin{subfigure}[b]{0.4\textwidth}
    \centering
    \begin{tikzpicture}
    \begin{axis}[ybar,
    symbolic x coords={$5^{th}$, $4^{th}$, $3^{rd}$, $2^{nd}$, $1^{st}$},
    xtick=data,
    nodes near coords,
    xlabel = {Eigenmode index},
    ylabel = {Percentage of varaince \\ explained by the structure},
    xlabel style = {font=\footnotesize},
    ylabel style = {font=\footnotesize},
    ylabel style = {align=center},
    yticklabel style = {font=\footnotesize},
    xticklabel style = {font=\footnotesize},
    x dir = reverse,
    width = 6.5cm,
    height = 6cm,
    ymin = 0,
    ymax = 80,
    nodes near coords align={vertical},
    bar width = 0.7cm,
    every node near coord/.append style={font=\scriptsize}]
        \addplot[blue!20!black,fill=blue!80!white] coordinates {($5^{th}$,1.93039) ($4^{th}$,2.3564) ($3^{rd}$,3.02039) ($2^{nd}$,6.43647) ($1^{st}$,69.7652)};
    \end{axis}
    \end{tikzpicture}
    \caption{\hspace*{-6em}}
    \label{fig:bar_pca_e}
    \end{subfigure}
    \begin{subfigure}[b]{0.5\textwidth}
    \centering
    \raisebox{0.25cm}{
    \includegraphics[width=0.65\textwidth]{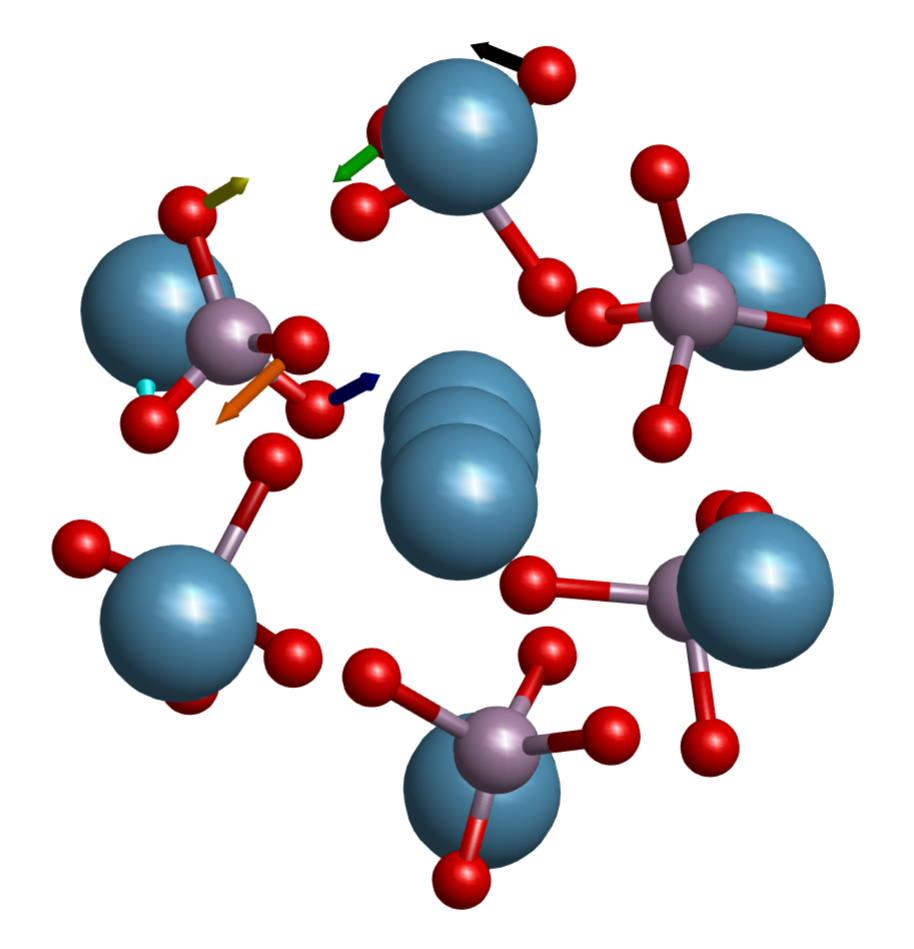}}
    \caption{}
    \label{fig:disp_e}
    \end{subfigure}
    %\hfill
    \\
    \begin{subfigure}[b]{0.4\textwidth}
    \centering
    \begin{tikzpicture}
    \begin{axis}[ybar,
    symbolic x coords={$5^{th}$, $4^{th}$, $3^{rd}$, $2^{nd}$, $1^{st}$},
    xtick=data,
    nodes near coords,
    xlabel = {Eigenmode index},
    ylabel = {Percentage of varaince \\ explained by the structure},
    xlabel style = {font=\footnotesize},
    ylabel style = {font=\footnotesize},
    ylabel style = {align=center},
    yticklabel style = {font=\footnotesize},
    xticklabel style = {font=\footnotesize},
    x dir = reverse,
    width = 6.5cm,
    height = 6cm,
    ymin = 0,
    ymax = 80,
    nodes near coords align={vertical},
    bar width = 0.7cm,
    every node near coord/.append style={font=\scriptsize}]
        \addplot[blue!20!black,fill=blue!80!white] coordinates {($5^{th}$,2.15) ($4^{th}$,3.12) ($3^{rd}$,5.70) ($2^{nd}$,8.83) ($1^{st}$,61.89)};
    \end{axis}
    \end{tikzpicture}
    \caption{\hspace*{-6em}}
    \label{fig:bar_pca_d}
    \end{subfigure}
    \begin{subfigure}[b]{0.5\textwidth}
    \centering
    \raisebox{0.25cm}{
    \includegraphics[width=0.65\textwidth]{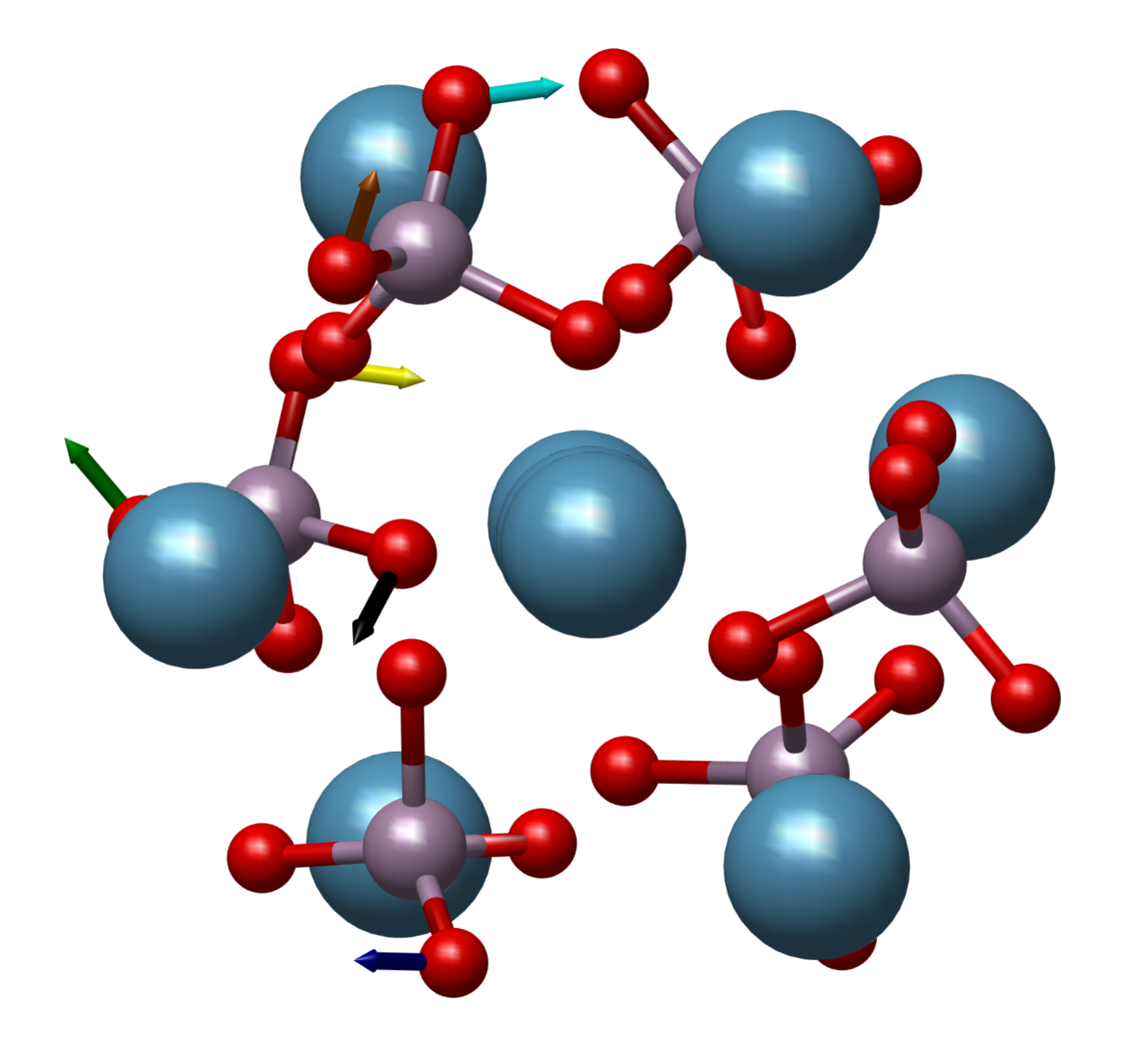}}
    \caption{}
    \label{fig:disp_d}
    \end{subfigure}
    \\
    \begin{subfigure}[b]{0.4\textwidth}
    \centering
    \begin{tikzpicture}
    \begin{axis}[ybar,
    symbolic x coords={$5^{th}$, $4^{th}$, $3^{rd}$, $2^{nd}$, $1^{st}$},
    xtick=data,
    nodes near coords,
    xlabel = {Eigenmode index},
    ylabel = {Percentage of varaince \\ explained by the structure},
    xlabel style = {font=\footnotesize},
    ylabel style = {font=\footnotesize},
    ylabel style = {align=center},
    yticklabel style = {font=\footnotesize},
    xticklabel style = {font=\footnotesize},
    x dir = reverse,
    width = 6.5cm,
    height = 6cm,
    ymin = 0,
    ymax = 80,
    nodes near coords align={vertical},
    bar width = 0.7cm,
    every node near coord/.append style={font=\scriptsize}]
        \addplot[blue!20!black,fill=blue!80!white] coordinates {($5^{th}$,2.33) ($4^{th}$,5.74) ($3^{rd}$,6.62) ($2^{nd}$,10.82) ($1^{st}$,53.98)};
    \end{axis}
    \end{tikzpicture}
    \caption{\hspace*{-6em}}
    \label{fig:bar_pca_b}
    \end{subfigure}
    \begin{subfigure}[b]{0.5\textwidth}
    \centering
    \raisebox{0.25cm}{
    \includegraphics[width=0.65\textwidth]{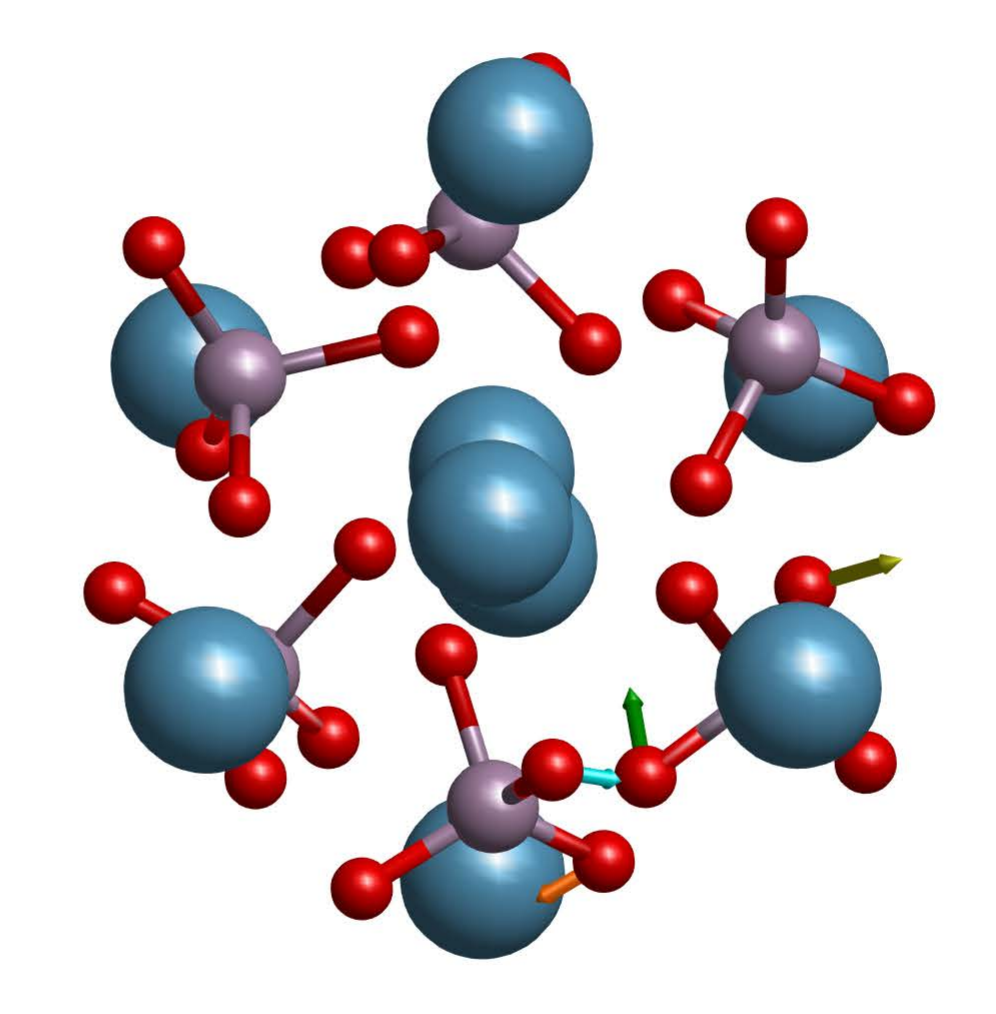}}
    \caption{}
    \label{fig:disp_b}
    \end{subfigure}
    \caption{\rev{(a), (c), (e) Eigenvalues resulting from PCA over a single dynamical run for configurations \textbf{E}, \textbf{D}, and \textbf{B} respectively. One predominant mode is observed. (b), (d), (f) Difference between the most dominant eigenmode and the time-average structure of the corresponding dynamical run. The arrows have been elongated threefold for clarity.}}
    \label{fig:disp_all}
\end{figure}

With our calculations unable to convincingly point us towards a PM structure with high symmetry, we performed PCA on the data from the dynamical runs. It was expected that, if the molecule exhibits any kind of high symmetry that might have been overlooked in our symmetry analysis methodology, the associated high symmetry structure would show up as one of the dominant eigenmodes of the PCA. However, none of the dominant eigenmodes displayed any kind of high symmetry. In fact, in all cases, the eigenmodes had either a \ce{C_i} or a \ce{C_s} point group symmetry. Moreover, as can be seen in Fig.~\ref{fig:bar_pca_e}, Fig.~\ref{fig:bar_pca_d}, and Fig.~\ref{fig:bar_pca_b}, each configuration yielded one predominant eigenmode. \rev{Similar plots and figures for all dynamical runs can be found in the SI.} While the maximum displacement of an atom in this mode from its corresponding position in the time-average structure over the entire dynamical run was small -- less than $0.3$\AA -- this mode, instead, reflected a more appreciable displacement of the phosphates due to their rotation. Structures corresponding to the first three dominant modes did not give way to higher symmetries. The small observed fluctuations further emphasize that considering an average structure over our dynamical data set was appropriate. A detailed PCA analysis for all other considered structures is provided in the SI. 

%At this point, it is relevant to notice the small deviation of these few eigenmodes from the time-average structure over the entire dynamical run, as shown in Fig.~\ref{fig:disp_all}. The maximum displacement of an atom is less than $0.3$\AA. Only a few atoms require appreciable displacement. This shows that considering an average structure over our dynamical data set is appropriate, because the structural variations resemble the dominant eigenmodes. Furthermore, upon close inspection, only the phosphates undergo a rotation. Thus, since the difference between the structures is relatively small -- as stated above -- it backs up the observation that there are at most three dominant eigemodes based on PCA of our data set, none of which represents a structure of high symmetry. 

%\rev{\sout{We next resorted to another powerful data analysis technique, namely \textit{k-}means clustering, to identify hidden patterns in the data generated from our dynamical runs that were not revealed by our earlier attempts. However, it was found that, for each of the eight starting configurations, the ideal number of clusters describing our data sets was $2$. Moreover, each of the associated mean structures had a low point group symmetry -- \ce{C_1} or \ce{C_i}. It was expected that the data clustering algorithm would yield as many clusters as the number of different point group symmetries that the structure went through in a single dynamical run, but the fact that only $2$ clusters are needed to optimally represent the data reinforces our claim that the importance and frequency of these high symmetry structures is low. This further agrees with the data that has already been presented above, and with our assessment that the PM mostly exists as an ensemble of low-symmetry structures. Clustering results for all eight configurations have been provided in the SI. We also performed vibrational spectrum calculations on our eight transition state structures, and compared them with an existing spectrum \cite{swift2018posner}, details of which can be found in the SI.}}

\rev{We next resorted to another powerful data analysis technique, namely \textit{k-}means clustering, to identify hidden patterns in the data generated from our dynamical runs that might not have been revealed by our earlier attempts. It was found that, for each of the eight starting configurations, the ideal number of clusters describing our data sets was $2$, and that the associated mean structure for each cluster had low point group symmetry -- \ce{C_1} or \ce{C_i}. This further agrees with the data and analyses that have already been presented above, and with our assessment that the PM mostly exists as an ensemble of low-symmetry structures. Clustering results for all eight configurations have been provided in the SI.}

%\rev{The following two paragaphs have been moved here from the end of the Introduction section verbatim, in line with the suggestions from both the reviewers.}

\rev{In summary, we have explored the dynamical structural properties of the PM. Our results do not indicate a predominant \ce{S6} molecular symmetry. Instead, we argue that the molecule exists within a spread of multiple lower-symmetry structures at room temperature. We also explored the most stable form of the PM in vacuum (concordant with earlier work) as well as in the presence of water (using a PCM), evaluated the point group symmetry in these scenarios, and came to conclusions corroborating our dynamical calculations. We used a variety of simulation techniques, including structural relaxation and molecular dynamics calculations --- both at the level of empirical force fields, and using density functional theory based on semi-local and hybrid exchange correlation functionals as well as dispersion corrected functionals. \cite{grimme2010consistent}. These calculations have been done assuming the isolated PM in vacuum or a dielectric continuum. Similar calculations in the presence of a non-homogenous intracellular fluid around the PM are expected to hinder the molecule's rotation and likely further reduce the cluster's point group symmetry, which will lead to faster relaxation of the \ce{^{31}P} coherences \cite{player2018posner}. To interpret the results of our simulations, we employed a variety of structural analysis tools, including vibrational spectra calculations and the evaluation of time persistence of symmetries, as well as a broad spectrum of data analysis tools, including structural averaging, Principal Component Analysis (PCA), and \textit{k-}means clustering. The exploration of thousands of individual molecular configurations, the focus on dynamical structural properties at room temperature, and the use of a wide variety of simulation and data analysis tools are all distinguishing features of our work, and together they serve to provide robustness to our conclusions.}

\rev{Our extensive analysis of the dynamical and structural properties of the Posner molecule suggests that it predominantly exists in low symmetry molecular structures such as \ce{C_s}, \ce{C_i} and \ce{C1} at room temperature, as opposed to the results of previous studies suggesting a prototypical \ce{S6} symmetric structure. Moreover, the initial configuration of the molecule often dictates the geometric configurations through which the molecule transitions during a dynamical run. Most of these transition structures exhibit low molecular point group symmetries; the high symmetry phases are found to be present only fleetingly and have thus been assumed to be unimportant. Average structures were also found to be of low symmetry. Our results indicate that that the molecule does not naturally exhibit a three-fold axis of rotation, such as present in \ce{S6}, in vacuum or in a homogenous solvent. Calculation of spin-spin coupling constants and spin coherence times for the structures explored by us constitutes ongoing and future work.}

\rev{Laslty, we suggest the possibility of experimental verifications of our results. At a very basic level, it may well be possible to establish the existence of Posner molecules in simulated body fluids by introducing the constituents at stoiciometric ratios. Imaging techniques such as Dynamic Light Scattering or Transmission Electron Microsopy and NMR spectroscopy could then be applied to observe/study the basic properties of these molecules \cite{fisher2015quantum}. However, we believe that more sophisticated techniques, possibly following established quantum optics protocols and using microfluidics, might be needed to confirm or refute the possibility of the Posner molecule sustaining \ce{^{31}P} qubit states.}

For structural relaxation, we used Quantum ESPRESSO \cite{QE-2009} -- with the Standard Solid-State Pseudopotentials library \cite{prandini2018precision, lejaeghere2016reproducibility} and the Perdew - Burke - Ernzerhof (PBE) exchange correlation functional -- and Q-Chem \cite{shao2015advances} -- with both PBE and B3LYP exchange-correlation functionals, and a basis set of 6-311G(d,p) for the atomic orbitals. Symmetries were analyzed using two toolkits, VMD and WebMO \cite{schmidt2020webmo}. Both toolkits gave the same symmetries. Further details about our methods can be found in the SI.

\begin{acknowledgement}
The authors would like to thank Michael W.\ Swift for insightful discussions, which helped with the preparation of the mansucript. The authors would also like to thank UCLA’s Institute for Digital Research and Education (IDRE) for making available the computing resources used in this work. DRK would like to thank the Office of Naval Research for financial support (ONR award number N62909-21-1-2018).
\end{acknowledgement}

\newpage

\begin{center}
    {\LARGE \textbf{Supporting Information}}
\end{center}

\tableofcontents

\newpage

\section{Methods}

For structural relaxation in our study, we used Quantum ESPRESSO and Q-Chem. Within Quantum ESPRESSO, we used the Standard Solid-State Pseudopotentials library \cite{prandini2018precision, lejaeghere2016reproducibility} and the Perdew - Burke - Ernzerhof (PBE) exchange correlation functional. The Broyden – Fletcher – Goldfarb – Shanno (BFGS) algorithm was used for optimization of the structure. A unit cell of more than twice the size of the initial structure was used, and its geometry was subsequently relaxed. The Polarizable Continuum Model (PCM) for solvation was also used to undertsand the effects of solvation, if any, on structural relaxation of the Posner molecule. We used the conductor-like PCM \cite{truong1995new,barone1998quantum} for solvation with the Switching/Gaussian method. \cite{lange2010smooth}. It was observed that solvation effects do not result in highly symmetric structures either. While working with Q-Chem for \textit{ab inito} molecular dynamics (AIMD) calculations, both PBE and B3LYP exchange-correlation functionals were used, along with a basis set of 6-311G(d,p) for the atomic orbitals, which uses polarized basis functions. Canonical NVT sampling was done using AIMD with the Langevin thermostat. Noting that semi-local density functionals do not capture dispersion interactions properly, we also used dispersion-corrected functionals (the DFT-D3(0) dispersion correction from Grimme et.\ al. \cite{grimme2010consistent}) for some of our simulations. Different self-consistent field (SCF) iteration algorithms, such as Direct Inversion in the Iterative Subspace (DIIS) \cite{pulay1980convergence, pulay1982improved} and Geometric Direct Minimization (GDM) \cite{van2002geometric} were used. Additionally, Pseudo-Fractional Occupation Number Method (pFON) was used, which is akin to introducing a finite electronic temperature, or smearing, in the system. For completeness, the symmetries were analyzed using two different toolkits, namely VMD and WebMO \cite{schmidt2020webmo}. No difference in the results was observed.

\newpage

\section{Calcium phosphate monomer}

\begin{figure}
    \centering
    \includegraphics[width=8cm]{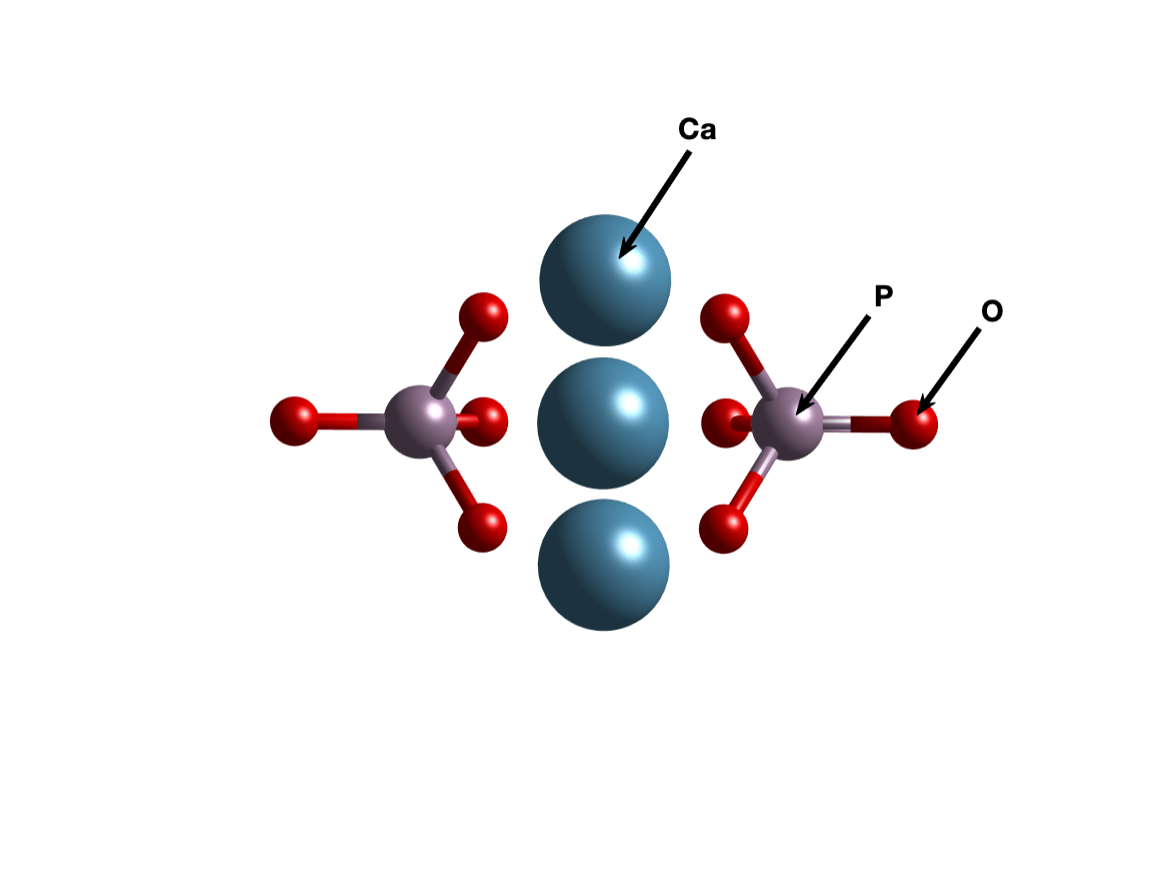}
    \caption{A calcium phosphate monomer exhibiting the expected \ce{D_{3h}} molecular point group symmetry. The result was obtained using DL POLY, Quantum ESPRESSO, and Q-Chem, and is consistent with previous studies. The blue, purple, and red spheres represent \ce{Ca}, \ce{P}, and \ce{O} atoms respectively.}
    \label{fig:monomer}
\end{figure}

A rough structure of a calcium phosphate monomer was first generated by hand. Once the monomer's structure was optimized using DL POLY \cite{smith2002dl_poly}, Quantum ESPRESSO, and Q-Chem, it always resulted in a structure with a molecular point group symmetry of \ce{D_{3h}} as shown in Fig.~\ref{fig:monomer}. This is in agreement with other studies \cite{treboux2000existence, kanzaki2001calcium} that have reported the geometry of a stable monomer structure.

\newpage

\section{Calcium phosphate dimers}

\begin{figure}[h!]
    \centering
    \begin{subfigure}[b]{0.3\textwidth}
      \centering
      \includegraphics[width=\textwidth]{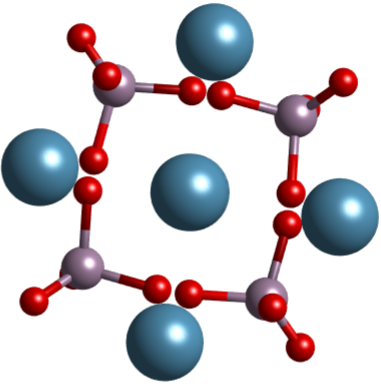}
      \captionsetup{font={small}}
      \caption{\ce{T_d} molecular symmetry}
      \label{fig:dimer_Td}
    \end{subfigure}
    \hfill
    \begin{subfigure}[b]{0.3\textwidth}
      \centering
      \includegraphics[width=\textwidth]{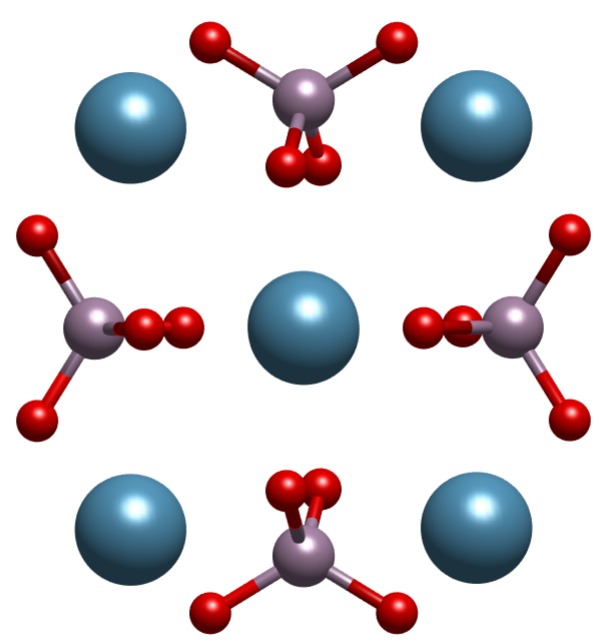}
      \captionsetup{font={small}}
      \caption{\ce{C_{2v}} molecular symmetry}
      \label{fig:dimer_C2v_10}
    \end{subfigure}
    \hfill 
    \begin{subfigure}[b]{0.3\textwidth}
      \centering
      \includegraphics[width=\textwidth]{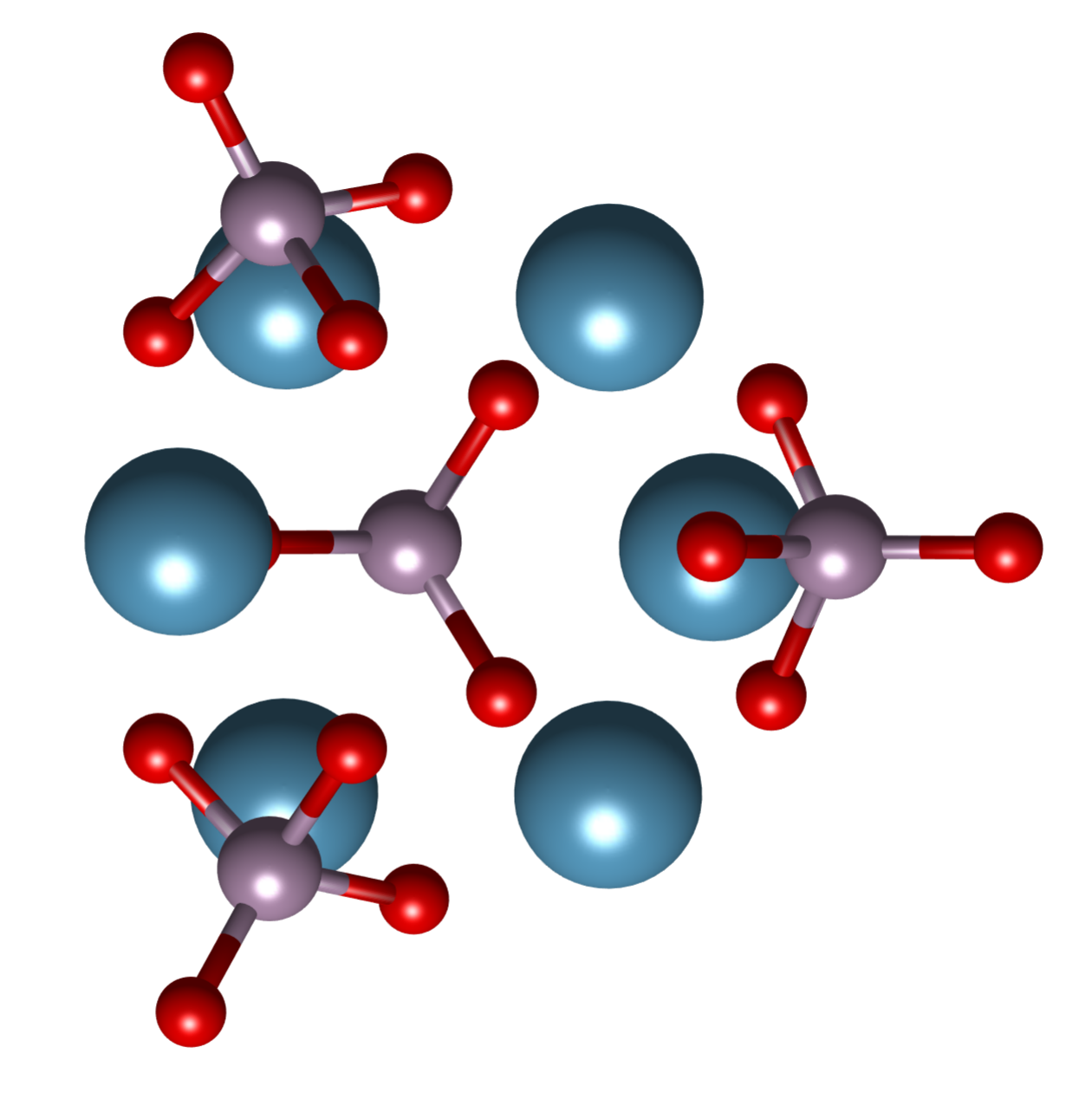}
      \captionsetup{font={small}}
      \caption{\ce{C_{s}} molecular symmetry}
      \label{fig:dimer_Cs_9}
    \end{subfigure}
    \hfill
    \\
    \begin{subfigure}[b]{0.3\textwidth}
      \centering
      \includegraphics[width=\textwidth]{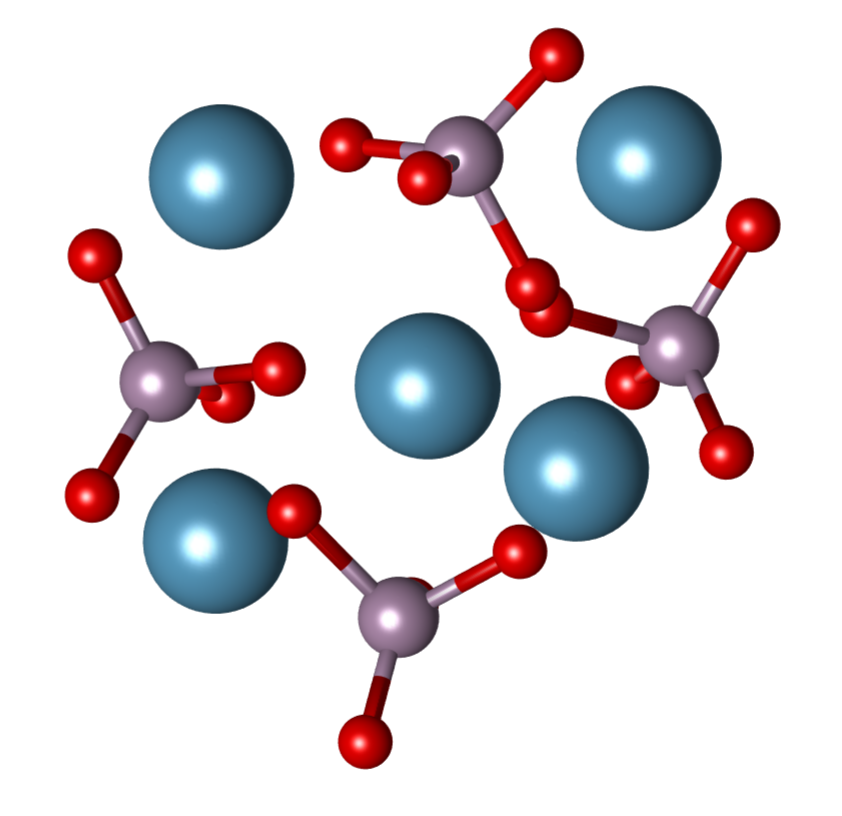}
      \captionsetup{font={small}}
      \caption{\ce{C2} molecular symmetry}
      \label{fig:dimer_C2_4}
    \end{subfigure}
    \hfill
    \begin{subfigure}[b]{0.3\textwidth}
      \centering
      \includegraphics[width=\textwidth]{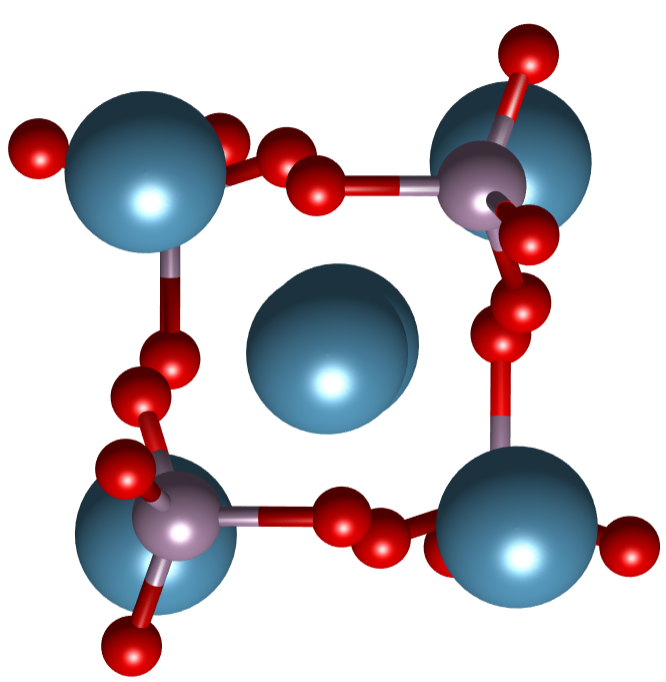}
      \captionsetup{font={small}}
      \caption{\ce{S4} molecular symmetry}
      \label{fig:dimer_S4_3}
    \end{subfigure}
    \hfill 
    \begin{subfigure}[b]{0.3\textwidth}
      \centering
      \includegraphics[width=\textwidth]{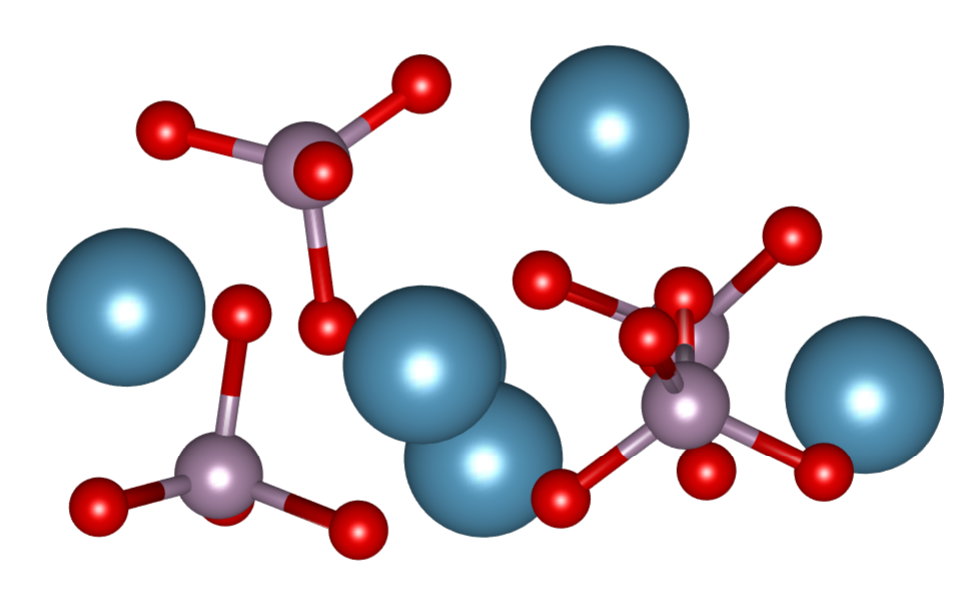}
      \vspace{0.7cm}
      \captionsetup{font={small}}
      \caption{\ce{C_{2}} molecular symmetry}
      \label{fig:dimer_C2_1}
    \end{subfigure}
    \hfill
    \vspace{0.5cm}
    \caption{Some calcium phosphate dimer relaxed geometries obtained in our study. The methodology to obtain these strctures was based on the results of Kanzaki et al.}
    \label{fig:dimers}
\end{figure}

When working with the dimer, which has a slightly more complicated structure than the monomer, some interesting features were observed. Earlier studies \cite{kanzaki2001calcium} have reported multiple possible geometries for the calcium phosphate dimer. To cover all possible geometries and starting configurations, we created over $100$ initial geometries that were then subjected to structural relaxation. The basic outline of our simulation setup was as follows. We first arrange the atoms in one out of several initial configurations. Based on suggestions in the work by Kanzaki et al. \cite{kanzaki2001calcium}, the most common of these was to arrange the calcium atoms and the phosphate groups symmetrically around a cube of appropriate dimensions. The length of the diagonal of this cube effectively describes the approximate diameter of the molecule and thus, the dimensions were chosen accordingly. A large number of starting configurations were  then obtained by independently rotating each phosphate group about the phosphorus atom in steps of $30$\textdegree\ while keeping the positions of the phosphorus and calcium atoms fixed.  Another route to generating an initial configuration is to arrange the atoms in such a way that it has a structure similar to earlier findings, and to then perturb the structure.  Subsequent to structural relaxation, we were able to obtain $6$ of the $11$ structures that have been reported before \cite{kanzaki2001calcium}, with \ce{C_s}, \ce{C_2}, \ce{T_d}, \ce{C_{2v}}, and \ce{D_{2h}} point group symmetries, as shown in Fig.~\ref{fig:dimers}. However, Kanzaki et al.\ reported additional dimer structures that were not found using our methods. This could be because of an incomplete consideration of the structure space of the molecule, as well as the crudeness in the angle of rotation for the phosphate groups. Different starting configurations often led to different optimized final geometries --- an artifact that was also later observed when studying the Posner molecule. Thus, we argue that a singular optimal structure does not exist, and that the calcium phosphate dimer possibly exists in a variety of symmetries. At this stage, it is also important to note that the relative energy differences between the structures obtained by us was not the same as have been reported by Kanzaki et al.\ For instance, even though the \ce{T_d} point group symmetry (in agreement with Kanzaki et al.) was found to be the most stable symmetry, the relative energies between that structure and other symmetries differed by about $0.023$ eV/atom on average. More details about this can be found in Section \ref{sec:dimer_table} of the SI.

\newpage

\section{Comparison of dimer data with results by Kanzaki et al.}
\label{sec:dimer_table}

\begin{center}
\centering
\begin{tabular}{|M{4.5cm}|M{4.5cm}|M{4.5cm}|}
 \hline
 \textbf{Structure symmetry (corresponding to Fig.~\ref{fig:dimers})} & \textbf{Relative energy reported by Kanzaki et.\ al (in eV)} & \textbf{Relative energy from this study (in eV)} \\
 \hline
 \ce{T_d} -- (a)  & 1.352  & 0.838  \\
 \hline
 \ce{C_{2v}} -- (b)  & 1.374  & 0.734  \\
 \hline
 \ce{C_s} -- (c)  & 1.3  & 0.495  \\
 \hline
 \ce{C2} -- (d)  & 0.797  & 0.076  \\
 \hline
 \ce{S4} -- (e)  & 0.702  & 0.032  \\
 \hline
 \ce{C2} -- (f)  & 0  & 0  \\
\hline
\end{tabular}
\captionof{table}{A comparison of the relative energies of the structures found in the current study and in the study done by Kanzaki et.\ al. The structure corresponding to the configuration (f) is used as the reference for relative energy differences.}
\end{center}

\newpage

\section{Methodology to create over 10,000 structures for structural relaxation}
\label{sec:10k_structs}

\rev{The technique used by Kanzaki et al.\ was closely followed. All atoms constituting the PM were arranged on a cube of appropriate dimensions. The nine \ce{Ca} atoms were placed in a body-centred cubic arrangement and the six \ce{PO4} groups in a face-centred cubic arrangement. The size of the cube was chosen such that the length of the diagonal was close to $9$\AA, which is the approximate diameter of the PM \cite{swift2018posner,player2018posner,roohani2021build}. Countless configurations of the cluster can be realized by rotating each \ce{PO4} group around its center. In order to extensively sample the configuration space, the phosphate groups were rotated in steps of $30$\textdegree\ in 3 dimensions to create over $2,800$ structures. Additionally, the coordinates of the calcium atoms and the \ce{PO_4} groups were scaled with reference to the central \ce{Ca} atom to account for the possibility of some previously built structures being over-strained. This resulted in over $10,000$ viable structures.}

\begin{figure}
    \centering
    \includegraphics[width=0.6\textwidth]{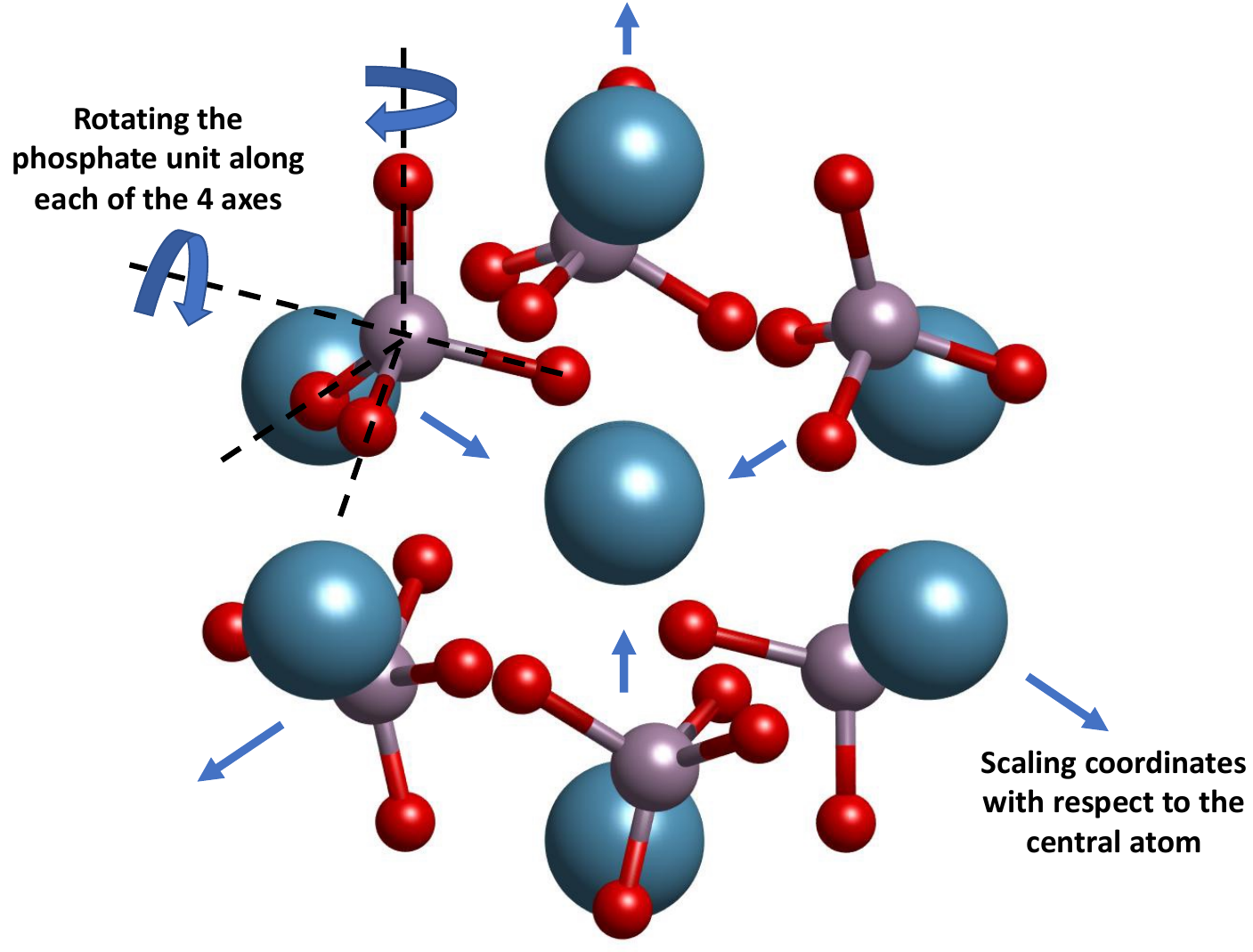}
    \vspace{0.8cm}
    \caption{\rev{The scheme used for creating over $10,000$ structures by rotating the phosphate units and scaling all the coordinates with respect to central atom. The blue, purple, and red spheres represent \ce{Ca}, \ce{P}, and \ce{O} atoms respectively.}}
    \label{fig:rotation_scaling_of_PM_supp}
\end{figure}

\newpage

\section[\texorpdfstring{\ce{S6}{ s}} symmetric structure created using a rigid phosphate tetrahedral geometry]{\texorpdfstring{\ce{S6}{ s}} symmetric structure created using a rigid \\ phosphate tetrahedral geometry}

\begin{figure}[htb]
    \centering
    \includegraphics[width=0.4\textwidth]{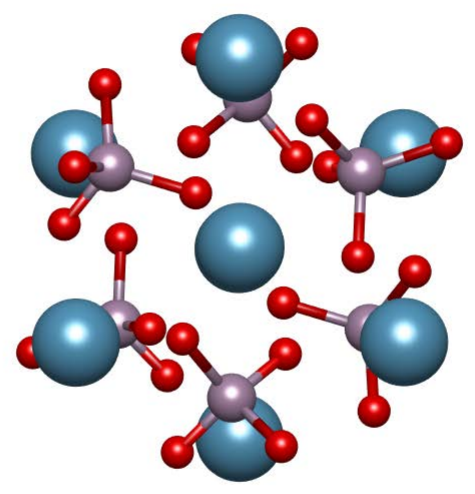}
    \caption{\ce{S6} molecular symmetry constructed using a rigid phosphate tetrahedral geometry. The blue, purple, and red spheres represent \ce{Ca}, \ce{P}, and \ce{O} atoms respectively.}
    \label{fig:ff_s6}
\end{figure}

The structure in Fig.~\ref{fig:ff_s6} was created by imposing an \ce{S6} symmetry in the molecule. The structure was described in terms of 10 parameters and a global minimization of the system's energy based on this parametrization was carried out.

\rev{Symmetry-constrained ab initio calculations on the above structure failed to reach self-consistency, and symmetry-unconstrained calculations did not yield stable structures. For comparison, the structure in Fig.~\ref{fig:ff_s6} had an energy which was approximately $5$ eV = $128.20$ meV/atom = $194.7\;k_{B}T$ larger than the mean energy in Fig.~\ref{fig:violin_new}. This is well beyond what may be thermally populated and further confirms the instability of the singular \ce{S6} structure.}

\newpage

\section{Time preservation of symmetries in a dynamical run}
\label{sec:Time_sym}

\rev{Here, we justify that the time frame considered for our dynamic simulations are sufficient for the subsequent analyses. Arguably, the relevant temporal quantity for structural relaxations is determined by molecular vibration. We refer to the calculation of the IR spectra of the two transition state structures considered in the SI in Figure \ref{fig:IR_comparison}. For both structures analysed, the lowest non-zero vibrational frequency mode lies at $100$ cm$^{-1}$. This wavenumber corresponds to a frequency of $3$ THz, which is associated with a time scale of $0.334$ ps. Thus, we argue that the relaxation time for the lowest vibrational frequency mode is well within the time frame of our simulation and is thus, adequate for structural averaging. The molecule is expected to go through its vibrational relaxation pathways within the time duration considered for our \textit{ab-initio} molecular dynamics simulations.  Note further that the energy differences suggested for \ce{S6}--\ce{C_i}, \ce{S6}--\ce{C2} and \ce{S6}--\ce{C1} transitions from Refs.\ \citenum{treboux2000existence,treboux2000symmetry,swift2018posner} are comparable to the thermal energy. Thus, the transition process is not expected to be a rare event, but is likely to proceed on the timescale of vibrations. Within these assumptions, the studied timescale is certainly sufficient to accommodate the envisaged structural reorganizations. A practical constraint that has to be noted is the balance between accessible simulation timescales and the computational resources consumed, due to the expensive nature of the ab initio calculations.}
\newpage

\subsection{At a temperature of 298 K}
\label{subsec:T_298}

\begin{figure}[h!]
    \centering
    \begin{subfigure}[b]{0.8\textwidth}
      \centering
      \includegraphics[height=8cm,width=\textwidth]{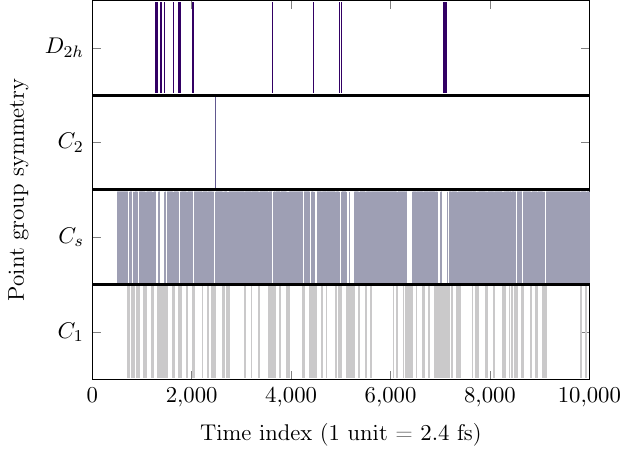}
      \caption{Configuration \textbf{A}}
      \label{fig:time_A}
    \end{subfigure}
    \\
    \vspace{0.5cm}
    \begin{subfigure}[b]{0.8\textwidth}
      \centering
      \includegraphics[height=8cm,width=\textwidth]{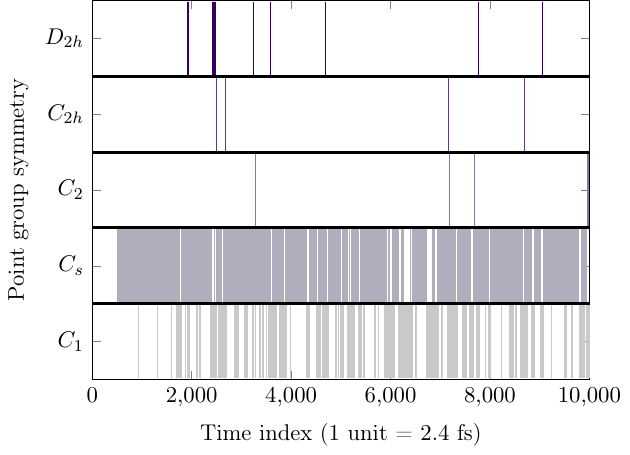}
      \caption{Configuration \textbf{B}}
      \label{fig:time_B}
    \end{subfigure}
\end{figure}

\newpage

\begin{figure}[h!]\ContinuedFloat
    \begin{subfigure}[b]{0.8\textwidth}
      \centering
      \includegraphics[height=8cm,width=\textwidth]{test-figure2.pdf}
      \caption{Configuration \textbf{C}}
      \label{fig:time_C}
    \end{subfigure}
    \\
    \vspace{0.5cm}
    \begin{subfigure}[b]{0.8\textwidth}
      \centering
      \includegraphics[height=8cm,width=\textwidth]{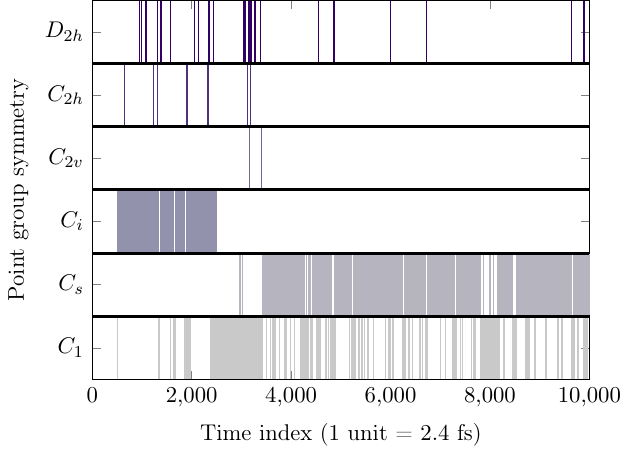}
      \caption{Configuration \textbf{D}}
      \label{fig:time_D}
    \end{subfigure}
\end{figure}

\newpage

\begin{figure}[h!]\ContinuedFloat
    \begin{subfigure}[b]{0.8\textwidth}
      \centering
      \includegraphics[height=8cm,width=\textwidth]{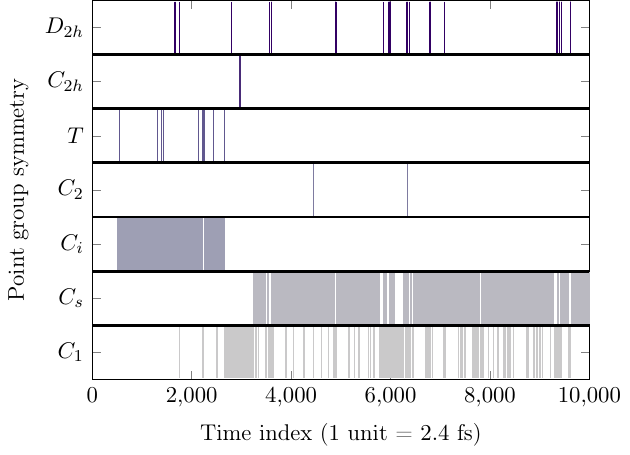}
      \caption{Configuration \textbf{E}}
      \label{fig:time_E}
    \end{subfigure}
    \\
    \vspace{0.5cm}
    \begin{subfigure}[b]{0.8\textwidth}
      \centering
      \includegraphics[height=8cm,width=\textwidth]{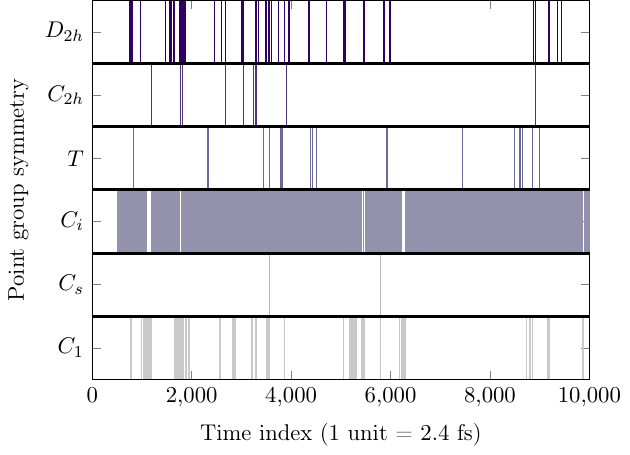}
      \caption{Configuration \textbf{F}}
      \label{fig:time_F}
    \end{subfigure}
\end{figure}

\newpage

\begin{figure}[h!]\ContinuedFloat
    \begin{subfigure}[b]{0.8\textwidth}
      \centering
      \includegraphics[height=8cm,width=\textwidth]{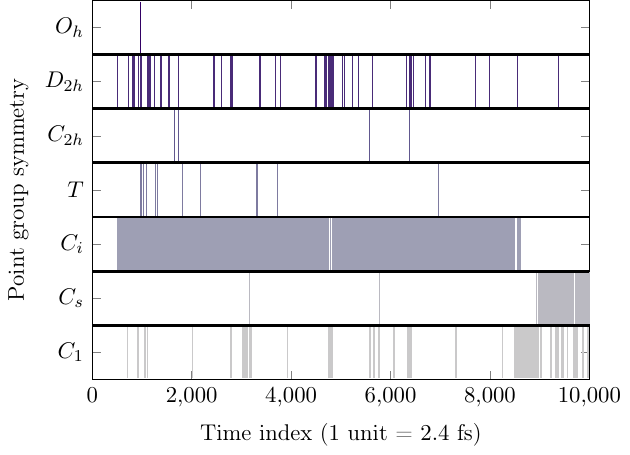}
      \caption{Configuration \textbf{G}}
      \label{fig:time_G}
    \end{subfigure}
    \\
    \vspace{0.5cm}
    \begin{subfigure}[b]{0.8\textwidth}
      \centering
      \includegraphics[height=8cm,width=\textwidth]{test-figure7.pdf}
      \caption{Configuration \textbf{H}}
      \label{fig:time_H}
    \end{subfigure}
    \vspace{0.5cm}
    \caption{Time persistence of symmetries for all the $8$ configurations at $T=298\;K$. The first $500$ instances in each case have not been plotted to account for rapid tumbling and stabilization of the molecule at the beginning of a dynamical run}
    \label{fig:time_sym}
\end{figure}
    
\newpage

\subsection{At a temperature of 315 K}

\begin{figure}[h!]
    \centering
    \begin{subfigure}[b]{0.8\textwidth}
      \centering
      \includegraphics[height=8cm,width=\textwidth]{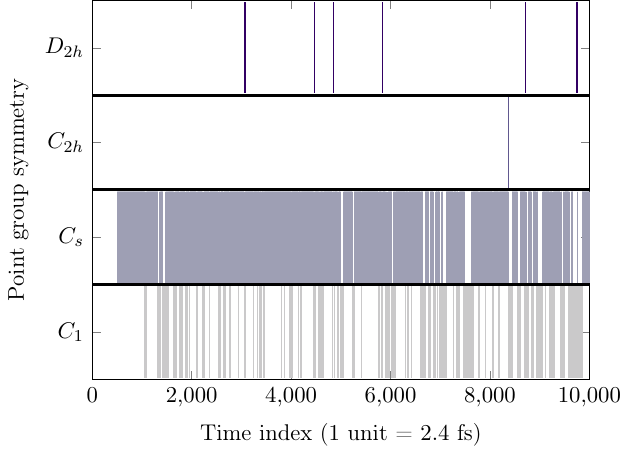}
      \caption{Configuration \textbf{A}}
      \label{fig:time_HT_A}
    \end{subfigure}
    \\
    \vspace{0.5cm}
    \begin{subfigure}[b]{0.8\textwidth}
      \centering
      \includegraphics[height=8cm,width=\textwidth]{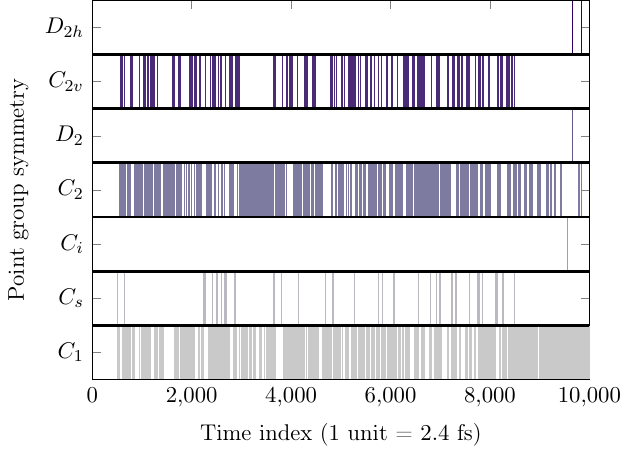}
      \caption{Configuration \textbf{B}}
      \label{fig:time_HT_B}
    \end{subfigure}
\end{figure}

\newpage

\begin{figure}[h!]\ContinuedFloat
    \begin{subfigure}[b]{0.8\textwidth}
      \centering
      \includegraphics[height=8cm,width=\textwidth]{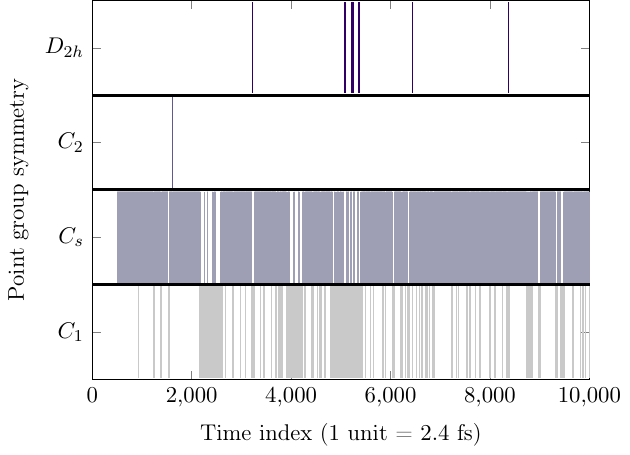}
      \caption{Configuration \textbf{C}}
      \label{fig:time_HT_C}
    \end{subfigure}
    \\
    \vspace{0.5cm}
    \begin{subfigure}[b]{0.8\textwidth}
      \centering
      \includegraphics[height=8cm,width=\textwidth]{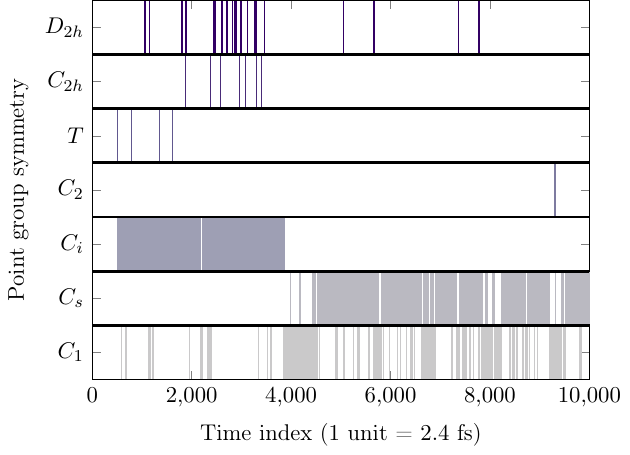}
      \caption{Configuration \textbf{D}}
      \label{fig:time_HT_D}
    \end{subfigure}
\end{figure}

\newpage

\begin{figure}[h!]\ContinuedFloat
    \begin{subfigure}[b]{0.8\textwidth}
      \centering
      \includegraphics[height=8cm,width=\textwidth]{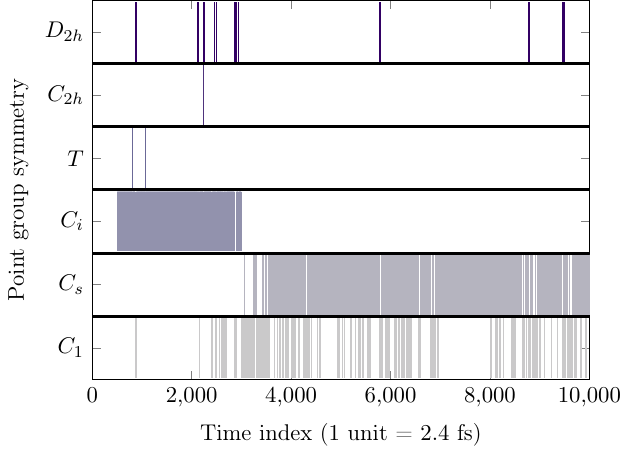}
      \caption{Configuration \textbf{E}}
      \label{fig:time_HT_E}
    \end{subfigure}
    \\
    \vspace{0.5cm}
    \begin{subfigure}[b]{0.8\textwidth}
      \centering
      \includegraphics[height=8cm,width=\textwidth]{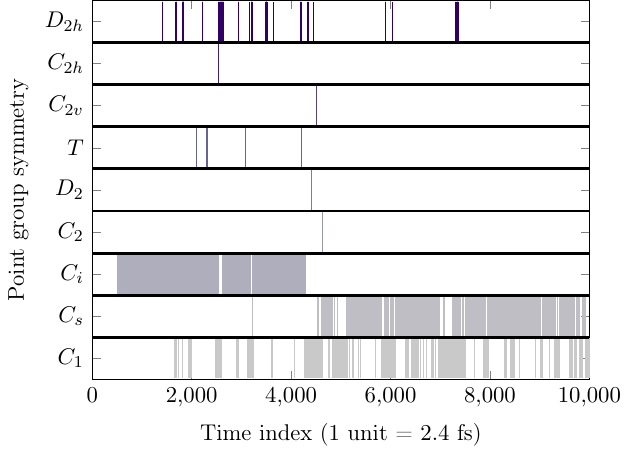}
      \caption{Configuration \textbf{F}}
      \label{fig:time_HT_F}
    \end{subfigure}
\end{figure}

\newpage

\begin{figure}[h!]\ContinuedFloat
    \begin{subfigure}[b]{0.8\textwidth}
      \centering
      \includegraphics[height=8cm,width=\textwidth]{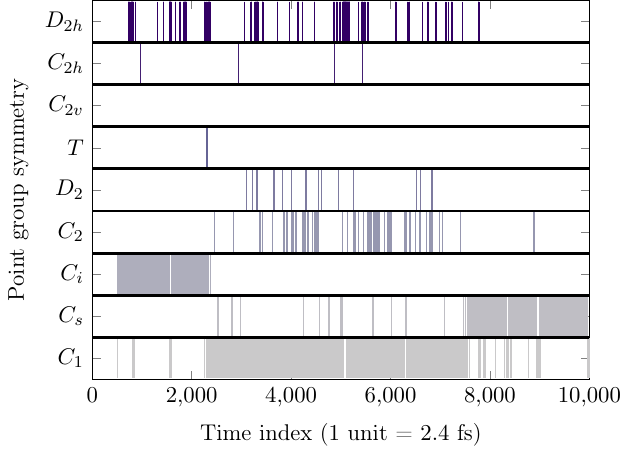}
      \caption{Configuration \textbf{G}}
      \label{fig:time_HT_G}
    \end{subfigure}
    \\
    \vspace{0.5cm}
    \begin{subfigure}[b]{0.8\textwidth}
      \centering
      \includegraphics[height=8cm,width=\textwidth]{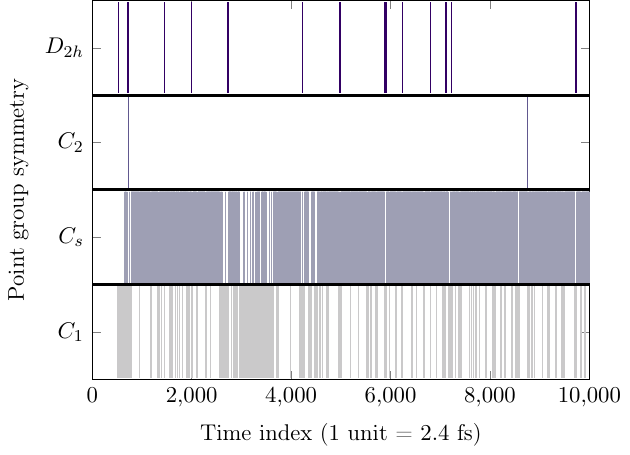}
      \caption{Configuration \textbf{H}}
      \label{fig:time_HT_H}
    \end{subfigure}
    \vspace{0.5cm}
    \caption{Time persistence of symmetries for all the $8$ configurations at $T=315\;K$. The first $500$ instances in each case have not been plotted to account for rapid tumbling and stabilization of the molecule at the beginning of a dynamical run}
    \label{fig:time_sym_hightemp}
\end{figure}
    
\newpage

\section{Percentage occurrence of point group symmetries in a dynamical run}

\begin{figure}[h!]
    \centering
    \begin{subfigure}[b]{0.48\textwidth}
    \centering
    \begin{tikzpicture}
    \begin{axis}[ybar,
    symbolic x coords={\ce{C_{1}}, \ce{C_{s}}, \ce{C_{2}}, \ce{C_{2h}}, \ce{D_{2h}}},
    xtick=data,
    xlabel = {},
    ylabel = {Percentage occurrence},
    xlabel style = {font=\tiny},
    ylabel style = {font=\tiny},
    yticklabel style = {font=\tiny},
    xticklabel style = {font=\tiny},
    ymin = 0,
    bar width = 0.3cm,
    legend style={font=\tiny}]
        \addplot[red!20!black,fill=red!80!white] coordinates { (\ce{C_{1}},20.96) (\ce{C_{s}},73.66) (\ce{C_{2}},0.0) (\ce{C_{2h}},0.01) (\ce{D_{2h}},0.37)};
        \addplot[blue!20!black,fill=blue!80!white] coordinates { (\ce{C_{1}},20.25) (\ce{C_{s}},73.72) (\ce{C_{2}},0.03) (\ce{C_{2h}},0.0) (\ce{D_{2h}},1.0)};
        \legend{$T=315\;K$, $T=298\;K$}
    \end{axis}
    \end{tikzpicture}
    \caption{Configuration \textbf{A}}
    \label{fig:bar_A}
    \end{subfigure}
    \hfill
    \begin{subfigure}[b]{0.48\textwidth}
    \centering
    \begin{tikzpicture}
    \begin{axis}[ybar,
    symbolic x coords={\ce{C_{1}}, \ce{C_{s}}, \ce{C_{i}}, \ce{C_{2}}, \ce{D_{2}}, \ce{C_{2v}}, \ce{C_{2h}}, \ce{D_{2h}}},
    xtick=data,
    xlabel = {},
    ylabel = {Percentage occurrence},
    xlabel style = {font=\tiny},
    ylabel style = {font=\tiny},
    yticklabel style = {font=\tiny},
    xticklabel style = {font=\tiny},
    ymin = 0,
    bar width = 0.3cm,
    legend style={font=\tiny}]
        \addplot[red!20!black,fill=red!80!white] coordinates { (\ce{C_{1}},44.09) (\ce{C_{s}},3.08) (\ce{C_{i}},0.02) (\ce{C_{2}},34.03) (\ce{D_{2}},0.01) (\ce{C_{2v}},13.72) (\ce{C_{2h}},0) (\ce{D_{2h}},0.05)};
        \addplot[blue!20!black,fill=blue!80!white] coordinates { (\ce{C_{1}},23.33) (\ce{C_{s}},70.89) (\ce{C_{i}},0) (\ce{C_{2}},0.13) (\ce{D_{2}},0) (\ce{C_{2v}},0) (\ce{C_{2h}},0.05) (\ce{D_{2h}},0.6)};
        \legend{$T=315\;K$, $T=298\;K$}
    \end{axis}
    \end{tikzpicture}
    \caption{Configuration \textbf{B}}
    \label{fig:bar_B}
    \end{subfigure}
    \\
    \vspace{0.5cm}
    \begin{subfigure}[b]{0.48\textwidth}
    \centering
    \begin{tikzpicture}
    \begin{axis}[ybar,
    symbolic x coords={\ce{C_{1}}, \ce{C_{s}}, \ce{C_{i}}, \ce{C_{2}}, \ce{T}, \ce{C_{2h}}, \ce{D_{2h}}},
    xtick=data,
    xlabel = {},
    ylabel = {Percentage occurrence},
    xlabel style = {font=\tiny},
    ylabel style = {font=\tiny},
    yticklabel style = {font=\tiny},
    xticklabel style = {font=\tiny},
    ymin = 0,
    bar width = 0.3cm,
    legend style={font=\tiny}]
        \addplot[red!20!black,fill=red!80!white] coordinates { (\ce{C_{1}},19.78) (\ce{C_{s}},74.88) (\ce{C_{i}},0) (\ce{C_{2}},0.06) (\ce{T},0) (\ce{C_{2h}},0) (\ce{D_{2h}},0.28)};
        \addplot[blue!20!black,fill=blue!80!white] coordinates { (\ce{C_{1}},30.41) (\ce{C_{s}},45.16) (\ce{C_{i}},17.45)  (\ce{C_{2}},0) (\ce{T},0.31)  (\ce{C_{2h}},0.33) (\ce{D_{2h}},1.34)};
        \legend{$T=315\;K$, $T=298\;K$}
    \end{axis}
    \end{tikzpicture}
    \caption{Configuration \textbf{C}}
    \label{fig:bar_C}
    \end{subfigure}
    \hfill
    \begin{subfigure}[b]{0.48\textwidth}
    \centering
    \begin{tikzpicture}
    \begin{axis}[ybar,
    symbolic x coords={\ce{C_{1}}, \ce{C_{s}}, \ce{C_{i}}, \ce{C_{2}}, \ce{T}, \ce{C_{2v}}, \ce{C_{2h}}, \ce{D_{2h}}},
    xtick=data,
    xlabel = {},
    ylabel = {Percentage occurrence},
    xlabel style = {font=\tiny},
    ylabel style = {font=\tiny},
    yticklabel style = {font=\tiny},
    xticklabel style = {font=\tiny},
    ymin = 0,
    bar width = 0.3cm,
    legend style={font=\tiny}]
        \addplot[red!20!black,fill=red!80!white] coordinates { (\ce{C_{1}},20.45) (\ce{C_{s}},42.23) (\ce{C_{i}},29.81) (\ce{C_{2}},0.19) (\ce{T},0.15) (\ce{C_{2v}},0) 
        (\ce{C_{2h}},0.36) (\ce{D_{2h}},1.81)};
        \addplot[blue!20!black,fill=blue!80!white] coordinates { (\ce{C_{1}},24.47)  (\ce{C_{s}},51)  (\ce{C_{i}},17.20)  (\ce{C_{2}},0) (\ce{T},0)  (\ce{C_{2v}},0.03)  (\ce{C_{2h}},0.39)  (\ce{D_{2h}},1.61)};
        \legend{$T=315\;K$, $T=298\;K$}
    \end{axis}
    \end{tikzpicture}
    \caption{Configuration \textbf{D}}
    \label{fig:bar_D}
    \end{subfigure}
\end{figure}

\newpage

\begin{figure}[h!]\ContinuedFloat
    \begin{subfigure}[b]{0.48\textwidth}
    \centering
    \begin{tikzpicture}
    \begin{axis}[ybar,
    symbolic x coords={\ce{C_{1}}, \ce{C_{s}}, \ce{C_{i}}, \ce{C_{2}}, \ce{T}, \ce{C_{2h}}, \ce{D_{2h}}},
    xtick=data,
    xlabel = {},
    ylabel = {Percentage occurrence},
    xlabel style = {font=\tiny},
    ylabel style = {font=\tiny},
    yticklabel style = {font=\tiny},
    xticklabel style = {font=\tiny},
    ymin = 0,
    bar width = 0.3cm,
    legend style={font=\tiny}]
        \addplot[red!20!black,fill=red!80!white] coordinates { (\ce{C_{1}},17.95) (\ce{C_{s}},52.95) (\ce{C_{i}},22.96) (\ce{C_{2}},0) (\ce{T},0.10) (\ce{C_{2h}},0.01) (\ce{D_{2h}},1.03)};
        \addplot[blue!20!black,fill=blue!80!white] coordinates { (\ce{C_{1}},18.66) (\ce{C_{s}},53.81) (\ce{C_{i}},20.73)  (\ce{C_{2}},0.02) (\ce{T},0.39) (\ce{C_{2h}},0.12) (\ce{D_{2h}},1.27)};
        \legend{$T=315\;K$, $T=298\;K$}
    \end{axis}
    \end{tikzpicture}
    \caption{Configuration \textbf{E}}
    \label{fig:bar_E}
    \end{subfigure}
    \hfill
    \begin{subfigure}[b]{0.48\textwidth}
    \centering
    \begin{tikzpicture}
    \begin{axis}[ybar,
    symbolic x coords={\ce{C_{1}}, \ce{C_{s}}, \ce{C_{i}}, \ce{C_{2}}, \ce{T}, \ce{D_{2}}, \ce{C_{2v}}, \ce{C_{2h}}, \ce{D_{2h}}},
    xtick=data,
    xlabel = {},
    ylabel = {Percentage occurrence},
    xlabel style = {font=\tiny},
    ylabel style = {font=\tiny},
    yticklabel style = {font=\tiny},
    xticklabel style = {font=\tiny},
    ymin = 0,
    bar width = 0.25cm,
    legend style={font=\tiny}]
        \addplot[red!20!black,fill=red!80!white] coordinates { (\ce{C_{1}},21.33) (\ce{C_{s}},38.04) (\ce{C_{i}},33.88) (\ce{C_{2}},0.01) (\ce{T},0.28) (\ce{D_{2}},0.05) (\ce{C_{2v}},0.02) (\ce{C_{2h}},0.05) (\ce{D_{2h}},1.34)};
        \addplot[blue!20!black,fill=blue!80!white] coordinates { (\ce{C_{1}},6.34) (\ce{C_{s}},0.03) (\ce{C_{i}},84.89)  (\ce{C_{2}},0) (\ce{T},0.63) (\ce{D_{2}},0) (\ce{C_{2v}},0)  (\ce{C_{2h}},0.24)  (\ce{D_{2h}},2.87)};
        \legend{$T=315\;K$, $T=298\;K$}
    \end{axis}
    \end{tikzpicture}
    \caption{Configuration \textbf{F}}
    \label{fig:bar_F}
    \end{subfigure}
    \\
    \vspace{0.5cm}
    \begin{subfigure}[b]{0.48\textwidth}
    \centering
    \begin{tikzpicture}
    \begin{axis}[ybar,
    symbolic x coords={\ce{C_{1}}, \ce{C_{s}}, \ce{C_{i}}, \ce{C_{2}}, \ce{T}, \ce{D_{2}}, \ce{C_{2v}}, \ce{C_{2h}}, \ce{D_{2h}}, \ce{O_{h}}},
    xtick=data,
    xlabel = {},
    ylabel = {Percentage occurrence},
    xlabel style = {font=\tiny},
    ylabel style = {font=\tiny},
    yticklabel style = {font=\tiny},
    xticklabel style = {font=\tiny},
    ymin = 0,
    bar width = 0.2cm,
    legend style={font=\tiny}]
        \addplot[red!20!black,fill=red!80!white] coordinates { (\ce{C_{1}},45.43) (\ce{C_{s}},23.20) (\ce{C_{i}},16.10) (\ce{C_{2}},4.5) (\ce{T},0.02) (\ce{D_{2}},0.4) (\ce{C_{2v}},0.43) (\ce{C_{2h}},0.06) (\ce{D_{2h}},4.86) (\ce{O_{h}},0)};
        \addplot[blue!20!black,fill=blue!80!white] coordinates { (\ce{C_{1}},9.20) (\ce{C_{s}},8.48) (\ce{C_{i}},73.94)  (\ce{C_{2}},0) (\ce{T},0.31) (\ce{D_{2}},0) (\ce{C_{2v}},0) (\ce{C_{2h}},0.14) (\ce{D_{2h}},2.91)  (\ce{O_{h}},0.01)};
        \legend{$T=315\;K$, $T=298\;K$}
    \end{axis}
    \end{tikzpicture}
    \caption{Configuration \textbf{G}}
    \label{fig:bar_G}
    \end{subfigure}
    \hfill
    \begin{subfigure}[b]{0.48\textwidth}
    \centering
    \begin{tikzpicture}
    \begin{axis}[ybar,
    symbolic x coords={\ce{C_{1}}, \ce{C_{s}}, \ce{C_{i}}, \ce{C_{2}}, \ce{C_{2v}}, \ce{C_{2h}}, \ce{D_{2h}}},
    xtick=data,
    xlabel = {},
    ylabel = {Percentage occurrence},
    xlabel style = {font=\tiny},
    ylabel style = {font=\tiny},
    yticklabel style = {font=\tiny},
    xticklabel style = {font=\tiny},
    ymin = 0,
    bar width = 0.3cm,
    legend style={font=\tiny}]
        \addplot[red!20!black,fill=red!80!white] coordinates { (\ce{C_{1}},21.21) (\ce{C_{s}},73.04) (\ce{C_{i}},0) (\ce{C_{2}},0.03) (\ce{C_{2v}},0) 
        (\ce{C_{2h}},0) (\ce{D_{2h}},0.72)};
        \addplot[blue!20!black,fill=blue!80!white] coordinates { (\ce{C_{1}},36.6)  (\ce{C_{s}},55.42)  (\ce{C_{i}},0.03)  (\ce{C_{2}},0.51)  (\ce{C_{2v}},0.03)  (\ce{C_{2h}},0.11)  (\ce{D_{2h}},2.27)};
        \legend{$T=315\;K$, $T=298\;K$}
    \end{axis}
    \end{tikzpicture}
    \caption{Configuration \textbf{H}}
    \label{fig:bar_H}
    \end{subfigure}
    
        \caption{Percentage occurrence of various symmetries during a dynamical run as shown in Section \ref{sec:Time_sym} of the SI}
        
    \label{fig:bar_and_time_for_sym_supp}
\end{figure}

\newpage

\section{Principal component analysis of each dynamical run at a temperature of 298 K}

\begin{figure}[h!]
    \centering
    \begin{subfigure}[b]{0.48\textwidth}
    \centering
    \begin{tikzpicture}
    \begin{axis}[ybar,
    symbolic x coords={$5^{th}$, $4^{th}$, $3^{rd}$, $2^{nd}$, $1^{st}$},
    xtick=data,
    nodes near coords,
    xlabel = {Eigenmode index},
    ylabel = {Percentage of variance explained by the structure},
    xlabel style = {font=\tiny},
    ylabel style = {font=\tiny},
    yticklabel style = {font=\tiny},
    xticklabel style = {font=\tiny},
    x dir = reverse,
    ymin = 0,
    ymax = 90,
    nodes near coords align={vertical},
    bar width = 0.8cm,
    every node near coord/.append style={font=\scriptsize}]
        \addplot[blue!20!black,fill=blue!80!white] coordinates {($5^{th}$,4.21574) ($4^{th}$,5.21843) ($3^{rd}$,5.55048) ($2^{nd}$,8.01959) ($1^{st}$,35.1755)};
    \end{axis}
    \end{tikzpicture}
    \caption{Configuration \textbf{A}}
    \label{fig:PCA_A}
    \end{subfigure}
    \hfill
    \begin{subfigure}[b]{0.48\textwidth}
    \centering
    \begin{tikzpicture}
    \begin{axis}[ybar,
    symbolic x coords={$5^{th}$, $4^{th}$, $3^{rd}$, $2^{nd}$, $1^{st}$},
    xtick=data,
    nodes near coords,
    xlabel = {Eigenmode index},
    ylabel = {Percentage of variance explained by the structure},
    xlabel style = {font=\tiny},
    ylabel style = {font=\tiny},
    yticklabel style = {font=\tiny},
    xticklabel style = {font=\tiny},
    x dir = reverse,
    ymin = 0,
    ymax = 90,
    nodes near coords align={vertical},
    bar width = 0.8cm,
    every node near coord/.append style={font=\scriptsize}]
        \addplot[blue!20!black,fill=blue!80!white] coordinates {($5^{th}$,2.32727) ($4^{th}$,5.742) ($3^{rd}$,6.61585) ($2^{nd}$,10.8172) ($1^{st}$,53.975)};
    \end{axis}
    \end{tikzpicture}
    \caption{Configuration \textbf{B}}
    \label{fig:PCA_B}
    \end{subfigure}
    \\
    \vspace{0.5cm}
    \centering
    \begin{subfigure}[b]{0.48\textwidth}
    \centering
    \begin{tikzpicture}
    \begin{axis}[ybar,
    symbolic x coords={$5^{th}$, $4^{th}$, $3^{rd}$, $2^{nd}$, $1^{st}$},
    xtick=data,
    nodes near coords,
    xlabel = {Eigenmode index},
    ylabel = {Percentage of variance explained by the structure},
    xlabel style = {font=\tiny},
    ylabel style = {font=\tiny},
    yticklabel style = {font=\tiny},
    xticklabel style = {font=\tiny},
    x dir = reverse,
    ymin = 0,
    ymax = 90,
    nodes near coords align={vertical},
    bar width = 0.8cm,
    every node near coord/.append style={font=\scriptsize}]
        \addplot[blue!20!black,fill=blue!80!white] coordinates {($5^{th}$,2.33184) ($4^{th}$,2.59562) ($3^{rd}$,4.16105) ($2^{nd}$,11.5626) ($1^{st}$,64.6462)};
    \end{axis}
    \end{tikzpicture}
    \caption{Configuration \textbf{C}}
    \label{fig:PCA_C}
    \end{subfigure}
    \hfill
    \begin{subfigure}[b]{0.48\textwidth}
    \centering
    \begin{tikzpicture}
    \begin{axis}[ybar,
    symbolic x coords={$5^{th}$, $4^{th}$, $3^{rd}$, $2^{nd}$, $1^{st}$},
    xtick=data,
    nodes near coords,
    xlabel = {Eigenmode index},
    ylabel = {Percentage of variance explained by the structure},
    xlabel style = {font=\tiny},
    ylabel style = {font=\tiny},
    yticklabel style = {font=\tiny},
    xticklabel style = {font=\tiny},
    x dir = reverse,
    ymin = 0,
    ymax = 90,
    nodes near coords align={vertical},
    bar width = 0.8cm,
    every node near coord/.append style={font=\scriptsize}]
        \addplot[blue!20!black,fill=blue!80!white] coordinates {($5^{th}$,2.15157) ($4^{th}$,3.12021) ($3^{rd}$,5.69516) ($2^{nd}$,8.8272) ($1^{st}$,61.8859)};
    \end{axis}
    \end{tikzpicture}
    \caption{Configuration \textbf{D}}
    \label{fig:PCA_D}
    \end{subfigure}
\end{figure}

\newpage

\begin{figure}[h!]\ContinuedFloat
    \centering
    \begin{subfigure}[b]{0.48\textwidth}
    \centering
    \begin{tikzpicture}
    \begin{axis}[ybar,
    symbolic x coords={$5^{th}$, $4^{th}$, $3^{rd}$, $2^{nd}$, $1^{st}$},
    xtick=data,
    nodes near coords,
    xlabel = {Eigenmode index},
    ylabel = {Percentage of variance explained by the structure},
    xlabel style = {font=\tiny},
    ylabel style = {font=\tiny},
    yticklabel style = {font=\tiny},
    xticklabel style = {font=\tiny},
    x dir = reverse,
    ymin = 0,
    ymax = 90,
    nodes near coords align={vertical},
    bar width = 0.8cm,
    every node near coord/.append style={font=\scriptsize}]
        \addplot[blue!20!black,fill=blue!80!white] coordinates {($5^{th}$,1.93039) ($4^{th}$,2.3564) ($3^{rd}$,3.02039) ($2^{nd}$,6.43647) ($1^{st}$,69.7652)};
    \end{axis}
    \end{tikzpicture}
    \caption{Configuration \textbf{E}}
    \label{fig:PCA_E}
    \end{subfigure}
    \hfill
    \begin{subfigure}[b]{0.48\textwidth}
    \centering
    \begin{tikzpicture}
    \begin{axis}[ybar,
    symbolic x coords={$5^{th}$, $4^{th}$, $3^{rd}$, $2^{nd}$, $1^{st}$},
    xtick=data,
    nodes near coords,
    xlabel = {Eigenmode index},
    ylabel = {Percentage of variance explained by the structure},
    xlabel style = {font=\tiny},
    ylabel style = {font=\tiny},
    yticklabel style = {font=\tiny},
    xticklabel style = {font=\tiny},
    x dir = reverse,
    ymin = 0,
    ymax = 90,
    nodes near coords align={vertical},
    bar width = 0.8cm,
    every node near coord/.append style={font=\scriptsize}]
        \addplot[blue!20!black,fill=blue!80!white] coordinates {($5^{th}$,3.73029) ($4^{th}$,5.50615) ($3^{rd}$,6.36377) ($2^{nd}$,16.6263) ($1^{st}$,27.8666)};
    \end{axis}
    \end{tikzpicture}
    \caption{Configuration \textbf{F}}
    \label{fig:PCA_F}
    \end{subfigure}
    \\
    \vspace{0.5cm}
    \centering
    \begin{subfigure}[b]{0.48\textwidth}
    \centering
    \begin{tikzpicture}
    \begin{axis}[ybar,
    symbolic x coords={$5^{th}$, $4^{th}$, $3^{rd}$, $2^{nd}$, $1^{st}$},
    xtick=data,
    nodes near coords,
    xlabel = {Eigenmode index},
    ylabel = {Percentage of variance explained by the structure},
    xlabel style = {font=\tiny},
    ylabel style = {font=\tiny},
    yticklabel style = {font=\tiny},
    xticklabel style = {font=\tiny},
    x dir = reverse,
    ymin = 0,
    ymax = 90,
    nodes near coords align={vertical},
    bar width = 0.8cm,
    every node near coord/.append style={font=\scriptsize}]
        \addplot[blue!20!black,fill=blue!80!white] coordinates {($5^{th}$,2.72621) ($4^{th}$,3.56308) ($3^{rd}$,6.38977) ($2^{nd}$,12.5805) ($1^{st}$,51.5009)};
    \end{axis}
    \end{tikzpicture}
    \caption{Configuration \textbf{G}}
    \label{fig:PCA_G}
    \end{subfigure}
    \hfill
    \begin{subfigure}[b]{0.48\textwidth}
    \centering
    \begin{tikzpicture}
    \begin{axis}[ybar,
    symbolic x coords={$5^{th}$, $4^{th}$, $3^{rd}$, $2^{nd}$, $1^{st}$},
    xtick=data,
    nodes near coords,
    xlabel = {Eigenmode index},
    ylabel = {Percentage of variance explained by the structure},
    xlabel style = {font=\tiny},
    ylabel style = {font=\tiny},
    yticklabel style = {font=\tiny},
    xticklabel style = {font=\tiny},
    x dir = reverse,
    ymin = 0,
    ymax = 90,
    nodes near coords align={vertical},
    bar width = 0.8cm,
    every node near coord/.append style={font=\scriptsize}]
        \addplot[blue!20!black,fill=blue!80!white] coordinates {($5^{th}$,2.36114) ($4^{th}$,3.49018) ($3^{rd}$,7.9203) ($2^{nd}$,22.8022) ($1^{st}$,47.9971)};
    \end{axis}
    \end{tikzpicture}
    \caption{Configuration \textbf{H}}
    \label{fig:PCA_H}
    \end{subfigure}
    
        \caption{Percentage of variance explained by the first $5$ dominant eigenmodes obtained from PCA of each of the $8$ dynamical runs at $T=298\;K$ as shown in Subsection \ref{subsec:T_298} of the SI}
        
    \label{fig:PCA}
\end{figure}

\newpage

\section[Energy distribution with symmetries shown at a lower threshold]{Energy distribution with symmetries shown at a \\ lower threshold}

We use the Visual Molecular Dynamics (VMD) toolkit to measure the molecular point group symmetry of every molecular structure. The software uses a tolerance to determine the symmetry. It is pertinent to note that relaxing this symmetry tolerance can naturally lead to more structures being classified as symmetric. For instance, given a molecular structure, a tolerance of $0.1$ units might identify the symmetry as \ce{X}, but raising the tolerance to $0.25$ might change the identified symmetry to \ce{Y} instead, where -- in general -- \ce{Y} is a larger point group than \ce{X}. It is easy to understand this behaviour intuitively based on the fact that VMD identifies the highest possible point group symmetry for a given structure, and that raising the symmetry tolerance allows more room for every atom to \textit{jiggle} as the optimal symmetry is being guessed.

\begin{figure}
  \centering
  \begin{tabular}{c}
%   \scalebox{0.8}{\input{all_md_higher}}
  \includegraphics[width=10.3cm]{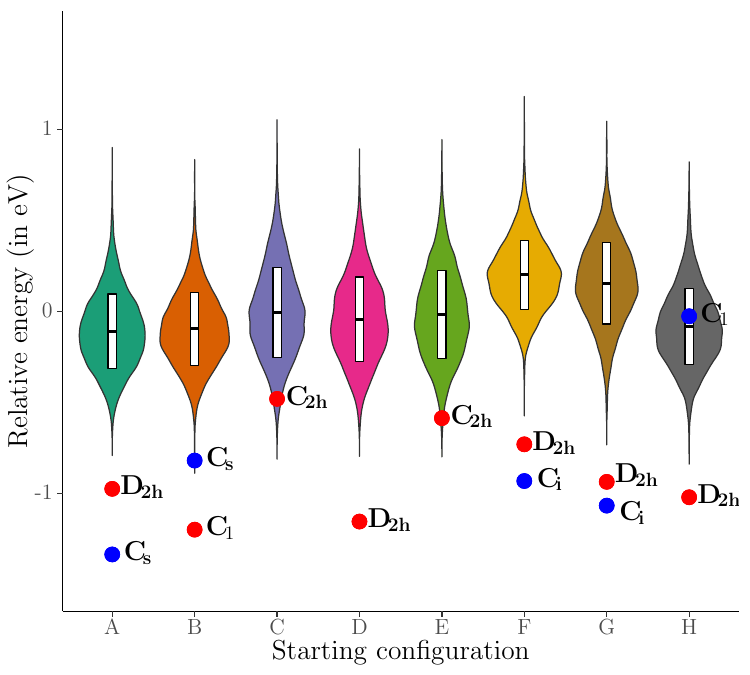}
  \\[-9.4cm]
%   \scalebox{0.8}{\input{seg_md}}
\includegraphics[width=10.3cm]{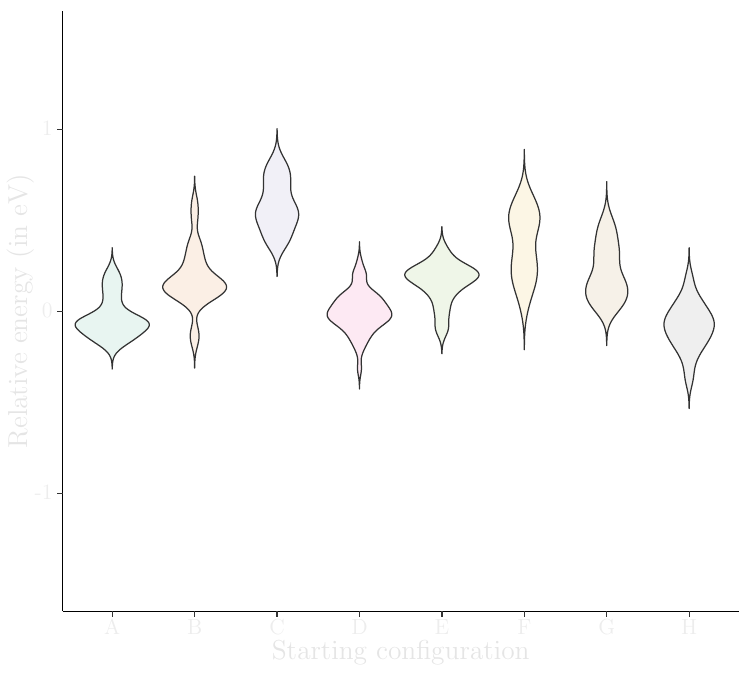}
  \end{tabular}
  \captionsetup{font={small}}
  \caption{Distribution of the energies of the ensemble of structures, but with the symmetries marked at a tolerance of $0.25$ -- instead of $0.1$, as presented in the paper. The blue dots represent the energy of the time-average structure over the entire dynamical run, and the red dot represents the energy of the time-average structure of the longest high symmetry phase during the run. The lighter color overlaps on each of the violin plots represent the energy distribution for the above high symmetry phase. Dots missing from the graph were found at energies higher than the scale of the figure. The plot also shows the mean and standard deviation of the energy spread within the shapes as white boxes. As is evident, some symmetries change but the overall most stable structure still has a very low \ce{C_s} symmetry}
  \label{fig:violin_new}
\end{figure}

To this effect, with a tolerance of $0.1$, the most stable molecular structure in the entire ensemble that we have considered, has a point group symmetry of \ce{C_1}, but it increases to a symmetry of \ce{C_s} upon specifying a higher tolerance of $0.25$. Even though both resultant symmetries in this particular case are low in general, and we do not expect this behaviour to change the overall narrative of the paper, we present Fig.~\ref{fig:violin_new} with symmetries calculated at a higher tolerance than what was presented in the paper, for the sake of completeness. The use of $0.1$ as the tolerance for identification of point group symmetries can be justfied by that fact that the normalized maximum displacement of an atom in successive molecular dynamics steps is less than the specified tolerance, as should be the case. Here, we define the nomarmalized displacement as the ratio of the atomic displacement and the molecular radius. In our case, the normalized maximum displacement for any atom between any two successive steps in any of the eight \textit{ab inito} molecular dynamics runs was $0.0047$, which is well with the specified tolerance. Thus, we argue that the used value of tolerance is appropriate.
\\
\\

\newpage

\section[Structural difference between the time-average structure and the most dominant mode from PCA of each dynamical run at T = 298 K]{Structural difference between the time-average \\ structure and the most dominant mode from PCA of each dynamical run at T = 298 K}

\begin{figure}[h!]
    \centering
    \begin{subfigure}[b]{0.5\textwidth}
    \centering
    \hspace{-1cm}
    \includegraphics[width=0.9\textwidth]{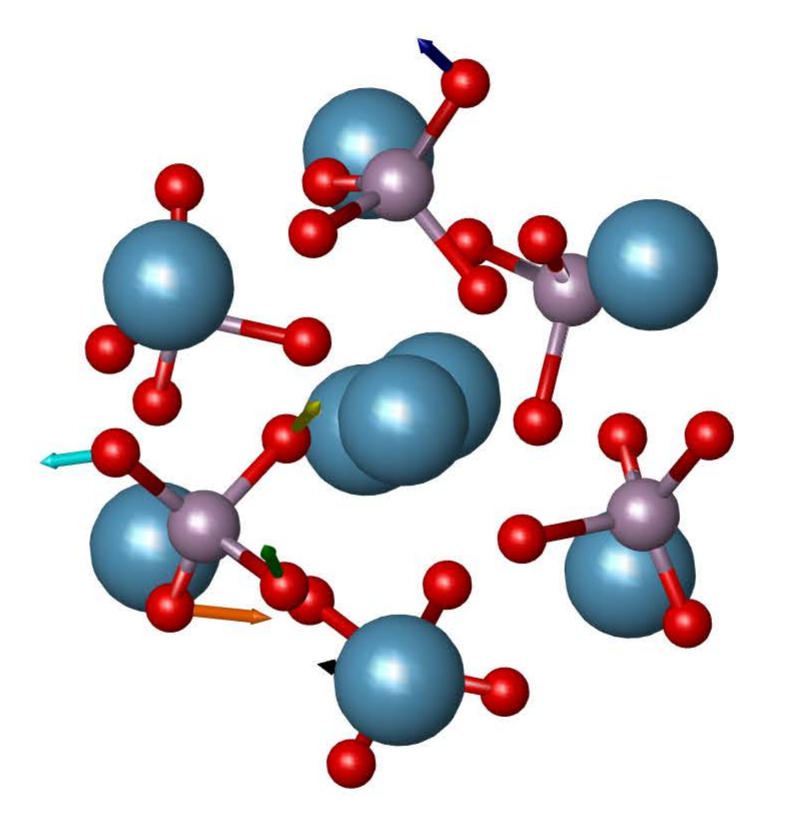}
    \caption{}
    \label{fig:disp_A}
    \end{subfigure}
    %\hfill
    \raisebox{4.5cm}{
    \begin{subfigure}[b]{0.4\textwidth}
    \centering
    \scriptsize
    \hspace{-1cm}
    \begin{tabular}{|c|l|}
    \hline
    \textbf{Arrow Color} & \textbf{Displacement Vector (in \AA)} \\ \hline
    \cellcolor{rgb,255:red,0; green,0; blue,0} & \hspace{2em} [0.047,\; -0.338,\; 0.018] \\
    \hline
    \cellcolor{rgb,255:red,0; green,179; blue,0} & \hspace{2em} [0.148,\; -0.024,\; 0.278] \\
    \hline
    \cellcolor{rgb,255:red,255; green,255; blue,0} & \hspace{2em} [0.08,\; 0.181,\; 0.271] \\
    \hline
    \cellcolor{rgb,255:red,0; green,255; blue,255} & \hspace{2em} [-0.01,\; -0.336,\; -0.14] \\
    \hline
    \cellcolor{rgb,255:red,255; green,102; blue,0} & \hspace{2em} [-0.199,\; 0.358,\; -0.077] \\
    \hline
    \cellcolor{rgb,255:red,0; green,0; blue,204} & \hspace{2em} [0.058,\; -0.185,\; 0.201] \\
    \hline
    \end{tabular}
%    \caption{}
    \label{fig:disp_table_A}
    \end{subfigure}
    }
    \\
    \vspace{0.5cm}
    \begin{subfigure}[b]{0.5\textwidth}
    \centering
    \hspace{-1cm}
    \includegraphics[width=0.9\textwidth]{B_pca_diff.pdf}
    \caption{}
    \label{fig:disp_B}
    \end{subfigure}
    %\hfill
    \raisebox{4.5cm}{
    \begin{subfigure}[b]{0.4\textwidth}
    \centering
    \scriptsize
    \hspace{-1cm}
    \begin{tabular}{|c|l|}
    \hline
    \textbf{Arrow Color} & \textbf{Displacement Vector (in \AA)} \\ \hline
    \cellcolor{rgb,255:red,0; green,179; blue,0} & \hspace{2em} [-0.252,\; 0.085,\; 0.214] \\
    \hline
    \cellcolor{rgb,255:red,255; green,255; blue,0} & \hspace{2em} [0.228,\; 0.224,\; 0.121] \\
    \hline
    \cellcolor{rgb,255:red,0; green,255; blue,255} & \hspace{2em} [-0.081,\; 0.295,\; -0.103] \\
    \hline
    \cellcolor{rgb,255:red,255; green,102; blue,0} & \hspace{2em} [-0.276,\; -0.092,\; -0.194] \\
    \hline
    \end{tabular}
%    \caption{}
    \label{fig:disp_table_B}
    \end{subfigure}
    }
\end{figure}

\newpage

\begin{figure}[h!]\ContinuedFloat
    \centering
    \begin{subfigure}[b]{0.5\textwidth}
    \centering
    \hspace{-1cm}
    \includegraphics[width=0.9\textwidth]{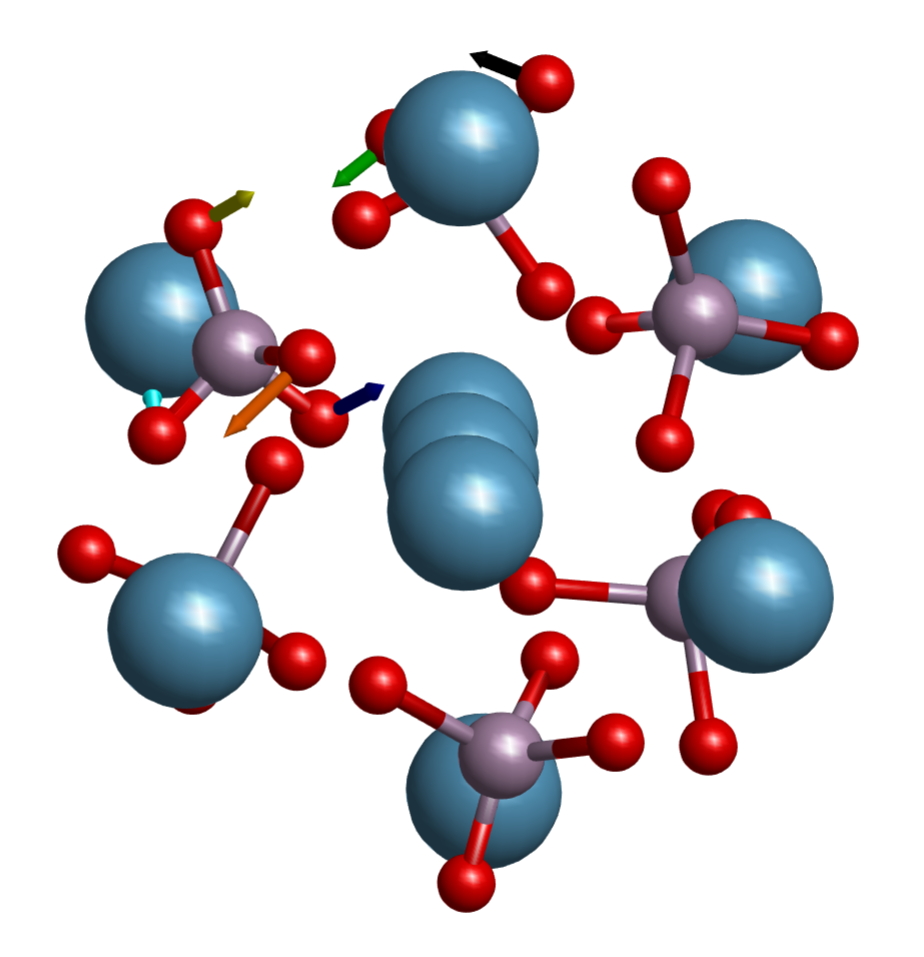}
    \caption{}
    \label{fig:disp_C}
    \end{subfigure}
    %\hfill
    \raisebox{4.5cm}{
    \begin{subfigure}[b]{0.4\textwidth}
    \centering
    \scriptsize
    \hspace{-1cm}
    \begin{tabular}{|c|l|}
    \hline
    \textbf{Arrow Color} & \textbf{Displacement Vector (in \AA)} \\ \hline
    \cellcolor{rgb,255:red,0; green,0; blue,0} & \hspace{2em} [-0.174,\; -0.191,\; 0.092] \\
    \hline
    \cellcolor{rgb,255:red,0; green,179; blue,0} & \hspace{2em} [-0.025,\; -0.232,\; -0.137] \\
    \hline
    \cellcolor{rgb,255:red,255; green,255; blue,0} & \hspace{2em} [0.196,\; 0.121,\; 0.173] \\
    \hline
    \cellcolor{rgb,255:red,0; green,255; blue,255} & \hspace{2em} [-0.361,\; 0.206,\; -0.003] \\
    \hline
    \cellcolor{rgb,255:red,255; green,102; blue,0} & \hspace{2em} [0.03,\; -0.35,\; -0.202] \\
    \hline
    \cellcolor{rgb,255:red,0; green,0; blue,204} & \hspace{2em} [0.253,\; 0.098,\; 0.19] \\
    \hline
    \end{tabular}
%    \caption{}
    \label{fig:disp_table_C}
    \end{subfigure}
    }
    \\
    \vspace{0.5cm}
    \begin{subfigure}[b]{0.5\textwidth}
    \centering
    \hspace{-1cm}
    \includegraphics[width=0.9\textwidth]{D_pca_diff.pdf}
    \caption{}
    \label{fig:disp_D}
    \end{subfigure}
    %\hfill
    \raisebox{4cm}{
    \begin{subfigure}[b]{0.4\textwidth}
    \centering
    \scriptsize
    \hspace{-1cm}
    \begin{tabular}{|c|l|}
    \hline
    \textbf{Arrow Color} & \textbf{Displacement Vector (in \AA)} \\ \hline
    \cellcolor{rgb,255:red,0; green,0; blue,0} & \hspace{2em} [0.293,\; -0.039,\; -0.233] \\
    \hline
    \cellcolor{rgb,255:red,0; green,179; blue,0} & \hspace{2em} [-0.042,\; -0.25,\; 0.249] \\
    \hline
    \cellcolor{rgb,255:red,255; green,255; blue,0} & \hspace{2em} [-0.226,\; 0.282,\; -0.058] \\
    \hline
    \cellcolor{rgb,255:red,0; green,255; blue,255} & \hspace{2em} [-0.143,\; 0.27,\; 0.039] \\
    \hline
    \cellcolor{rgb,255:red,255; green,102; blue,0} & \hspace{2em} [0.002,\; 0.1,\; 0.289] \\
    \hline
    \cellcolor{rgb,255:red,0; green,0; blue,204} & \hspace{2em} [0.213,\; -0.221,\; 0.02] \\
    \hline
    \end{tabular}
%    \caption{}
    \label{fig:disp_table_D}
    \end{subfigure}
    }
\end{figure}

\newpage

\begin{figure}[h!]\ContinuedFloat
    \centering
    \begin{subfigure}[b]{0.5\textwidth}
    \centering
    \hspace{-1cm}
    \includegraphics[width=0.9\textwidth]{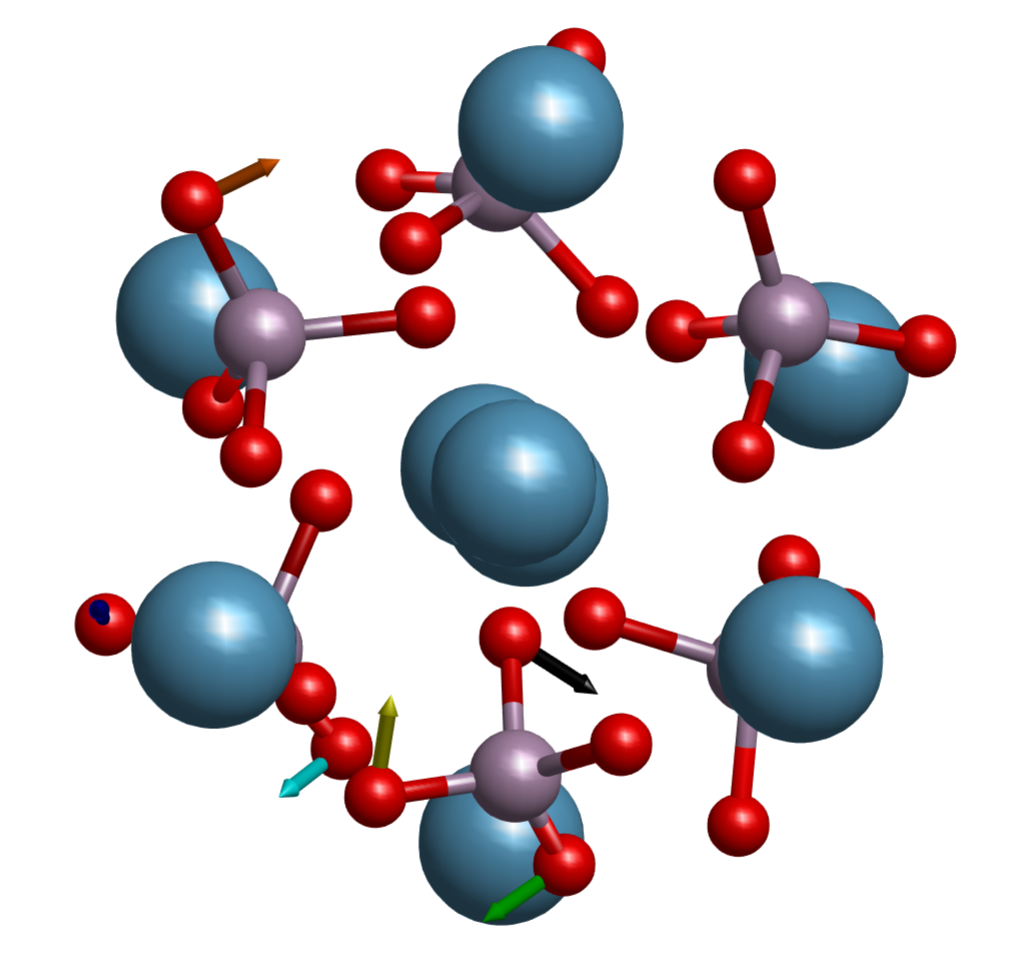}
    \caption{}
    \label{fig:disp_E}
    \end{subfigure}
    %\hfill
    \raisebox{4cm}{
    \begin{subfigure}[b]{0.4\textwidth}
    \centering
    \scriptsize
    \hspace{-1cm}
    \begin{tabular}{|c|l|}
    \hline
    \textbf{Arrow Color} & \textbf{Displacement Vector (in \AA)} \\ \hline
    \cellcolor{rgb,255:red,0; green,0; blue,0} & \hspace{2em} [0.029,\; 0.295,\; -0.219] \\
    \hline
    \cellcolor{rgb,255:red,0; green,179; blue,0} & \hspace{2em} [0.008,\; -0.292,\; -0.145] \\
    \hline
    \cellcolor{rgb,255:red,255; green,255; blue,0} & \hspace{2em} [-0.047,\; 0.080,\; 0.311] \\
    \hline
    \cellcolor{rgb,255:red,0; green,255; blue,255} & \hspace{2em} [0.104,\; -0.274,\; -0.093] \\
    \hline
    \cellcolor{rgb,255:red,255; green,102; blue,0} & \hspace{2em} [0.053,\; 0.283,\; 0.116] \\
    \hline
    \cellcolor{rgb,255:red,0; green,0; blue,204} & \hspace{2em} [0.230,\; -0.145,\; 0.146] \\
    \hline
    \end{tabular}
%    \caption{}
    \label{fig:disp_table_E}
    \end{subfigure}
    }
    \\          
    \vspace{0.5cm}
    \begin{subfigure}[b]{0.5\textwidth}
    \centering
    \hspace{-1cm}
    \includegraphics[width=0.9\textwidth]{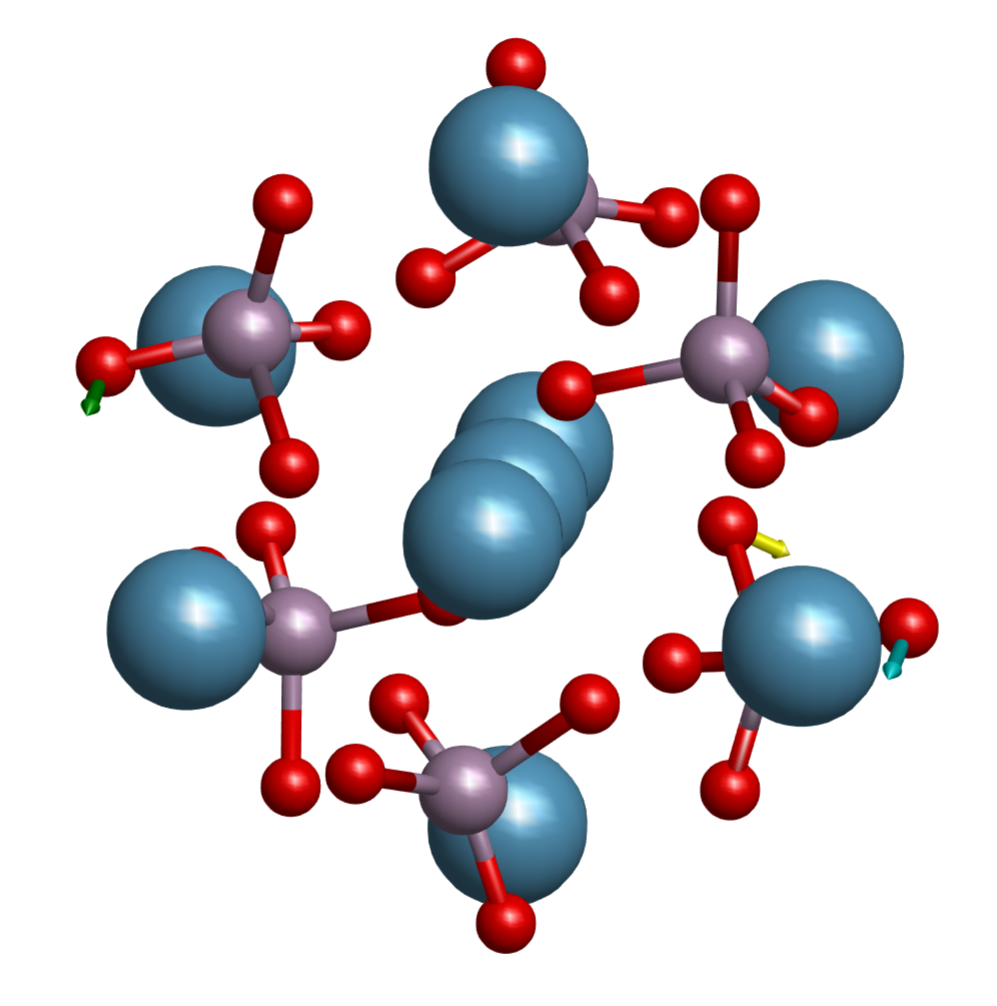}
    \caption{}
    \label{fig:disp_F}
    \end{subfigure}
    %\hfill
    \raisebox{4cm}{
    \begin{subfigure}[b]{0.4\textwidth}
    \centering
    \scriptsize
    \hspace{-1cm}
    \begin{tabular}{|c|l|}
    \hline
    \textbf{Arrow Color} & \textbf{Displacement Vector (in \AA)} \\ \hline
    \cellcolor{rgb,255:red,0; green,179; blue,0} & \hspace{2em} [0.184,\; -0.163,\; -0.083] \\
    \hline
    \cellcolor{rgb,255:red,255; green,255; blue,0} & \hspace{2em} [-0.007,\; 0.232,\; -0.142] \\
    \hline
    \cellcolor{rgb,255:red,0; green,255; blue,255} & \hspace{2em} [0.222,\; -0.187,\; -0.075] \\
    \hline
    \end{tabular}
%    \caption{}
    \label{fig:disp_table_F}
    \end{subfigure}
    }
\end{figure}

\newpage

\begin{figure}[h!]\ContinuedFloat
    \centering
    \begin{subfigure}[b]{0.5\textwidth}
    \centering
    \hspace{-1cm}
    \includegraphics[width=0.9\textwidth]{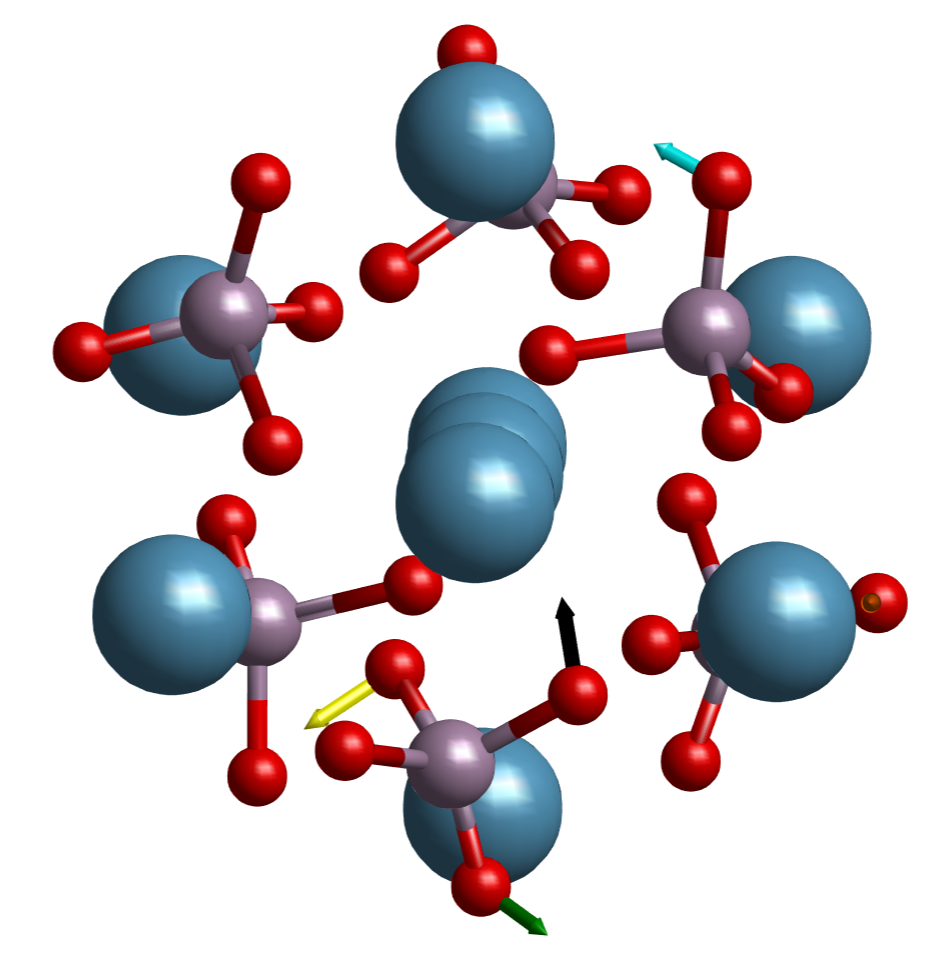}
    \caption{}
    \label{fig:disp_G}
    \end{subfigure}
    %\hfill
    \raisebox{4.5cm}{
    \begin{subfigure}[b]{0.4\textwidth}
    \centering
    \scriptsize
    \hspace{-1cm}
    \begin{tabular}{|c|l|}
    \hline
    \textbf{Arrow Color} & \textbf{Displacement Vector (in \AA)} \\ \hline
    \cellcolor{rgb,255:red,0; green,0; blue,0} & \hspace{2em} [-0.16,\; 0.088,\; 0.279] \\
    \hline
    \cellcolor{rgb,255:red,0; green,179; blue,0} & \hspace{2em} [0.242,\; 0.067,\; -0.127] \\
    \hline
    \cellcolor{rgb,255:red,255; green,255; blue,0} & \hspace{2em} [-0.304,\; -0.117,\; -0.255] \\
    \hline
    \cellcolor{rgb,255:red,0; green,255; blue,255} & \hspace{2em} [-0.253,\; -0.062,\; 0.085] \\
    \hline
    \cellcolor{rgb,255:red,255; green,102; blue,0} & \hspace{2em} [0.151,\; -0.191,\; 0.086] \\
    \hline
    \end{tabular}
%    \caption{}
    \label{fig:disp_table_G}
    \end{subfigure}
    }
    \\              
    \vspace{0.5cm}
    \begin{subfigure}[b]{0.5\textwidth}
    \centering
    \hspace{-1cm}
    \includegraphics[width=0.9\textwidth]{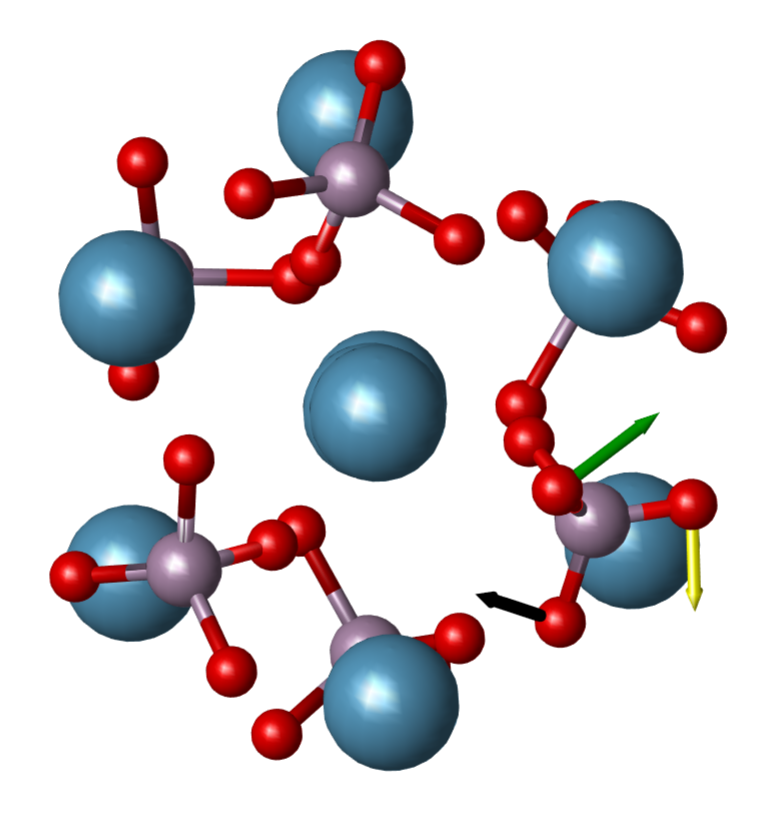}
    \caption{}
    \label{fig:disp_H}
    \end{subfigure}
    %\hfill
    \raisebox{4.5cm}{
    \begin{subfigure}[b]{0.4\textwidth}
    \centering
    \scriptsize
    \hspace{-1cm}
    \begin{tabular}{|c|l|}
    \hline
    \textbf{Arrow Color} & \textbf{Displacement Vector (in \AA)} \\ \hline
    \cellcolor{rgb,255:red,0; green,0; blue,0} & \hspace{2em} [0.335,\; -0.134,\; 0.255] \\
    \hline
    \cellcolor{rgb,255:red,0; green,179; blue,0} & \hspace{2em} [-0.411,\; 0.155,\; 0.225] \\
    \hline
    \cellcolor{rgb,255:red,255; green,255; blue,0} & \hspace{2em} [0.004,\; 0.020,\; -0.509] \\
    \hline
    \end{tabular}
%    \caption{}
    \label{fig:disp_table_H}
    \end{subfigure}
    }
    \caption{The difference between the time-averaged structure (shown) and the structure corresponding to the most dominant PCA eigenmode for each of the $8$ dynamical runs at $T=298\;K$ as shown in Subsection \ref{subsec:T_298} of the SI. The arrows have been elongated threefold for clarity. }
\end{figure}

\newpage

\section{k-means clustering for each dynamical run}

\begin{figure}[h!]
    \centering
    \begin{subfigure}[b]{0.45\textwidth}
      \centering
      \includegraphics[width=\textwidth]{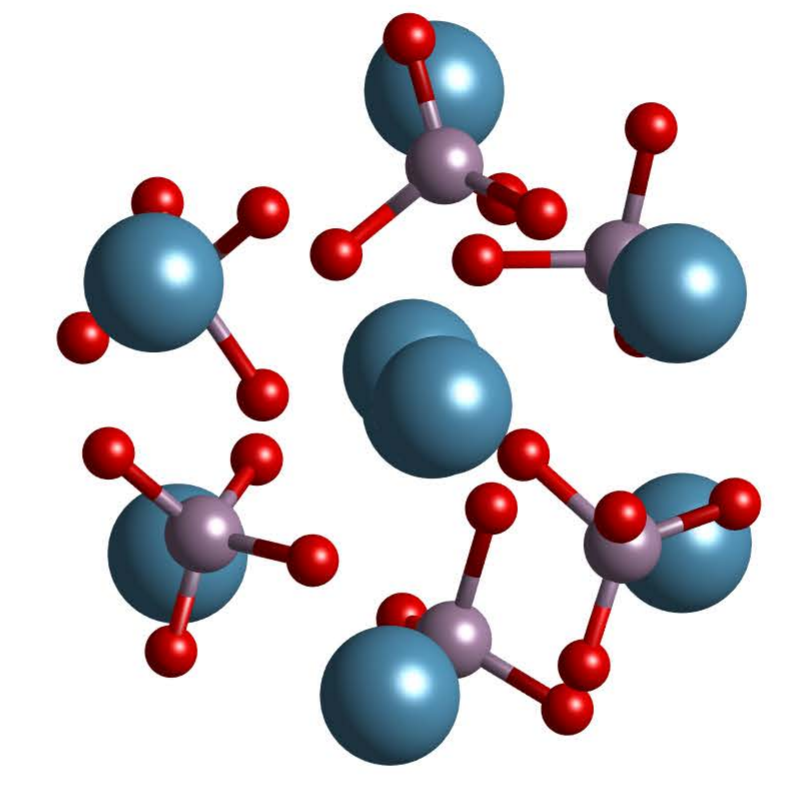}
      \captionsetup{font={small}}
      \caption{Configuration \textbf{A} -- Cluster 1 (\ce{C1})}
      \label{fig:kmeans_A_1}
    \end{subfigure}
    \hfill
    \begin{subfigure}[b]{0.45\textwidth}
      \centering
      \includegraphics[width=\textwidth]{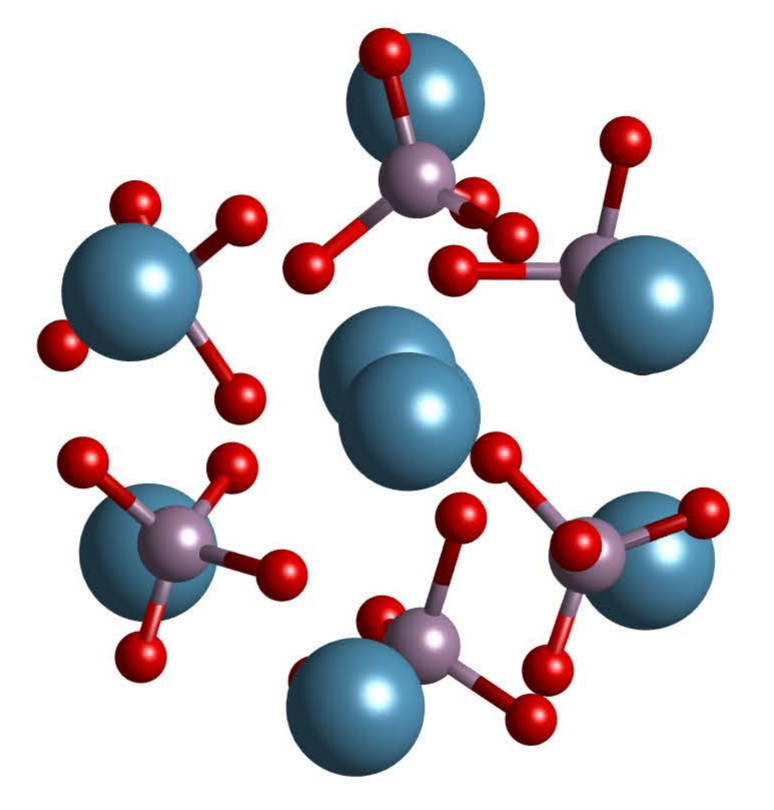}
      \captionsetup{font={small}}
      \caption{Configuration \textbf{A} -- Cluster 2 (\ce{C1})}
      \label{fig:kmeans_A_2}
    \end{subfigure}
    \\
    \vspace{0.5cm}
    \begin{subfigure}[b]{0.45\textwidth}
      \centering
      \includegraphics[width=\textwidth]{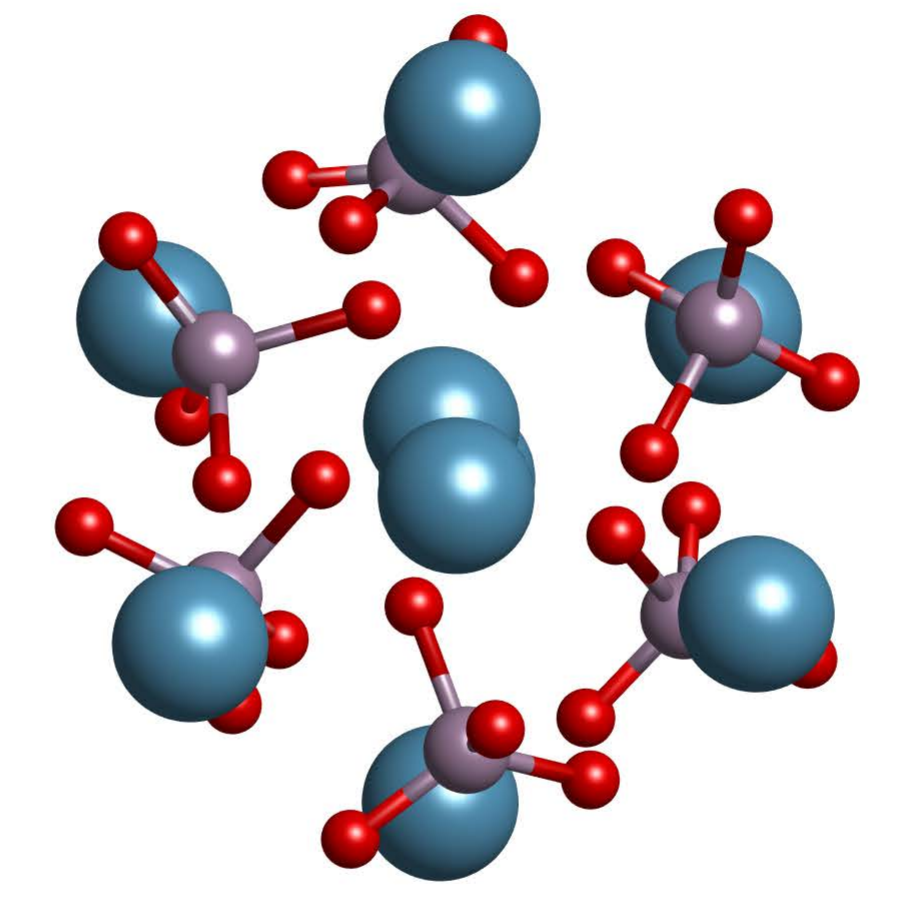}
      \captionsetup{font={small}}
      \caption{Configuration \textbf{B} -- Cluster 1 (\ce{C1})}
      \label{fig:kmeans_B_1}
    \end{subfigure}
    \hfill
    \begin{subfigure}[b]{0.45\textwidth}
      \centering
      \includegraphics[width=\textwidth]{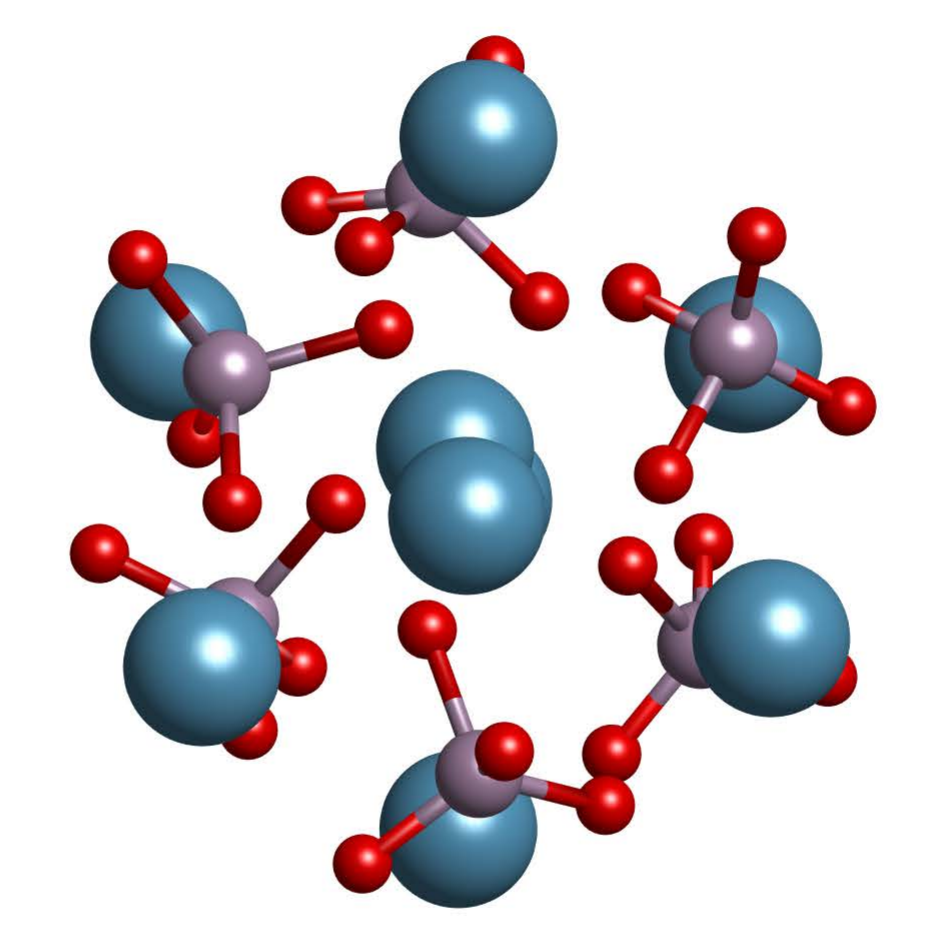}
      \captionsetup{font={small}}
      \caption{Configuration \textbf{B} -- Cluster 2 (\ce{C1})}
      \label{fig:kmeans_B_2}
    \end{subfigure}
\end{figure}

\newpage

\begin{figure}[h!]\ContinuedFloat
    \centering
    \begin{subfigure}[b]{0.45\textwidth}
      \centering
      \includegraphics[width=\textwidth]{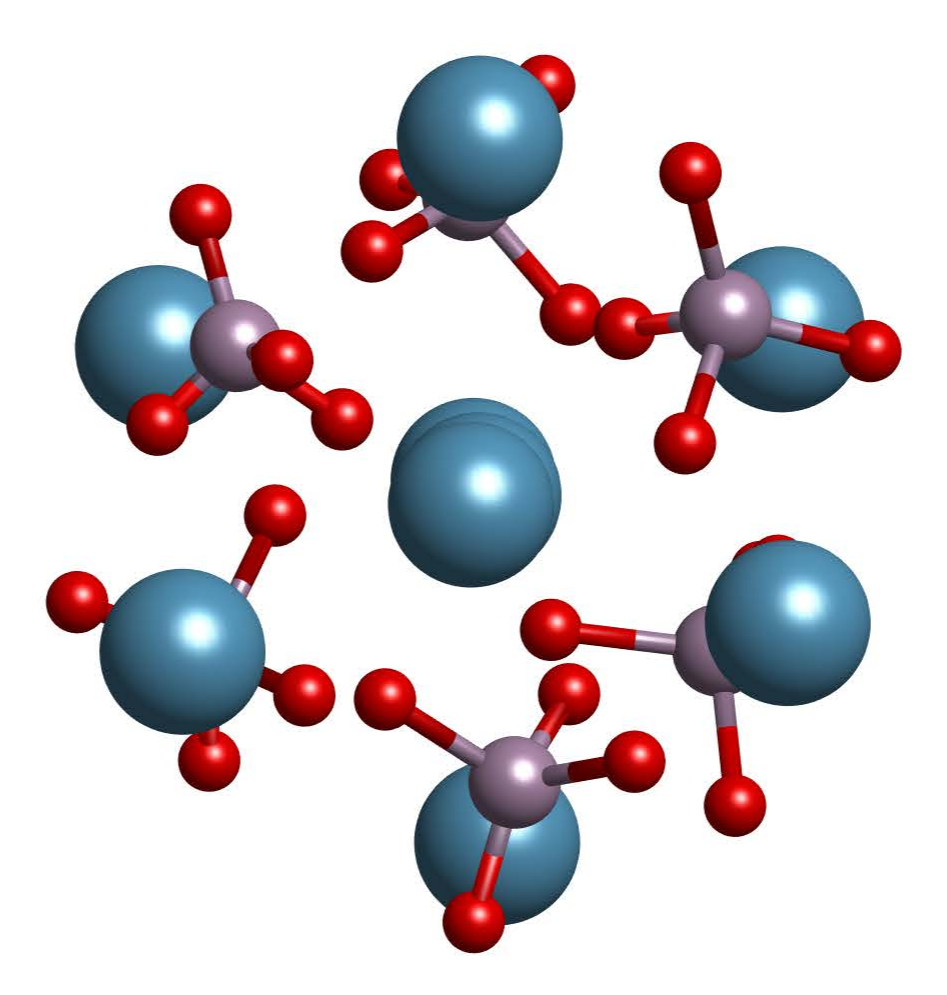}
      \captionsetup{font={small}}
      \caption{Configuration \textbf{C} -- Cluster 1 (\ce{C1})}
      \label{fig:kmeans_C_1}
    \end{subfigure}
    \hfill
    \begin{subfigure}[b]{0.45\textwidth}
      \centering
      \includegraphics[width=\textwidth]{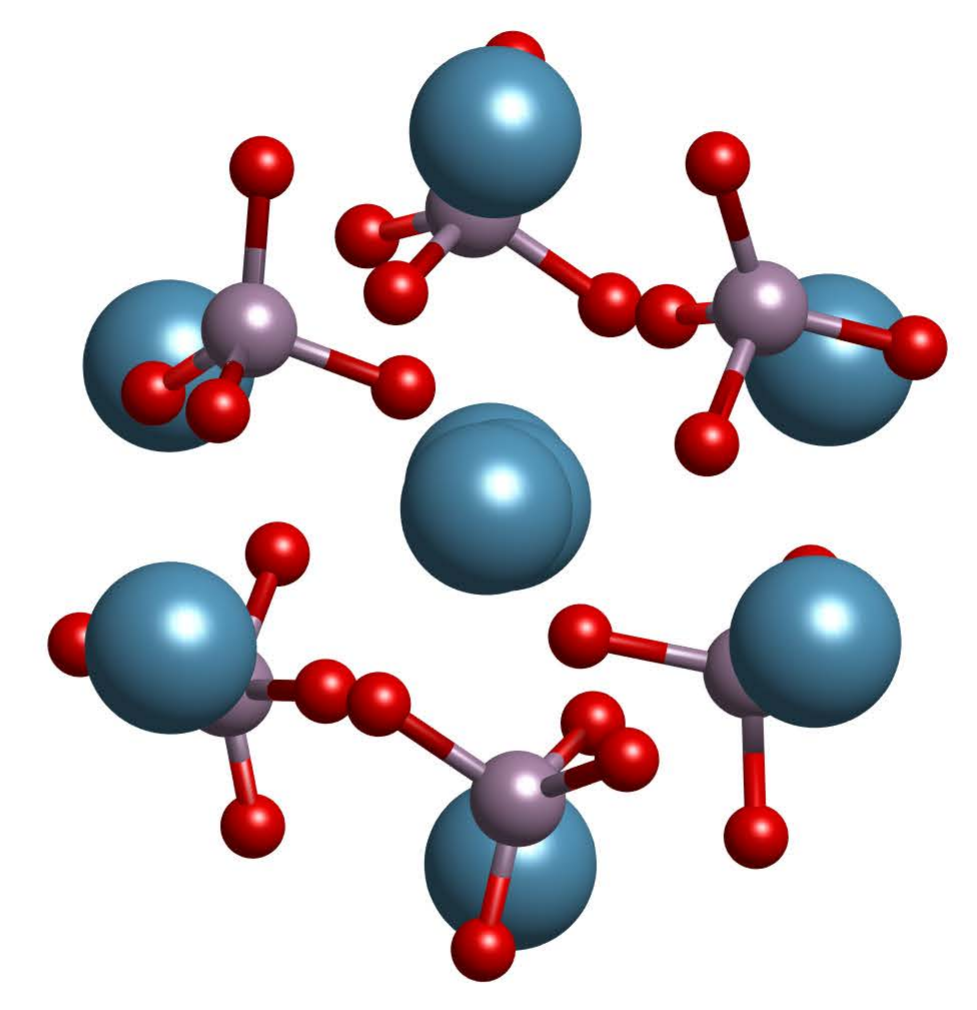}
      \captionsetup{font={small}}
      \caption{Configuration \textbf{C} -- Cluster 2 (\ce{C_i})}
      \label{fig:kmeans_C_2}
    \end{subfigure}
    \\
    \vspace{0.5cm}
    \begin{subfigure}[b]{0.45\textwidth}
      \centering
      \includegraphics[width=\textwidth]{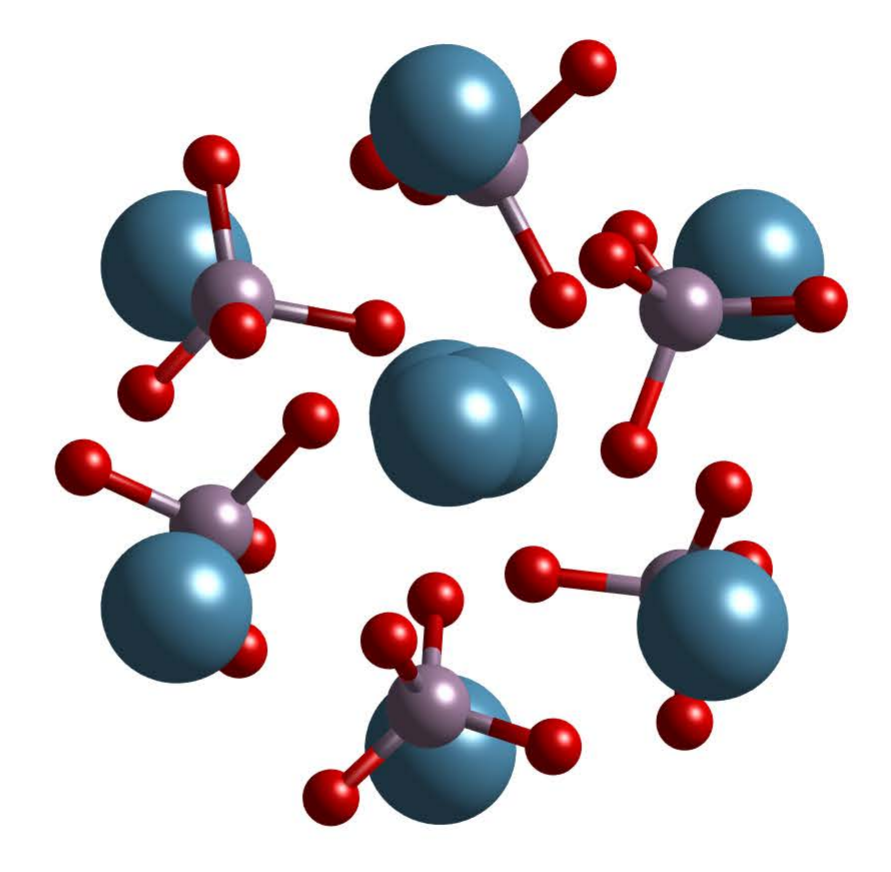}
      \captionsetup{font={small}}
      \caption{Configuration \textbf{D} -- Cluster 1 (\ce{C1})}
      \label{fig:kmeans_D_1}
    \end{subfigure}
    \hfill
    \begin{subfigure}[b]{0.45\textwidth}
      \centering
      \includegraphics[width=\textwidth]{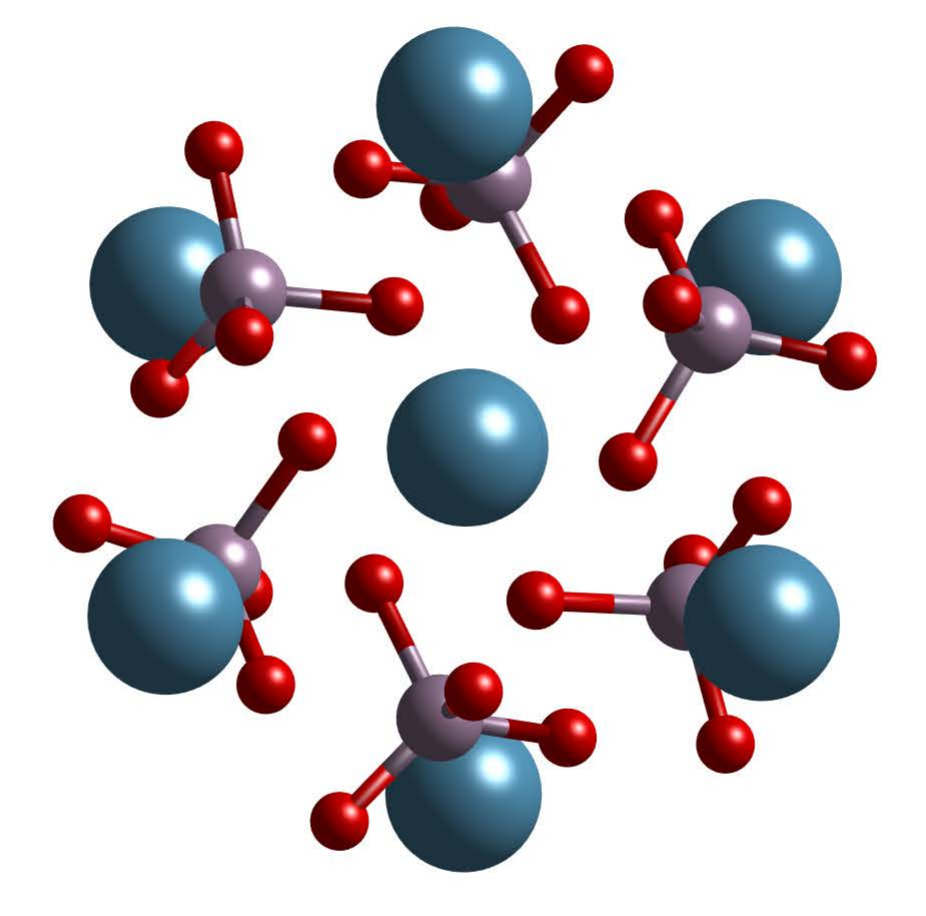}
      \captionsetup{font={small}}
      \caption{Configuration \textbf{D} -- Cluster 2 (\ce{C_i})}
      \label{fig:kmeans_D_2}
    \end{subfigure}
\end{figure}

\newpage

\begin{figure}[h!]\ContinuedFloat
    \centering
    \begin{subfigure}[b]{0.45\textwidth}
      \centering
      \includegraphics[width=\textwidth]{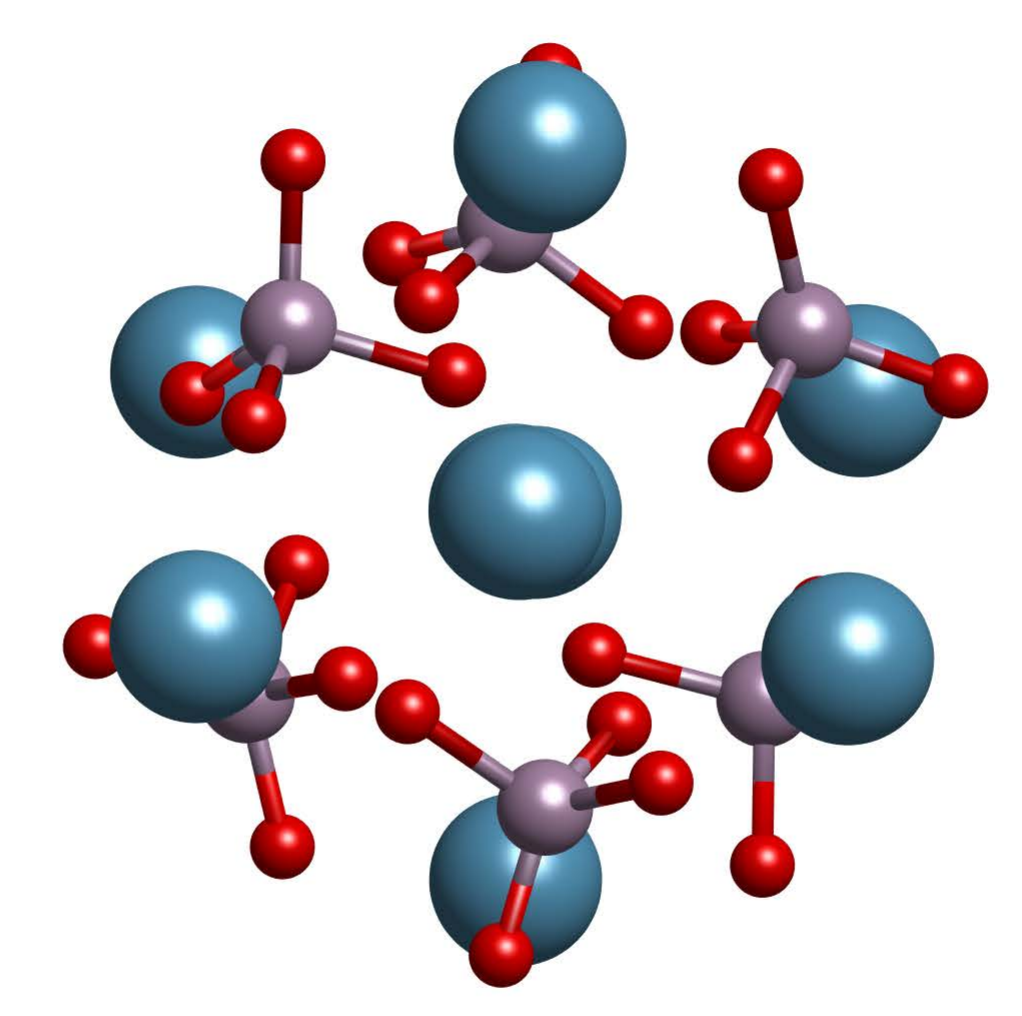}
      \captionsetup{font={small}}
      \caption{Configuration \textbf{E} -- Cluster 1 (\ce{C_i})}
      \label{fig:kmeans_E_1}
    \end{subfigure}
    \hfill
    \begin{subfigure}[b]{0.45\textwidth}
      \centering
      \includegraphics[width=\textwidth]{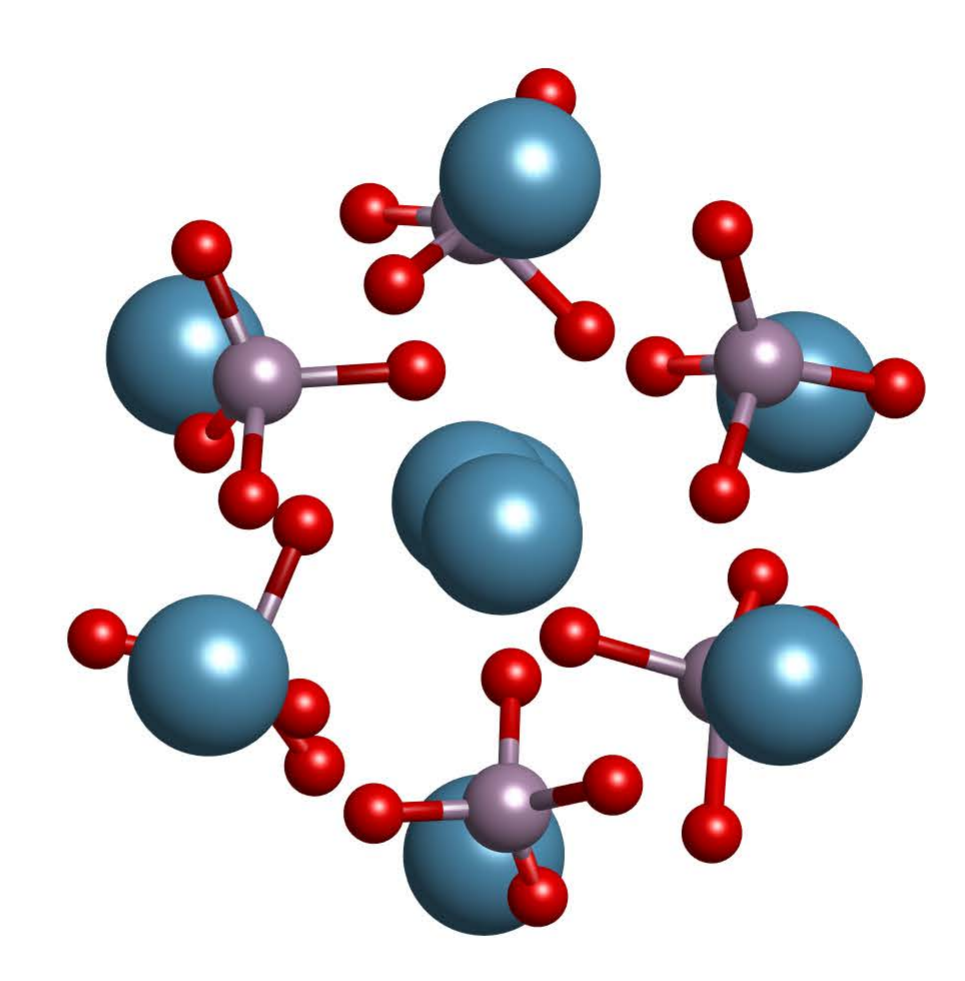}
      \captionsetup{font={small}}
      \caption{Configuration \textbf{E} -- Cluster 2 (\ce{C1})}
      \label{fig:kmeans_E_2}
    \end{subfigure}
    \\
    \vspace{0.5cm}
    \begin{subfigure}[b]{0.45\textwidth}
      \centering
      \includegraphics[width=\textwidth]{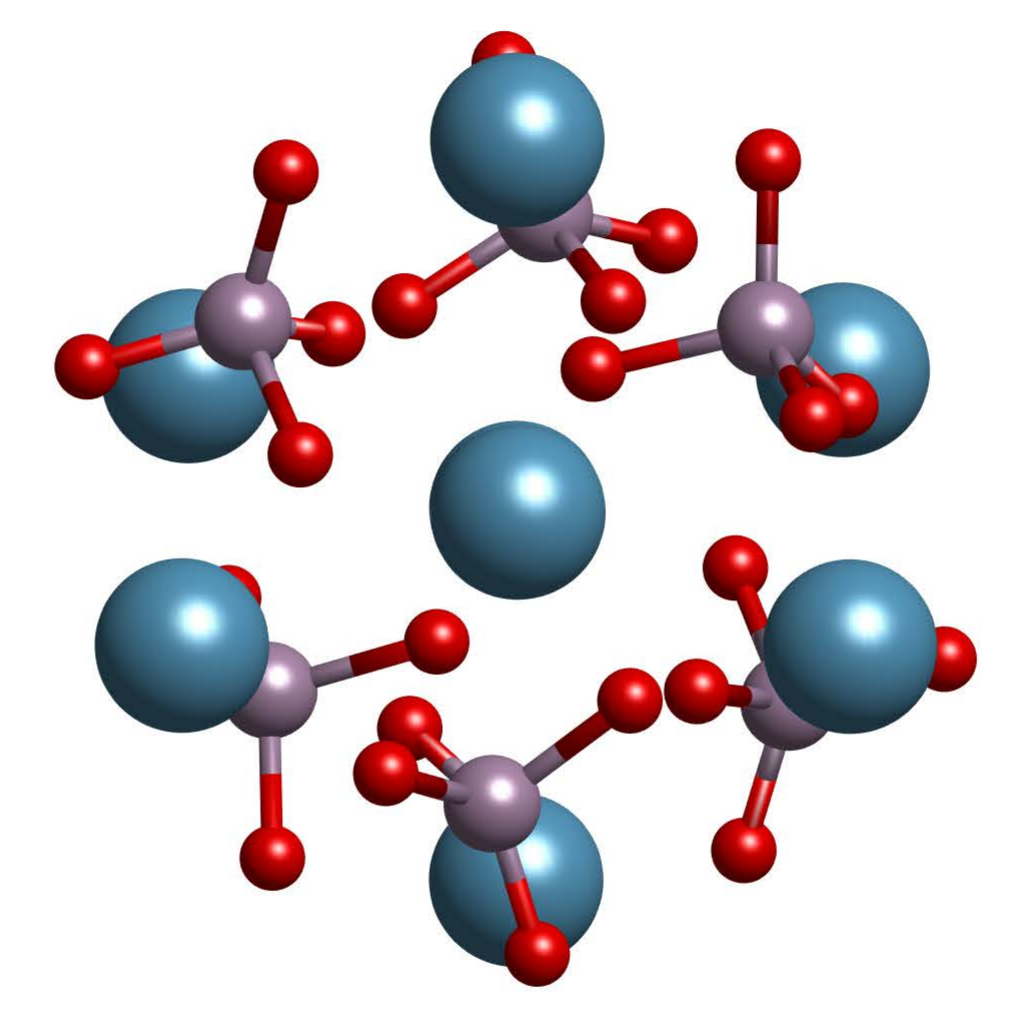}
      \captionsetup{font={small}}
      \caption{Configuration \textbf{F} -- Cluster 1 (\ce{C_i})}
      \label{fig:kmeans_F_1}
    \end{subfigure}
    \hfill
    \begin{subfigure}[b]{0.45\textwidth}
      \centering
      \includegraphics[width=\textwidth]{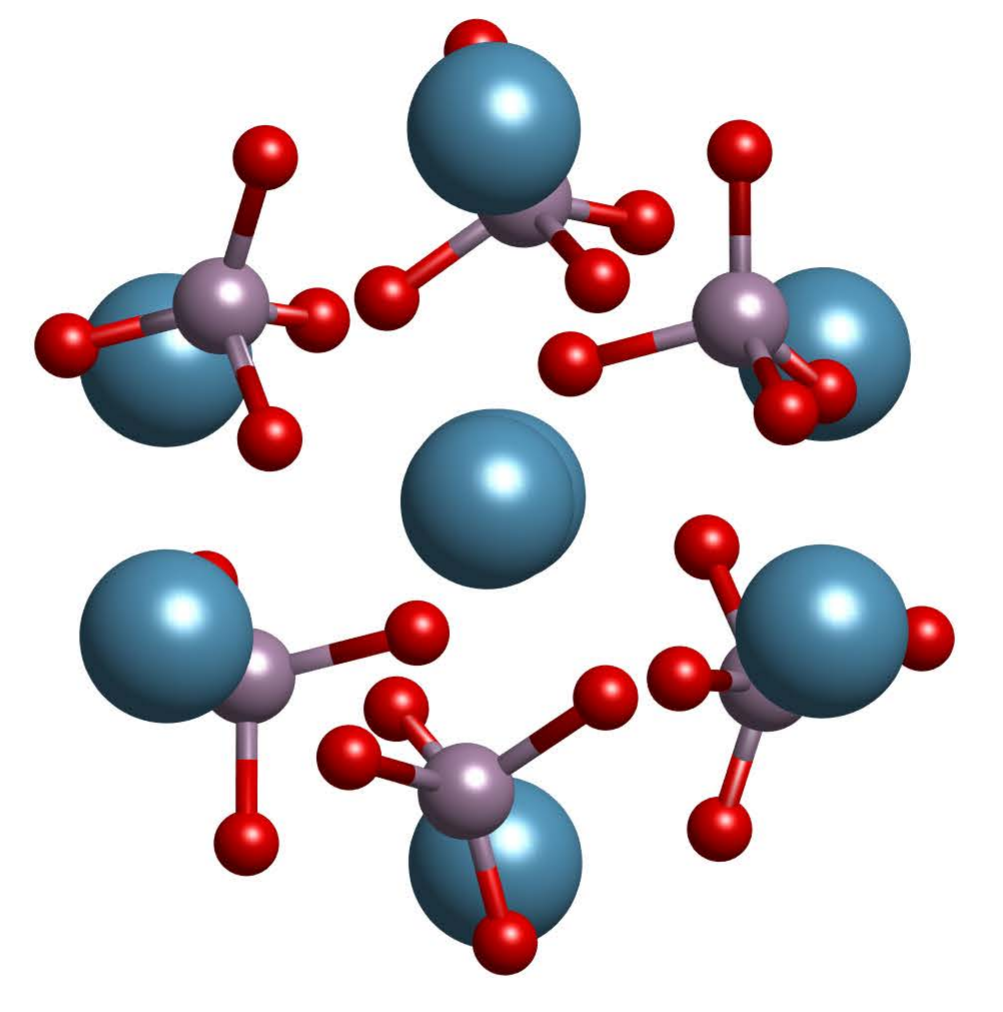}
      \captionsetup{font={small}}
      \caption{Configuration \textbf{F} -- Cluster 2 (\ce{C_i})}
      \label{fig:kmeans_F_2}
    \end{subfigure}
\end{figure}

\newpage

\begin{figure}[h!]\ContinuedFloat
    \centering
    \begin{subfigure}[b]{0.45\textwidth}
      \centering
      \includegraphics[width=\textwidth]{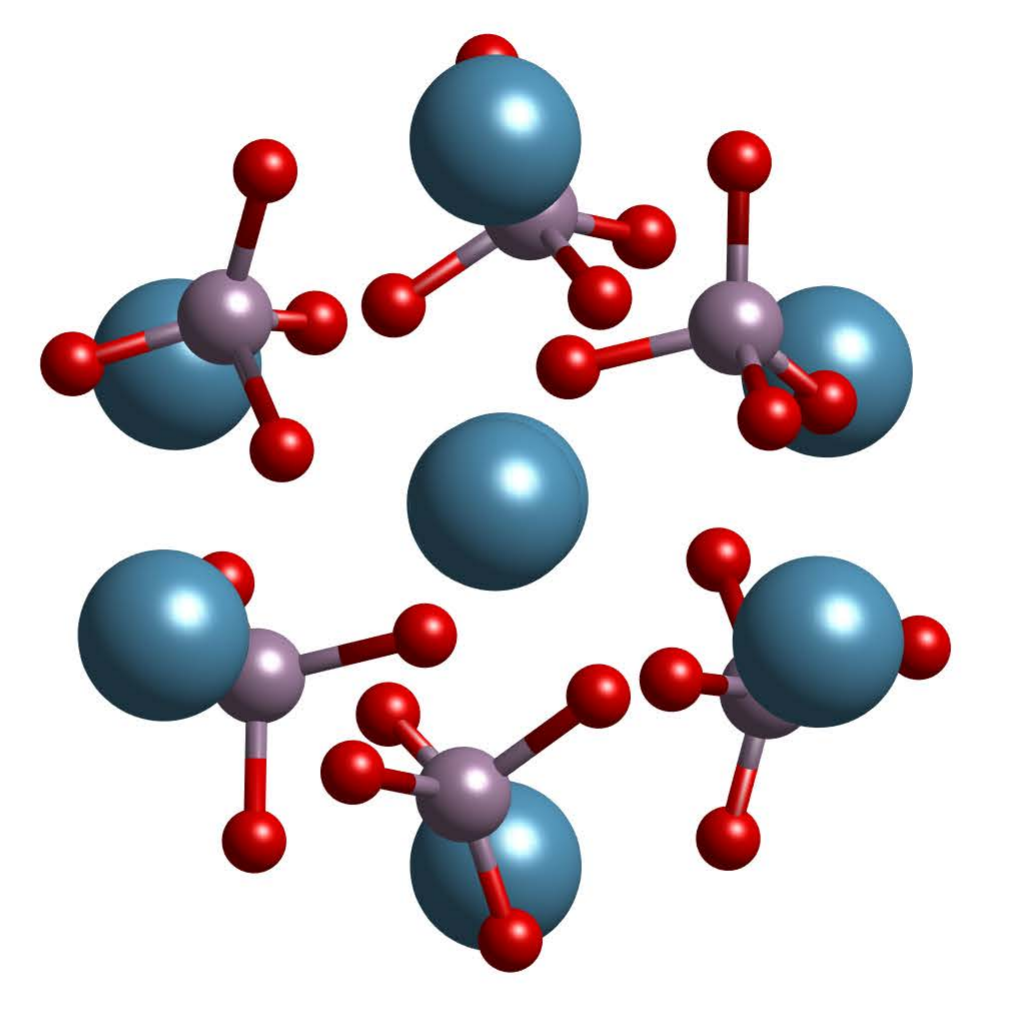}
      \captionsetup{font={small}}
      \caption{Configuration \textbf{G} -- Cluster 1 (\ce{C_i})}
      \label{fig:kmeans_G_1}
    \end{subfigure}
    \hfill
    \begin{subfigure}[b]{0.45\textwidth}
      \centering
      \includegraphics[width=\textwidth]{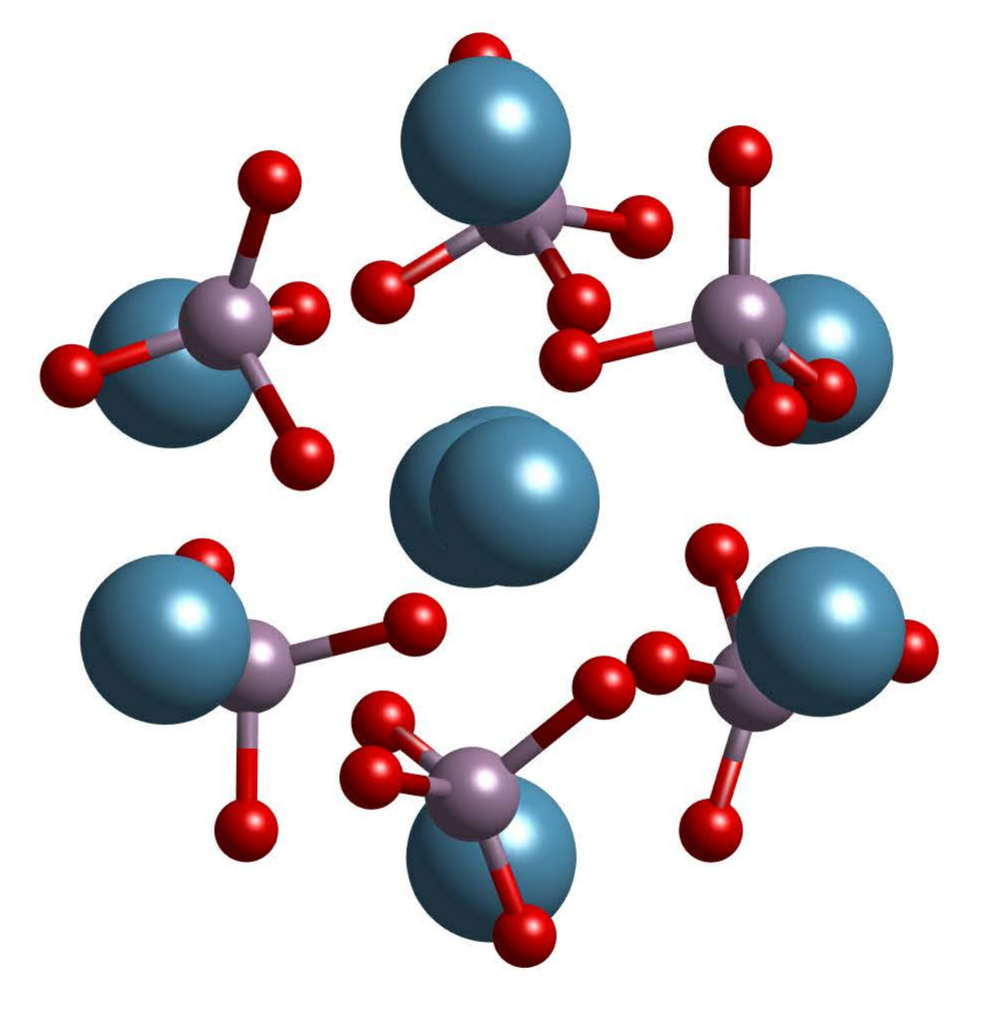}
      \captionsetup{font={small}}
      \caption{Configuration \textbf{G} -- Cluster 2 (\ce{C_i})}
      \label{fig:kmeans_G_2}
    \end{subfigure}
    \\
    \vspace{0.5cm}
    \begin{subfigure}[b]{0.45\textwidth}
      \centering
      \includegraphics[width=\textwidth]{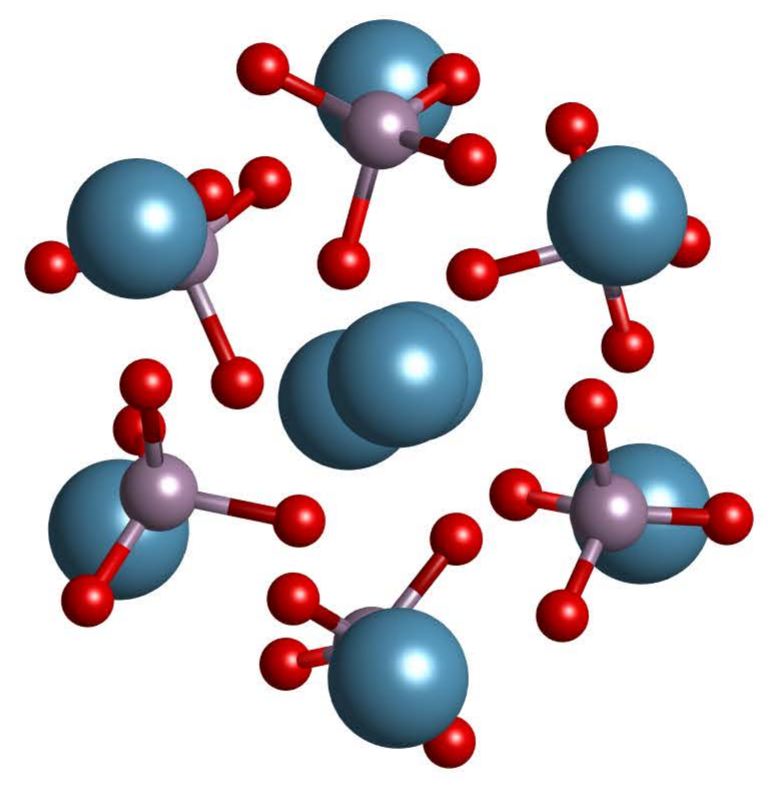}
      \captionsetup{font={small}}
      \caption{Configuration \textbf{H} -- Cluster 1 (\ce{C1})}
      \label{fig:kmeans_H_1}
    \end{subfigure}
    \hfill
    \begin{subfigure}[b]{0.45\textwidth}
      \centering
      \includegraphics[width=\textwidth]{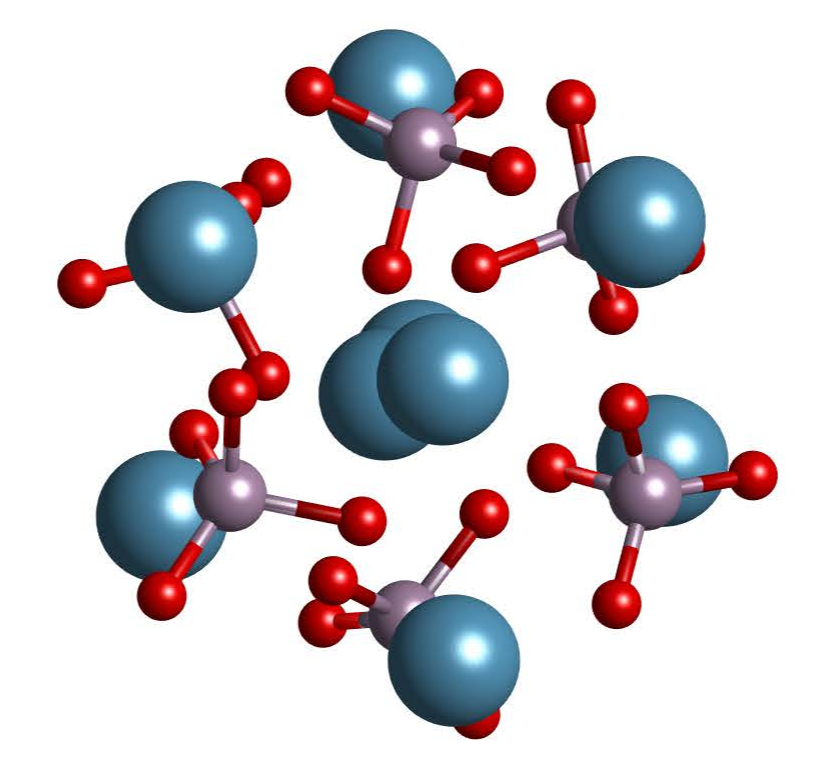}
      \captionsetup{font={small}}
      \caption{Configuration \textbf{H} -- Cluster 2 (\ce{C1})}
      \label{fig:kmeans_H_2}
    \end{subfigure}
    \caption{Structures, and their point group symmetries, corresponding to each unique cluster based on \textit{k}-means clustering for all of the $8$ dynamical runs at $T=298$K, as shown in Subsection \ref{subsec:T_298} of the SI}
    \label{fig:kmeans}
\end{figure}

\newpage

\section{Vibrational spectrum calculation on the 8 transition structures}

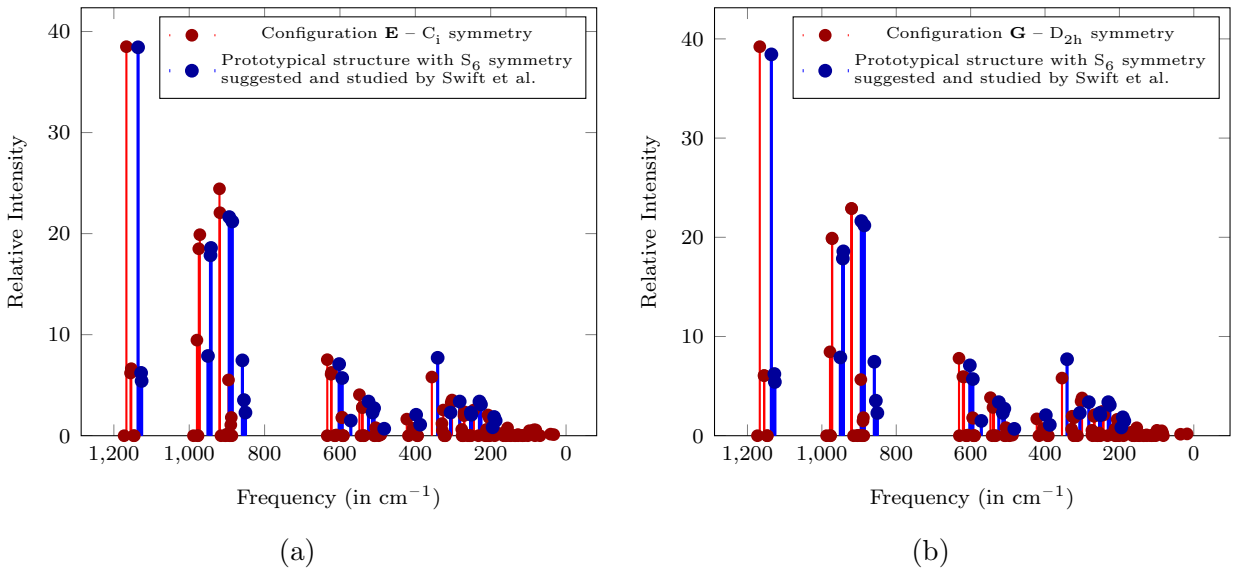
\begin{figure}[htb]
    \centering
    \begin{subfigure}[b]{0.49\textwidth}
    \centering
    \begin{tikzpicture}
        \begin{axis}[ylabel = {Relative Intensity},
        xlabel = {Frequency (in \ce{cm^{-1}})},
        xlabel style = {font=\scriptsize},
        ylabel style = {font=\scriptsize},
        yticklabel style = {font=\scriptsize},
        xticklabel style = {font=\scriptsize},
        x dir = reverse,
        ymin = 0,
        legend style={font=\tiny}]
            \addplot [ycomb, red, thick, mark=*, mark options={red!60!black}] table[x = freq, y expr = \thisrow{int}/45.1878, col sep = comma]{s6_v3_ir.csv};
            \addplot [ycomb, blue, very thick, mark=*, mark options={blue!60!black}] table[x = freq, y = int, col sep = comma]{swift.csv};
            \addlegendentry[align=left]{Configuration \textbf{E} -- \ce{C_i} symmetry}
            \addlegendentry[align=left]{Prototypical structure with \ce{S6} symmetry \\ suggested and studied by Swift et al.}
        \end{axis}
        \end{tikzpicture}
        \caption{ }
        \label{fig:IR_E}
    \end{subfigure}
    \hfill
    \begin{subfigure}[b]{0.49\textwidth}
    \centering
    \begin{tikzpicture}
        \begin{axis}[ylabel = {Relative Intensity},
        xlabel = {Frequency (in \ce{cm^{-1}})},
        xlabel style = {font=\scriptsize},
        ylabel style = {font=\scriptsize},
        yticklabel style = {font=\scriptsize},
        xticklabel style = {font=\scriptsize},
        x dir = reverse,
        ymin = 0,
        legend style={font=\tiny}]
            \addplot [ycomb, red, thick, mark=*, mark options={red!60!black}] table[x = freq, y expr = \thisrow{int}/45.1878, col sep = comma]{s6im1m_C3_ir_2.csv};
            \addplot [ycomb, blue, very thick, mark=*, mark options={blue!60!black}] table[x = freq, y = int, col sep = comma]{swift.csv};
            \addlegendentry[align=left]{Configuration \textbf{G} -- \ce{D_{2h}} symmetry}
            \addlegendentry[align=left]{Prototypical structure with \ce{S6} symmetry \\ suggested and studied by Swift et al.}
        \end{axis}
        \end{tikzpicture}
        \caption{ }
        \label{fig:IR_G}
    \end{subfigure}
    \caption{Comparison between the IR spectra of the structural configuration \textbf{E} and \textbf{G} used in this study, and of that obtained by Swift et al..}
    \label{fig:IR_comparison}
\end{figure}

We performed vibrational spectrum calculations on our eight transition state structures, and compared them with an existing spectrum \cite{swift2018posner} reported in a study that considered the \ce{S6} symmetric structure to be the prototypical structure for the Posner molecule. The results can be seen in Fig.~\ref{fig:IR_comparison}. The two spectra match closely up to a constant value of shift on the frequency axis. This shows that the structures considered in our study and in the Swift et al. study are quite similar in nature, up to the extent of bond stretching and rotation. However, the replication of the IR spectrum was only found in two cases. Moreover, the similarity in IR spectra is not necessarily translated to a similarity in the spin properties of the corresponding structures. It remains to be seen if and how the coupling network in these structures vary.

\newpage

\section{The most stable structure for the Posner molecule}

Based on our methods, the most energetically stable structure for the Ponser molecule is shown below, along with its atomic coordinates. The point group symmetry for the molecule is \ce{C1}. The atomic coordinates of the structure follow.

\begin{figure}[h!]
    \centering
    \includegraphics[height=7cm, width=7cm]{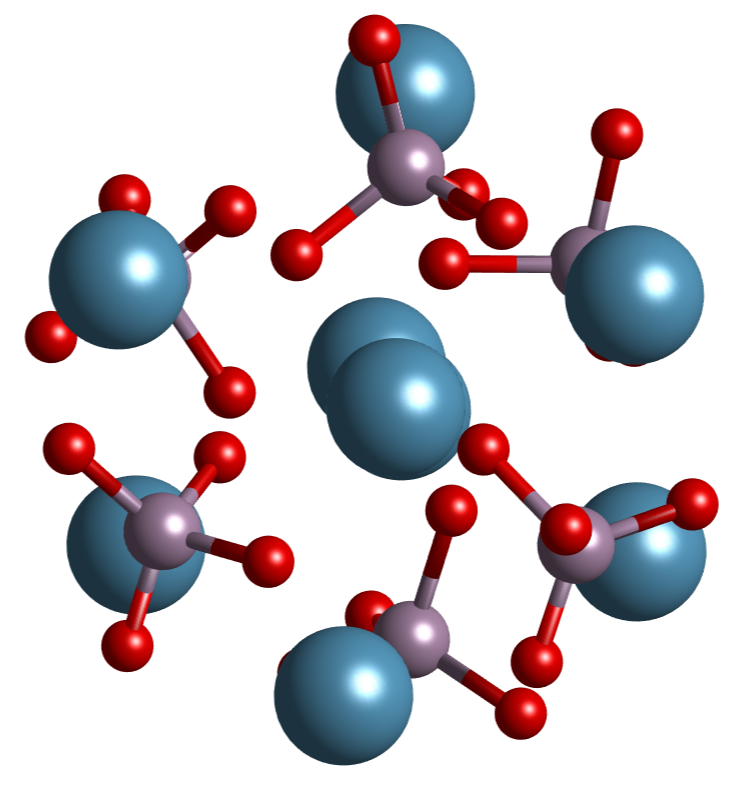}
    \caption{The most stable structure within the dynamical ensemble of the Posner molecule. It exhibits a \ce{C1} point group symmetry}
    \label{fig:most_stable}
\end{figure}

\noindent Ca        -0.280934         0.347084         0.193340 \\
Ca         1.508502        -2.344606         2.155304 \\
Ca         2.646092         1.155743        -1.879618 \\
Ca        -1.279140         2.740540         2.103531 \\
Ca         1.041487        -2.717414        -1.929450 \\
Ca        -2.579467        -1.519796         1.827644 \\
Ca        -1.187323         2.490355        -2.265939 \\
Ca         2.596607         1.188289         1.784528 \\
Ca        -2.380648        -1.463709        -1.958496 \\
P         0.683044         3.473472        -0.095908 \\
P        -0.985339        -3.305952         0.050447 \\
P         0.248536         0.089694         3.458331 \\
P         0.276538         0.155894        -3.391309 \\
P         3.180207        -1.159108         0.016301 \\
P        -3.426851         0.731476        -0.027032 \\
O         3.912056         0.210075        -0.049241 \\
O         1.720708        -0.718128         0.519375 \\
O         3.600734        -2.123039         1.067616  \\
O         2.970469        -1.686830        -1.389827 \\
O        -2.449428         0.802433         1.274461 \\
O        -2.429012         0.750779        -1.296764 \\
O        -4.416518         1.830870        -0.045555 \\
O        -3.879526        -0.782325        -0.013427 \\
O         1.886943         2.421146        -0.000008 \\
O        -0.690138         2.608707        -0.030545 \\
O         0.591547         4.294176         1.181998 \\
O         0.643123         4.031899        -1.517736 \\
O         0.527988        -3.536376         0.364236 \\
O        -1.153207        -1.700790        -0.047700 \\
O        -1.966341        -3.714708         1.136021 \\
O        -1.225738        -3.679952        -1.440521 \\
O         0.344310         1.235539         2.317993 \\
O        -0.470036        -1.145747         2.756416 \\
O         1.752509        -0.357924         3.540197 \\
O        -0.485926         0.652306         4.609227 \\
O         0.426364         0.813548        -1.924014 \\
O        -0.299597        -1.291276        -3.108499 \\
O         1.723365         0.127225        -3.870137 \\
O        -0.755655         1.046397        -4.075240

\newpage

%\bibliography{arXiv}

\begin{mcitethebibliography}{45}
\providecommand*\natexlab[1]{#1}
\providecommand*\mciteSetBstSublistMode[1]{}
\providecommand*\mciteSetBstMaxWidthForm[2]{}
\providecommand*\mciteBstWouldAddEndPuncttrue
  {\def\EndOfBibitem{\unskip.}}
\providecommand*\mciteBstWouldAddEndPunctfalse
  {\let\EndOfBibitem\relax}
\providecommand*\mciteSetBstMidEndSepPunct[3]{}
\providecommand*\mciteSetBstSublistLabelBeginEnd[3]{}
\providecommand*\EndOfBibitem{}
\mciteSetBstSublistMode{f}
\mciteSetBstMaxWidthForm{subitem}{(\alph{mcitesubitemcount})}
\mciteSetBstSublistLabelBeginEnd
  {\mcitemaxwidthsubitemform\space}
  {\relax}
  {\relax}

\bibitem[Posner and Betts(1975)Posner, and Betts]{posner1975synthetic}
Posner,~A.~S.; Betts,~F. Synthetic amorphous calcium phosphate and its relation
  to bone mineral structure. \emph{Accounts of Chemical Research}
  \textbf{1975}, \emph{8}, 273--281\relax
\mciteBstWouldAddEndPuncttrue
\mciteSetBstMidEndSepPunct{\mcitedefaultmidpunct}
{\mcitedefaultendpunct}{\mcitedefaultseppunct}\relax
\EndOfBibitem
\bibitem[Treboux \latin{et~al.}(2000)Treboux, Layrolle, Kanzaki, Onuma, and
  Ito]{treboux2000symmetry}
Treboux,~G.; Layrolle,~P.; Kanzaki,~N.; Onuma,~K.; Ito,~A. Symmetry of Posner's
  cluster. \emph{Journal of the American Chemical Society} \textbf{2000},
  \emph{122}, 8323--8324\relax
\mciteBstWouldAddEndPuncttrue
\mciteSetBstMidEndSepPunct{\mcitedefaultmidpunct}
{\mcitedefaultendpunct}{\mcitedefaultseppunct}\relax
\EndOfBibitem
\bibitem[Onuma and Ito(1998)Onuma, and Ito]{onuma1998cluster}
Onuma,~K.; Ito,~A. Cluster growth model for hydroxyapatite. \emph{Chemistry of
  materials} \textbf{1998}, \emph{10}, 3346--3351\relax
\mciteBstWouldAddEndPuncttrue
\mciteSetBstMidEndSepPunct{\mcitedefaultmidpunct}
{\mcitedefaultendpunct}{\mcitedefaultseppunct}\relax
\EndOfBibitem
\bibitem[Yin and Stott(2003)Yin, and Stott]{yin2003biological}
Yin,~X.; Stott,~M.~J. Biological calcium phosphates and Posner’s cluster.
  \emph{The Journal of Chemical Physics} \textbf{2003}, \emph{118},
  3717--3723\relax
\mciteBstWouldAddEndPuncttrue
\mciteSetBstMidEndSepPunct{\mcitedefaultmidpunct}
{\mcitedefaultendpunct}{\mcitedefaultseppunct}\relax
\EndOfBibitem
\bibitem[Du \latin{et~al.}(2013)Du, Bian, Gou, Jiang, Huang, Gao, Zhao, Wen,
  Zhang, and Wang]{du2013structure}
Du,~L.-W.; Bian,~S.; Gou,~B.-D.; Jiang,~Y.; Huang,~J.; Gao,~Y.-X.; Zhao,~Y.-D.;
  Wen,~W.; Zhang,~T.-L.; Wang,~K. Structure of clusters and formation of
  amorphous calcium phosphate and hydroxyapatite: from the perspective of
  coordination chemistry. \emph{Crystal growth \& design} \textbf{2013},
  \emph{13}, 3103--3109\relax
\mciteBstWouldAddEndPuncttrue
\mciteSetBstMidEndSepPunct{\mcitedefaultmidpunct}
{\mcitedefaultendpunct}{\mcitedefaultseppunct}\relax
\EndOfBibitem
\bibitem[Dey \latin{et~al.}(2010)Dey, Bomans, M{\"u}ller, Will, and
  Frederik]{dey2010g}
Dey,~A.; Bomans,~P.~H.; M{\"u}ller,~F.~A.; Will,~J.; Frederik,~P.~M. G. de With
  and NAJM Sommerdijk. \emph{Nat. Mater} \textbf{2010}, \emph{9},
  1010--1014\relax
\mciteBstWouldAddEndPuncttrue
\mciteSetBstMidEndSepPunct{\mcitedefaultmidpunct}
{\mcitedefaultendpunct}{\mcitedefaultseppunct}\relax
\EndOfBibitem
\bibitem[Wang \latin{et~al.}(2012)Wang, Li, Ruiz-Agudo, Putnis, and
  Putnis]{wang2012posner}
Wang,~L.; Li,~S.; Ruiz-Agudo,~E.; Putnis,~C.~V.; Putnis,~A. Posner's cluster
  revisited: direct imaging of nucleation and growth of nanoscale calcium
  phosphate clusters at the calcite-water interface. \emph{CrystEngComm}
  \textbf{2012}, \emph{14}, 6252--6256\relax
\mciteBstWouldAddEndPuncttrue
\mciteSetBstMidEndSepPunct{\mcitedefaultmidpunct}
{\mcitedefaultendpunct}{\mcitedefaultseppunct}\relax
\EndOfBibitem
\bibitem[Fisher(2015)]{fisher2015quantum}
Fisher,~M.~P. Quantum cognition: The possibility of processing with nuclear
  spins in the brain. \emph{Annals of Physics} \textbf{2015}, \emph{362},
  593--602\relax
\mciteBstWouldAddEndPuncttrue
\mciteSetBstMidEndSepPunct{\mcitedefaultmidpunct}
{\mcitedefaultendpunct}{\mcitedefaultseppunct}\relax
\EndOfBibitem
\bibitem[Weingarten \latin{et~al.}(2016)Weingarten, Doraiswamy, and
  Fisher]{weingarten2016new}
Weingarten,~C.~P.; Doraiswamy,~P.~M.; Fisher,~M. A new spin on neural
  processing: quantum cognition. \emph{Frontiers in human neuroscience}
  \textbf{2016}, \emph{10}, 541\relax
\mciteBstWouldAddEndPuncttrue
\mciteSetBstMidEndSepPunct{\mcitedefaultmidpunct}
{\mcitedefaultendpunct}{\mcitedefaultseppunct}\relax
\EndOfBibitem
\bibitem[Swift \latin{et~al.}(2018)Swift, Van~de Walle, and
  Fisher]{swift2018posner}
Swift,~M.~W.; Van~de Walle,~C.~G.; Fisher,~M.~P. Posner molecules: from atomic
  structure to nuclear spins. \emph{Physical Chemistry Chemical Physics}
  \textbf{2018}, \emph{20}, 12373--12380\relax
\mciteBstWouldAddEndPuncttrue
\mciteSetBstMidEndSepPunct{\mcitedefaultmidpunct}
{\mcitedefaultendpunct}{\mcitedefaultseppunct}\relax
\EndOfBibitem
\bibitem[Player and Hore(2018)Player, and Hore]{player2018posner}
Player,~T.~C.; Hore,~P. Posner qubits: spin dynamics of entangled Ca9 (PO4) 6
  molecules and their role in neural processing. \emph{Journal of the Royal
  Society Interface} \textbf{2018}, \emph{15}, 20180494\relax
\mciteBstWouldAddEndPuncttrue
\mciteSetBstMidEndSepPunct{\mcitedefaultmidpunct}
{\mcitedefaultendpunct}{\mcitedefaultseppunct}\relax
\EndOfBibitem
\bibitem[Buckingham and Love(1970)Buckingham, and Love]{buckingham1970theory}
Buckingham,~A.; Love,~I. Theory of the anisotropy of nuclear spin coupling.
  \emph{Journal of Magnetic Resonance (1969)} \textbf{1970}, \emph{2},
  338--351\relax
\mciteBstWouldAddEndPuncttrue
\mciteSetBstMidEndSepPunct{\mcitedefaultmidpunct}
{\mcitedefaultendpunct}{\mcitedefaultseppunct}\relax
\EndOfBibitem
\bibitem[Perras and Bryce(2013)Perras, and Bryce]{perras2013symmetry}
Perras,~F.~A.; Bryce,~D.~L. Symmetry-amplified J splittings for quadrupolar
  spin pairs: a solid-state NMR probe of homoatomic covalent bonds.
  \emph{Journal of the American Chemical Society} \textbf{2013}, \emph{135},
  12596--12599\relax
\mciteBstWouldAddEndPuncttrue
\mciteSetBstMidEndSepPunct{\mcitedefaultmidpunct}
{\mcitedefaultendpunct}{\mcitedefaultseppunct}\relax
\EndOfBibitem
\bibitem[Annabestani and Cory(2018)Annabestani, and
  Cory]{annabestani2018dipolar}
Annabestani,~R.; Cory,~D.~G. Dipolar relaxation mechanism of long-lived states
  of methyl groups. \emph{Quantum information processing} \textbf{2018},
  \emph{17}, 1--25\relax
\mciteBstWouldAddEndPuncttrue
\mciteSetBstMidEndSepPunct{\mcitedefaultmidpunct}
{\mcitedefaultendpunct}{\mcitedefaultseppunct}\relax
\EndOfBibitem
\bibitem[Lidar(2012)]{lidar2012review}
Lidar,~D.~A. Review of decoherence free subspaces, noiseless subsystems, and
  dynamical decoupling. \emph{arXiv preprint arXiv:1208.5791} \textbf{2012},
  \relax
\mciteBstWouldAddEndPunctfalse
\mciteSetBstMidEndSepPunct{\mcitedefaultmidpunct}
{}{\mcitedefaultseppunct}\relax
\EndOfBibitem
\bibitem[Feng \latin{et~al.}(2014)Feng, Theis, Wu, Claytor, and
  Warren]{feng2014long}
Feng,~Y.; Theis,~T.; Wu,~T.-L.; Claytor,~K.; Warren,~W.~S. Long-lived
  polarization protected by symmetry. \emph{The Journal of chemical physics}
  \textbf{2014}, \emph{141}, 134307\relax
\mciteBstWouldAddEndPuncttrue
\mciteSetBstMidEndSepPunct{\mcitedefaultmidpunct}
{\mcitedefaultendpunct}{\mcitedefaultseppunct}\relax
\EndOfBibitem
\bibitem[Vinogradov and Grant(2007)Vinogradov, and Grant]{vinogradov2007long}
Vinogradov,~E.; Grant,~A.~K. Long-lived states in solution NMR: Selection rules
  for intramolecular dipolar relaxation in low magnetic fields. \emph{Journal
  of Magnetic Resonance} \textbf{2007}, \emph{188}, 176--182\relax
\mciteBstWouldAddEndPuncttrue
\mciteSetBstMidEndSepPunct{\mcitedefaultmidpunct}
{\mcitedefaultendpunct}{\mcitedefaultseppunct}\relax
\EndOfBibitem
\bibitem[Stevanato \latin{et~al.}(2015)Stevanato, Roy, Hill-Cousins, Kuprov,
  Brown, Brown, Pileio, and Levitt]{stevanato2015long}
Stevanato,~G.; Roy,~S.~S.; Hill-Cousins,~J.; Kuprov,~I.; Brown,~L.~J.;
  Brown,~R.~C.; Pileio,~G.; Levitt,~M.~H. Long-lived nuclear spin states far
  from magnetic equivalence. \emph{Physical Chemistry Chemical Physics}
  \textbf{2015}, \emph{17}, 5913--5922\relax
\mciteBstWouldAddEndPuncttrue
\mciteSetBstMidEndSepPunct{\mcitedefaultmidpunct}
{\mcitedefaultendpunct}{\mcitedefaultseppunct}\relax
\EndOfBibitem
\bibitem[Treboux \latin{et~al.}(2000)Treboux, Layrolle, Kanzaki, Onuma, and
  Ito]{treboux2000existence}
Treboux,~G.; Layrolle,~P.; Kanzaki,~N.; Onuma,~K.; Ito,~A. Existence of
  Posner's cluster in vacuum. \emph{The Journal of Physical Chemistry A}
  \textbf{2000}, \emph{104}, 5111--5114\relax
\mciteBstWouldAddEndPuncttrue
\mciteSetBstMidEndSepPunct{\mcitedefaultmidpunct}
{\mcitedefaultendpunct}{\mcitedefaultseppunct}\relax
\EndOfBibitem
\bibitem[Lin and Chiu(2018)Lin, and Chiu]{lin2018structures}
Lin,~T.-J.; Chiu,~C.-C. Structures and infrared spectra of calcium phosphate
  clusters by ab initio methods with implicit solvation models. \emph{Physical
  Chemistry Chemical Physics} \textbf{2018}, \emph{20}, 345--356\relax
\mciteBstWouldAddEndPuncttrue
\mciteSetBstMidEndSepPunct{\mcitedefaultmidpunct}
{\mcitedefaultendpunct}{\mcitedefaultseppunct}\relax
\EndOfBibitem
\bibitem[Swift(2020)]{Swift2020pvtcomms}
Swift,~M. Private Communication. 2020\relax
\mciteBstWouldAddEndPuncttrue
\mciteSetBstMidEndSepPunct{\mcitedefaultmidpunct}
{\mcitedefaultendpunct}{\mcitedefaultseppunct}\relax
\EndOfBibitem
\bibitem[Kanzaki \latin{et~al.}(2001)Kanzaki, Treboux, Onuma, Tsutsumi, and
  Ito]{kanzaki2001calcium}
Kanzaki,~N.; Treboux,~G.; Onuma,~K.; Tsutsumi,~S.; Ito,~A. Calcium phosphate
  clusters. \emph{Biomaterials} \textbf{2001}, \emph{22}, 2921--2929\relax
\mciteBstWouldAddEndPuncttrue
\mciteSetBstMidEndSepPunct{\mcitedefaultmidpunct}
{\mcitedefaultendpunct}{\mcitedefaultseppunct}\relax
\EndOfBibitem
\bibitem[Mancardi \latin{et~al.}(2017)Mancardi, Tamargo, Di~Tommaso, and
  De~Leeuw]{mancardi2017detection}
Mancardi,~G.; Tamargo,~C. E.~H.; Di~Tommaso,~D.; De~Leeuw,~N.~H. Detection of
  Posner's clusters during calcium phosphate nucleation: a molecular dynamics
  study. \emph{Journal of Materials Chemistry B} \textbf{2017}, \emph{5},
  7274--7284\relax
\mciteBstWouldAddEndPuncttrue
\mciteSetBstMidEndSepPunct{\mcitedefaultmidpunct}
{\mcitedefaultendpunct}{\mcitedefaultseppunct}\relax
\EndOfBibitem
\bibitem[Stokely and Votapka(2019)Stokely, and Votapka]{stokely2019molecular}
Stokely,~A.; Votapka,~L. Molecular dynamics investigation into the effect of
  phosphorus nuclear spin state on the pyrophosphatase-catalyzed hydrolysis of
  pyrophosphate. \textbf{2019}, \relax
\mciteBstWouldAddEndPunctfalse
\mciteSetBstMidEndSepPunct{\mcitedefaultmidpunct}
{}{\mcitedefaultseppunct}\relax
\EndOfBibitem
\bibitem[Ainsworth \latin{et~al.}(2012)Ainsworth, Tommaso, Christie, and
  de~Leeuw]{ainsworth2012polarizable}
Ainsworth,~R.~I.; Tommaso,~D.~D.; Christie,~J.~K.; de~Leeuw,~N.~H. Polarizable
  force field development and molecular dynamics study of phosphate-based
  glasses. \emph{The Journal of chemical physics} \textbf{2012}, \emph{137},
  234502\relax
\mciteBstWouldAddEndPuncttrue
\mciteSetBstMidEndSepPunct{\mcitedefaultmidpunct}
{\mcitedefaultendpunct}{\mcitedefaultseppunct}\relax
\EndOfBibitem
\bibitem[Demichelis \latin{et~al.}(2018)Demichelis, Garcia, Raiteri,
  Innocenti~Malini, Freeman, Harding, and Gale]{demichelis2018simulation}
Demichelis,~R.; Garcia,~N.~A.; Raiteri,~P.; Innocenti~Malini,~R.;
  Freeman,~C.~L.; Harding,~J.~H.; Gale,~J.~D. Simulation of calcium phosphate
  species in aqueous solution: force field derivation. \emph{The Journal of
  Physical Chemistry B} \textbf{2018}, \emph{122}, 1471--1483\relax
\mciteBstWouldAddEndPuncttrue
\mciteSetBstMidEndSepPunct{\mcitedefaultmidpunct}
{\mcitedefaultendpunct}{\mcitedefaultseppunct}\relax
\EndOfBibitem
\bibitem[Truong and Stefanovich(1995)Truong, and Stefanovich]{truong1995new}
Truong,~T.~N.; Stefanovich,~E.~V. A new method for incorporating solvent effect
  into the classical, ab initio molecular orbital and density functional theory
  frameworks for arbitrary shape cavity. \emph{Chemical Physics Letters}
  \textbf{1995}, \emph{240}, 253--260\relax
\mciteBstWouldAddEndPuncttrue
\mciteSetBstMidEndSepPunct{\mcitedefaultmidpunct}
{\mcitedefaultendpunct}{\mcitedefaultseppunct}\relax
\EndOfBibitem
\bibitem[Barone and Cossi(1998)Barone, and Cossi]{barone1998quantum}
Barone,~V.; Cossi,~M. Quantum calculation of molecular energies and energy
  gradients in solution by a conductor solvent model. \emph{The Journal of
  Physical Chemistry A} \textbf{1998}, \emph{102}, 1995--2001\relax
\mciteBstWouldAddEndPuncttrue
\mciteSetBstMidEndSepPunct{\mcitedefaultmidpunct}
{\mcitedefaultendpunct}{\mcitedefaultseppunct}\relax
\EndOfBibitem
\bibitem[Grimme \latin{et~al.}(2010)Grimme, Antony, Ehrlich, and
  Krieg]{grimme2010consistent}
Grimme,~S.; Antony,~J.; Ehrlich,~S.; Krieg,~H. A consistent and accurate ab
  initio parametrization of density functional dispersion correction (DFT-D)
  for the 94 elements H-Pu. \emph{The Journal of chemical physics}
  \textbf{2010}, \emph{132}, 154104\relax
\mciteBstWouldAddEndPuncttrue
\mciteSetBstMidEndSepPunct{\mcitedefaultmidpunct}
{\mcitedefaultendpunct}{\mcitedefaultseppunct}\relax
\EndOfBibitem
\bibitem[Roohani \latin{et~al.}(2021)Roohani, Cheong, and
  Wang]{roohani2021build}
Roohani,~I.; Cheong,~S.; Wang,~A. How to Build a Bone?-Hydroxyapatite or
  Posners Clusters as Bone Minerals. \emph{Open Ceramics} \textbf{2021},
  100092\relax
\mciteBstWouldAddEndPuncttrue
\mciteSetBstMidEndSepPunct{\mcitedefaultmidpunct}
{\mcitedefaultendpunct}{\mcitedefaultseppunct}\relax
\EndOfBibitem
\bibitem[Humphrey \latin{et~al.}(1996)Humphrey, Dalke, and Schulten]{HUMP96}
Humphrey,~W.; Dalke,~A.; Schulten,~K. {VMD} -- {V}isual {M}olecular {D}ynamics.
  \emph{Journal of Molecular Graphics} \textbf{1996}, \emph{14}, 33--38\relax
\mciteBstWouldAddEndPuncttrue
\mciteSetBstMidEndSepPunct{\mcitedefaultmidpunct}
{\mcitedefaultendpunct}{\mcitedefaultseppunct}\relax
\EndOfBibitem
\bibitem[Schmidt and Polik(2020)Schmidt, and Polik]{schmidt2020webmo}
Schmidt,~J.; Polik,~W. WebMO Enterprise, version 20.0. \emph{WebMO, LLC,
  Holland, MI, USA. http://www. webmo. net (accessed 2020). Google Scholar
  There is no corresponding record for this reference} \textbf{2020}, \relax
\mciteBstWouldAddEndPunctfalse
\mciteSetBstMidEndSepPunct{\mcitedefaultmidpunct}
{}{\mcitedefaultseppunct}\relax
\EndOfBibitem
\bibitem[Weigend and Ahlrichs(2005)Weigend, and Ahlrichs]{weigend2005balanced}
Weigend,~F.; Ahlrichs,~R. Balanced basis sets of split valence, triple zeta
  valence and quadruple zeta valence quality for H to Rn: Design and assessment
  of accuracy. \emph{Physical Chemistry Chemical Physics} \textbf{2005},
  \emph{7}, 3297--3305\relax
\mciteBstWouldAddEndPuncttrue
\mciteSetBstMidEndSepPunct{\mcitedefaultmidpunct}
{\mcitedefaultendpunct}{\mcitedefaultseppunct}\relax
\EndOfBibitem
\bibitem[Weigend(2006)]{weigend2006accurate}
Weigend,~F. Accurate Coulomb-fitting basis sets for H to Rn. \emph{Physical
  chemistry chemical physics} \textbf{2006}, \emph{8}, 1057--1065\relax
\mciteBstWouldAddEndPuncttrue
\mciteSetBstMidEndSepPunct{\mcitedefaultmidpunct}
{\mcitedefaultendpunct}{\mcitedefaultseppunct}\relax
\EndOfBibitem
\bibitem[Giannozzi \latin{et~al.}(2009)Giannozzi, Baroni, Bonini, Calandra,
  Car, Cavazzoni, Ceresoli, Chiarotti, Cococcioni, Dabo, {Dal Corso},
  de~Gironcoli, Fabris, Fratesi, Gebauer, Gerstmann, Gougoussis, Kokalj,
  Lazzeri, Martin-Samos, Marzari, Mauri, Mazzarello, Paolini, Pasquarello,
  Paulatto, Sbraccia, Scandolo, Sclauzero, Seitsonen, Smogunov, Umari, and
  Wentzcovitch]{QE-2009}
Giannozzi,~P. \latin{et~al.}  QUANTUM ESPRESSO: a modular and open-source
  software project for quantum simulations of materials. \emph{Journal of
  Physics: Condensed Matter} \textbf{2009}, \emph{21}, 395502 (19pp)\relax
\mciteBstWouldAddEndPuncttrue
\mciteSetBstMidEndSepPunct{\mcitedefaultmidpunct}
{\mcitedefaultendpunct}{\mcitedefaultseppunct}\relax
\EndOfBibitem
\bibitem[Prandini \latin{et~al.}(2018)Prandini, Marrazzo, Castelli, Mounet, and
  Marzari]{prandini2018precision}
Prandini,~G.; Marrazzo,~A.; Castelli,~I.~E.; Mounet,~N.; Marzari,~N. Precision
  and efficiency in solid-state pseudopotential calculations. \emph{npj
  Computational Materials} \textbf{2018}, \emph{4}, 1--13\relax
\mciteBstWouldAddEndPuncttrue
\mciteSetBstMidEndSepPunct{\mcitedefaultmidpunct}
{\mcitedefaultendpunct}{\mcitedefaultseppunct}\relax
\EndOfBibitem
\bibitem[Lejaeghere \latin{et~al.}(2016)Lejaeghere, Bihlmayer, Bj{\"o}rkman,
  Blaha, Bl{\"u}gel, Blum, Caliste, Castelli, Clark, Dal~Corso, \latin{et~al.}
  others]{lejaeghere2016reproducibility}
Lejaeghere,~K.; Bihlmayer,~G.; Bj{\"o}rkman,~T.; Blaha,~P.; Bl{\"u}gel,~S.;
  Blum,~V.; Caliste,~D.; Castelli,~I.~E.; Clark,~S.~J.; Dal~Corso,~A.,
  \latin{et~al.}  Reproducibility in density functional theory calculations of
  solids. \emph{Science} \textbf{2016}, \emph{351}\relax
\mciteBstWouldAddEndPuncttrue
\mciteSetBstMidEndSepPunct{\mcitedefaultmidpunct}
{\mcitedefaultendpunct}{\mcitedefaultseppunct}\relax
\EndOfBibitem
\bibitem[Shao \latin{et~al.}(2015)Shao, Gan, Epifanovsky, Gilbert, Wormit,
  Kussmann, Lange, Behn, Deng, Feng, \latin{et~al.} others]{shao2015advances}
Shao,~Y.; Gan,~Z.; Epifanovsky,~E.; Gilbert,~A.~T.; Wormit,~M.; Kussmann,~J.;
  Lange,~A.~W.; Behn,~A.; Deng,~J.; Feng,~X., \latin{et~al.}  Advances in
  molecular quantum chemistry contained in the Q-Chem 4 program package.
  \emph{Molecular Physics} \textbf{2015}, \emph{113}, 184--215\relax
\mciteBstWouldAddEndPuncttrue
\mciteSetBstMidEndSepPunct{\mcitedefaultmidpunct}
{\mcitedefaultendpunct}{\mcitedefaultseppunct}\relax
\EndOfBibitem
\bibitem[Lange and Herbert(2010)Lange, and Herbert]{lange2010smooth}
Lange,~A.~W.; Herbert,~J.~M. A smooth, nonsingular, and faithful discretization
  scheme for polarizable continuum models: The switching/Gaussian approach.
  \emph{The Journal of chemical physics} \textbf{2010}, \emph{133},
  244111\relax
\mciteBstWouldAddEndPuncttrue
\mciteSetBstMidEndSepPunct{\mcitedefaultmidpunct}
{\mcitedefaultendpunct}{\mcitedefaultseppunct}\relax
\EndOfBibitem
\bibitem[Pulay(1980)]{pulay1980convergence}
Pulay,~P. Convergence acceleration of iterative sequences. The case of SCF
  iteration. \emph{Chemical Physics Letters} \textbf{1980}, \emph{73},
  393--398\relax
\mciteBstWouldAddEndPuncttrue
\mciteSetBstMidEndSepPunct{\mcitedefaultmidpunct}
{\mcitedefaultendpunct}{\mcitedefaultseppunct}\relax
\EndOfBibitem
\bibitem[Pulay(1982)]{pulay1982improved}
Pulay,~P. Improved SCF convergence acceleration. \emph{Journal of Computational
  Chemistry} \textbf{1982}, \emph{3}, 556--560\relax
\mciteBstWouldAddEndPuncttrue
\mciteSetBstMidEndSepPunct{\mcitedefaultmidpunct}
{\mcitedefaultendpunct}{\mcitedefaultseppunct}\relax
\EndOfBibitem
\bibitem[Van~Voorhis and Head-Gordon(2002)Van~Voorhis, and
  Head-Gordon]{van2002geometric}
Van~Voorhis,~T.; Head-Gordon,~M. A geometric approach to direct minimization.
  \emph{Molecular Physics} \textbf{2002}, \emph{100}, 1713--1721\relax
\mciteBstWouldAddEndPuncttrue
\mciteSetBstMidEndSepPunct{\mcitedefaultmidpunct}
{\mcitedefaultendpunct}{\mcitedefaultseppunct}\relax
\EndOfBibitem
\bibitem[Smith \latin{et~al.}(2002)Smith, Yong, and Rodger]{smith2002dl_poly}
Smith,~W.; Yong,~C.; Rodger,~P. DL\_POLY: Application to molecular simulation.
  \emph{Molecular Simulation} \textbf{2002}, \emph{28}, 385--471\relax
\mciteBstWouldAddEndPuncttrue
\mciteSetBstMidEndSepPunct{\mcitedefaultmidpunct}
{\mcitedefaultendpunct}{\mcitedefaultseppunct}\relax
\EndOfBibitem
\end{mcitethebibliography}
\providecommand{\latin}[1]{#1}
\makeatletter
\providecommand{\doi}
  {\begingroup\let\do\@makeother\dospecials
  \catcode`\{=1 \catcode`\}=2 \doi@aux}
\providecommand{\doi@aux}[1]{\endgroup\texttt{#1}}
\makeatother
\providecommand*\mcitethebibliography{\thebibliography}
\csname @ifundefined\endcsname{endmcitethebibliography}
  {\let\endmcitethebibliography\endthebibliography}{}

\end{document}